\documentclass[a4paper,11pt]{article}
\pdfoutput=1 

\usepackage{jheppub} 

\usepackage{soul}
\usepackage{color}

\usepackage{empheq}
\usepackage{mathrsfs} 
\usepackage{url}
\usepackage[utf8]{inputenc}
\usepackage[toc,page]{appendix}
\usepackage[autostyle]{csquotes}
\usepackage{slashed}
\usepackage{amssymb}
\usepackage{graphicx}
\usepackage{longtable}
\usepackage{array}
\usepackage{bm} 
\usepackage{amsmath}
\usepackage{nccmath}
\usepackage{cancel}
\usepackage{mathtools}
\usepackage{amsfonts}
\usepackage{amssymb}
\usepackage[labelfont=bf]{caption}
\usepackage{paralist} 
\usepackage{hyperref}
\usepackage{cleveref}
\DeclareMathAlphabet{\mathpzc}{OT1}{pzc}{m}{it} 
\usepackage{tensor}
\usepackage{stmaryrd}
\title{Logarithmic correction to the entropy of a Kerr-Newman family of black holes in $U(1)^2$-charged STU supergravity models}

\author[a]{Sudip Karan,}
\author[b]{Gurmeet Singh Punia}
\author[b]{and Surajit Biswas}
\affiliation[a]{Department of Physics, Indian Institute of Technology Guwahati, Guwahati 781039, Assam, India\\}
\affiliation[b]{Indian Institute of Science Education and Research Bhopal,\\ Bhopal Bypass Road, Bhauri, Bhopal 420066, India\\}

\emailAdd{sudip.karan@iitg.ac.in}
\emailAdd{gurmeet17@iiserb.ac.in}
\emailAdd{surajit18@iiserb.ac.in}

\abstract{
The leading quantum-gravitational correction to the black hole entropy is known to be a universal logarithmic term. In this study, we investigate the logarithmic corrections for the black holes in the STU supergravity models, which are a bosonic truncation into a specific class of $U(1)^2$-charged Einstein-Maxwell-dilaton theory. We demonstrate how the entire Kerr-Newman-AdS and Kerr-Newman family of black holes can be recovered within the gauged and ungauged STU supergravity models as special embedding choices in 4D. Logarithmic corrections are computed using two distinct Euclidean quantum gravity setups for extremal and non-extremal limits of all embedded rotating, static, charged, and neutral black holes. Our calculations employ the on-shell heat kernel method based Seeley-DeWitt expansion computations. Notably, all the AdS$_4$ results exhibit a confirmed non-topological nature as compared to the flat counterparts, offering a natural and more comprehensive ``infrared window into the microstates'' of black holes.}


\arxivnumber{2403.11823}

\dedicated{\href{https://doi.org/10.1103/PhysRevD.111.066016}{https://doi.org/10.1103/PhysRevD.111.066016}}

\begin{document} 
\maketitle
\flushbottom


\section{Introduction}\label{intro}
In the realm of Einstein's gravity, the entropy of black holes is universally attributed as one-quarter of the horizon area in the semi-classical limit \cite{Bekenstein:1973jb,Hawking:1975sh}. This establishes the seminal Bekenstein-Hawking area law (BHAL), depicted as $S_{\text{BH}}$ in \eqref{int1}. However, the applicability of BHAL is not solely restricted to Einstein's gravity; it extends to any self-consistent quantum gravity theory at the tree level, followed by additional corrections emerging in the presence of quantum fluctuations at the Planck scale. 

The leading-order quantum-gravitational correction to BHAL is universally found to be a logarithmic term of the horizon area \cite{Solodukhin:1995na,Solodukhin:1995nb,Kaul:2000rk,Carlip:2000nv,Banerjee:2008cf,Banerjee:2009fz,Majhi:2009gi,Cai:2010ua,Banerjee:2011oo,Banerjee:2011pp,Sen:2012qq,Sen:2012rr,Bhattacharyya:2012ss,Chowdhury:2014np,Gupta:2014ns,Jeon:2017ij,Karan:2019sk,Sen:2013ns,Keeler:2014nn,Charles:2015nn,Larsen:2015nx,Castro:2018tg,Banerjee:2020wbr,Karan:2020sk,Karan:2021teq,Banerjee:2021pdy,David:2021eoq,Karan:2022dfy,Bobev:2023dwx,El-Menoufi:2015cqw,Xiao:2021zly,Calmet:2021lny,Delgado:2022pcc,Pourhassan:2022auo,H:2023qko,Anupam:2023yns}, as indicated by $\ln S_{\text{BH}}$ in \eqref{int1}. 
Over the last few decades, these logarithmic corrections have emerged as a gateway to quantum gravity, serving as a trendy litmus test. This test asserts that the macroscopic results of quantum-corrected black hole entropy, computed in the low-energy effective theory of (super-)gravity, i.e., the IR side, must be matched by any precise enumeration of the microstate data, such as the Strominger-Vafa counting \cite{Strominger:1996sh} available within the corresponding UV-complete quantum gravity candidate. Sen and collaborators have extensively analyzed this test for various examples of asymptotically-flat BPS black holes in string theory, and all have passed the test with flying colors \cite{Sen:2014aja,Banerjee:2011oo,Banerjee:2011pp,Sen:2012qq,Sen:2012rr}.  
To date, string theory has successfully enumerated microstates underlying the entropy of a wide class of asymptotically-flat and asymptotically-AdS black holes \cite{Strominger:1996sh,Maldacena:1996gb,Horowitz:1996fn,Emparan:2006it,Sen:2014aja,Belin:2016knb,Benini:2019dyp,Gang:2019uay,PandoZayas:2020iqr,Liu:2017vll,Benini:2015eyy,Liu:2018bac}.\footnote{The list is not exhaustive -- interested readers are referred to the references and citations therein.} 
Thus, the computation of logarithmic corrections for all such cases, as well as future examples when available, appears to be a robust ``infrared window'' into black hole microstates for providing a non-trivial consistency check. Conversely, one can also verify whether any concerned (super-)gravity model is indeed a low-energy effective limit of the UV-complete microscopic counterpart. In this paper, we aim to progress toward the macroscopic or IR end of the aforementioned line.

Extensive studies \cite{Solodukhin:1995na,Solodukhin:1995nb,Kaul:2000rk,Carlip:2000nv,Banerjee:2008cf,Banerjee:2009fz,Majhi:2009gi,Cai:2010ua,Banerjee:2011oo,Banerjee:2011pp,Sen:2012qq,Sen:2012rr,Bhattacharyya:2012ss,Chowdhury:2014np,Gupta:2014ns,Jeon:2017ij,Karan:2019sk,Sen:2013ns,Keeler:2014nn,Charles:2015nn,Larsen:2015nx,Castro:2018tg,Banerjee:2020wbr,Karan:2020sk,Karan:2021teq,Banerjee:2021pdy,David:2021eoq,Karan:2022dfy,Bobev:2023dwx,El-Menoufi:2015cqw,Xiao:2021zly,Calmet:2021lny,Delgado:2022pcc,Pourhassan:2022auo,H:2023qko,Anupam:2023yns} have shown that the quantum-corrected entropy of black holes can be expressed in the following general form (in natural units: $c= \hbar = k_B = 1$)
\begin{align}\label{int1}
	S_{\text{bh}}(\mathcal{A}_H) =  S_{\text{BH}} + \mathcal{C}_{\text{log}}\ln S_{\text{BH}} + \mathcal{O} {\left(\frac{1}{ S_{\text{BH}}}\right)}+ \cdots, \enspace S_{\text{BH}}=\frac{\mathcal{A}_H}{4G_N},
\end{align}
where $\mathcal{A}_H$ refers to horizon area, $G_N$ is Newton's gravitational constant and $\mathcal{C}_{\text{log}}$ is a constant. The first term, $S_{\text{BH}}$, corresponds to the Bekenstein-Hawking area law and constitutes the leading tree-level contribution in the entropy formula \eqref{int1}. The subsequent terms account for various quantum gravitational corrections to the BHAL, stemming from perturbative or non-perturbative frameworks.\footnote{It is important to note that the ellipsis in formula \eqref{int1} includes additional non-perturbative exponential correction terms \cite{Chatterjee:2020iuf,Dabholkar:2014ema} of the form $\Delta S_{\text{BH,non-per}} = \eta e^{-S_{\text{BH}}}$, where $\eta$ controls the relative strength. Moreover, recent arguments \cite{Iliesiu:2022onk,Banerjee:2023quv,Kapec:2023ruw,Rakic:2023vhv,Banerjee:2023gll,Maulik:2024dwq} suggest the presence of a typical correction proportional to the logarithmic of Hawking temperature for the near-extremal black holes at very low temperatures.}. These include the so-called logarithmic and power-law corrections, denoted as the second and third terms, respectively, capturing the perturbative contributions in formula \eqref{int1}. These perturbative contributions correspond to different-order quantum loop contributions ($g_s$), arising when evaluating the Euclidean gravitational path integral considering quantum fluctuations around any generic black hole saddle point:
\begin{align}\label{int2}
	\Delta S_{\text{BH,per}} \approx  \mathcal{C}_{\text{log}}\ln \left(\frac{\mathcal{A}_H}{G_N}\right) + \sum_{n\geq1} \kappa_n {\left(\frac{\mathcal{A}_H}{G_N}\right)}^{-n+1} ,
\end{align}
where $n$ denotes the order of quantum loops and $\kappa_n$ represents constant values that control the relative strengths of the loop or power-law corrections.
In the case of quantum fluctuations in a higher-derivative modified gravity model, the saddle-point contribution extends the BHAL into the Bekenstein-Hawking-Wald formula \cite{Wald:1993rw} by incorporating higher-derivative ($\alpha^\prime$) corrections. However, the modified form of the perturbative correction formula \eqref{int2} remains the same. Remarkably, the logarithmic correction appears as a special class of one-loop contributions explicitly induced from the two-derivative sector of the theory and remains unaffected by the higher-derivative and power-law corrections \cite{Banerjee:2011oo,Sen:2013ns}.

The logarithmic corrections have garnered significant attention due to their intriguing properties, offering crucial yet non-trivial insights into the nature of quantum gravity. Their ubiquitous form $\propto\ln\mathcal{A}_H$ is not specific to any particular model or theory; rather, it emerges universally for all types of black holes evolving within the framework of different quantum gravity approaches. These approaches encompass conical singularity \cite{Solodukhin:1995na,Solodukhin:1995nb}, quantum geometry \cite{Kaul:2000rk}, Cardy formula \cite{Carlip:2000nv}, quantum tunneling \cite{Banerjee:2008cf,Banerjee:2009fz,Majhi:2009gi}, conformal anomaly \cite{Cai:2010ua}, Euclidean effective action method \cite{Banerjee:2011oo,Banerjee:2011pp,Sen:2012qq,Sen:2012rr,Bhattacharyya:2012ss,Chowdhury:2014np,Gupta:2014ns,Jeon:2017ij,Karan:2019sk,Sen:2013ns,Keeler:2014nn,Charles:2015nn,Larsen:2015nx,Castro:2018tg,Banerjee:2020wbr,Karan:2020sk,Karan:2021teq,Banerjee:2021pdy,David:2021eoq,Karan:2022dfy,Bobev:2023dwx}, supersymmetric index \cite{H:2023qko,Anupam:2023yns}, non-local quantum gravity \cite{El-Menoufi:2015cqw,Xiao:2021zly,Calmet:2021lny,Delgado:2022pcc,Pourhassan:2022auo}, and more. In the context of (super-)gravity models being regarded as a low-energy limit of string theory or other quantum gravity theories, it is convenient to impose the so-called large-charge limit\footnote{The large-charge limit scales the black hole geometric parameters while keeping their dimensionless ratios fixed, resulting in a black hole horizon area much larger than the Planck area (e.g., see \cite{Sen:2013ns}). This ensures that the study of quantum gravitational effects within the macroscopic Einstein's gravity frameworks remains under good semi-classical control.} on black hole backgrounds, rendering BHAL a valid choice at the leading order in black hole entropy \eqref{int1}. Consequently, the logarithmic correction becomes the most dominant subleading quantum correction contribution, suppressing the power-law corrections. Hence, any true quantum gravity candidate or microstate counting data that fails to reproduce the leading-order logarithmic-corrected macroscopic black hole entropy is deemed incorrect. Notably, the quantum-gravity corrections to black hole entropy typically depend on the specifics of the UV completion, i.e., the contributions from various massive modes in the quantum loops. Interestingly, the logarithmic corrections are independent of the UV completion and can be computed using the massless fluctuations (i.e., low-energy or infrared modes) running in the one-loop \cite{Banerjee:2011oo,Banerjee:2011pp,Sen:2012qq,Sen:2012rr,Bhattacharyya:2012ss,Chowdhury:2014np,Gupta:2014ns,Jeon:2017ij,Karan:2019sk,Sen:2013ns,Keeler:2014nn,Charles:2015nn,Larsen:2015nx,Castro:2018tg,Banerjee:2020wbr,Karan:2020sk,Karan:2021teq,Banerjee:2021pdy,David:2021eoq,Karan:2022dfy}.

The prefactor $\mathcal{C}_{\text{log}}$ is a dimensionless constant that exhibits theory-specific behavior, unlike the universal $\ln\mathcal{A}_H$ part in the logarithmic entropy correction. In a given theory, the value and sign of $\mathcal{C}_{\text{log}}$ are completely determined by the geometric parameters (such as mass, charge, angular momentum, etc.) characterizing the relevant black hole, as well as the quantum fluctuation data (e.g., conformal anomaly, central charges, etc.) of the entire field content of the theory. In general, $\mathcal{C}_{\text{log}}$ are found to be a complex function of different dimensionless ratios of black hole parameters \cite{Bhattacharyya:2012ss,Sen:2013ns,Castro:2018tg,Karan:2019sk,Karan:2020sk,Banerjee:2020wbr,David:2021eoq,Karan:2022dfy}. However, in certain special or limiting cases, as seen in \cite{Banerjee:2011pp,Sen:2012qq,Sen:2012rr,Gupta:2014ns,Keeler:2014nn,Larsen:2015nx,Charles:2015nn}, $\mathcal{C}_{\text{log}}$ exhibits topological values (i.e., pure numbers), giving rise to a fully universal logarithmic correction within the specific theory of choice. Investigating such universal or topological vs. non-topological nature of $\mathcal{C}_{\text{log}}$ is not only fascinating but also highly sensitive to microstate counting data within the realm of quantum gravity. In this paper, we aim to compute and analyze $\mathcal{C}_{\text{log}}$ results for a well-studied family of AdS and flat black holes embedded within the so-called STU supergravity models in 4D.


Let us elucidate the motivation behind investigating quantum black holes within the framework of STU supergravity models. Supergravities are the popular low-energy (small curvature expansion) limit of superstring theories, with one well-studied example of string theory compactifications down to four dimensions being the $\mathcal{N}=8$ supergravity \cite{Cremmer:1978ds,Cremmer:1979up}. They feature many $U(1)$ gauge field strengths and scalar moduli fields in their bosonic sector. Although numerous truncations of $\mathcal{N}=8$ supergravity have been reported over the last few decades in search of a general family of black hole solutions, many of them have proven elusive. However, there exists a consistent truncation of $\mathcal{N}=8$ supergravity (for both the ungauged and gauged versions) into a model popularly known as the STU supergravity (e.g., see \cite{Cremmer:1984hj,Duff:1995sm}), which can be viewed as a pure $\mathcal{N}=2$ supergravity multiplet coupled with three vector multiplets \cite{Duff:1995sm}. Notably, the equations of motion in supergravity are technically more intricate than those in general relativity, requiring sophisticated solving procedures. However, STU solutions employ such a distinctive procedure, which relies on global symmetries (i.e., $U$-dualities) inherent to string theory, generating the most general black holes of $\mathcal{N}=8$ supergravity via truncation to a system with only four $U(1)$ gauge fields and three complex scalar fields \cite{Sen:1994eb,Cvetic:1996zq}. Moreover, STU supergravity solutions find wide application in generating black holes within all $\mathcal{N}\geq 2$ supergravity theories \cite{Cvetic:1996zq}. Therefore, STU supergravity models serve as fundamental and universal building blocks to structure the central bosonic sector of various low-energy effective superstring theories or supergravities in four dimensions.

STU supergravity models admit a generalized yet non-trivial family of black hole solutions with higher-dimensional origins in string theory. In this study, our objective is to recover their previously known general-relativistic trivial subcases, specifically the simplest four-dimensional Einstein-Maxwell (EM) and Einstein-Maxwell-AdS (EM-AdS) backgrounds \cite{Carter:1968ks,Plebanski:1976gy,Caldarelli:1999x,Adamo:2014lk} within STU models. This motivation stems from the intriguing properties and broader implications of the EM solutions, which already have a solid microscopic foundation within string theory \cite{Sen:2014aja,Banerjee:2011oo,Banerjee:2011pp,Sen:2012qq,Sen:2012rr}. The four-dimensional EM theory involves a single $U(1)$ Maxwell field coupled minimally to the metric field and is known to admit supersymmetric black hole solutions by structuring the bosonic sector of pure $\mathcal{N}=2$ supergravity \cite{Freedman:2012xp}. However, while the STU supergravity we focus on is a more generalized theory, there is no direct and consistent truncation that reduces STU models to the EM theory. In this paper, we detail how the entire asymptotically-AdS$_4$ and flat$_4$ counterparts of the Kerr-Newman family of black holes can be systematically recovered within the gauged and ungauged versions of STU supergravity models as special embedding choices. Notably, this EM embedding process is not straightforward, but it is conceptually and technically well-understood.

This work primarily regards the STU supergravity models as a distinct class of $U(1)^4$-charged Einstein-Maxwell-dilaton (EMD) theory with four Maxwell and three dilaton fields, admitting the generic four-charge black hole solutions \cite{Cvetic:1999xp,Clement:2013fc,Chow:2014cca,Cvetic:2014vsa,Cvetic:2021lss,Anabalon:2022aig}. However, there exist two consistent scenarios in which the $U(1)^4$-charged EMD system, intersecting with the bosonic sector of STU supergravity models, can be further reduced to two distinct versions of $U(1)^2$-charged EMD systems \cite{Clement:2013fc,Chow:2014cca,Cvetic:2014vsa,Lu:2013eoa,Mai:2021yny}. In these truncated STU versions, a single dilaton is non-minimally coupled to two Maxwell fields via two separate exponential coupling functions,\footnote{For details, please refer to the action setup \eqref{mod2}.} where the values of dilaton coupling coefficients are fixed during the compactification of superstring theories. The non-minimal nature of Maxwell-dilaton couplings prevents all charged EM backgrounds from directly solving the equations of motion of $U(1)^2$-charged EMD theories. However, we identify a specific case where suitable constraints on the two $U(1)$ or Maxwell charges can effectively decouple the two non-minimal Maxwell-dilaton couplings in the $U(1)^2$-charged EMD models. Consequently, the modified STU equations of motion exhibit a vanishing dilaton background, thereby reducing to the field equations governing a class of single-charged black hole solutions within the EM theory. Notably, the EM embedding procedure also extends to $U(1)^2$-charged EMD systems with a negative cosmological constant, which are referred to as $U(1)^2$-charged EMD-AdS theories. Following this approach, we successfully embed the Kerr-Newman-AdS and Kerr-Newman, Kerr-AdS and Kerr, Reissner-Nordstr\"om-AdS and Reissner-Nordstr\"om, and Schwarzschild-AdS and Schwarzschild black holes into $U(1)^2$-charged EMD-AdS and EMD theories intersecting with gauged and ungauged STU supergravities in 4D. Ultimately, our goal is to compute the logarithmic correction to the entropy of all these embedded asymptotically AdS$_4$ and flat$_4$ black holes in both their non-extremal and extremal temperature limits.

We employ the traditional heat kernel method \cite{Hawking:1977te,Denardo:1982tb,Avramidi:1994th,Barvinsky:2015} within the framework of Euclidean quantum gravity \cite{Gibbons:1977ta,Hawking:1978td} to compute the logarithmic entropy corrections for the black holes addressed in this paper. Our methodology combines Sen's quantum entropy function formalism \cite{Sen:2008wa,Sen:2009wb,Sen:2009wc} with techniques developed in \cite{Sen:2013ns}, allowing us to extract the essential ``logarithmic'' component from the Euclideanized one-loop quantum effective action associated with the entropy of extremal and non-extremal black holes, respectively. The computation of the one-loop effective action involves expressing it in terms of the heat kernel of the kinetic operator governing only one-loop or quadratic fluctuations and subsequently expanding it using the well-known Seeley-DeWitt expansion \cite{Seeley:1966tt,Seeley:1969uu,DeWitt:1965ff,DeWitt:1967gg,DeWitt:1967hh,DeWitt:1967ii}. The working formula for logarithmic entropy corrections involves integrating a specific Seeley-DeWitt coefficient around the relevant part of the black hole geometries. For the four-dimensional black holes considered in this study, we only require to compute the third-order heat kernel expansion coefficient, denoted as $a_4(x)$ in \eqref{set14}, which we accomplish by following Gilkey's approach \cite{Vassilevich:2003ll}. This approach proves to be highly efficient and universally applicable, enabling us to investigate the quantum entropy of any charged, neutral, static, rotating, non-supersymmetric, supersymmetric or BPS, extremal, and non-extremal black holes within a unified framework without limitations. 

In contrast, many other established approaches, such as the eigenfunction expansion of the heat kernel operator employed in \cite{Banerjee:2011oo,Banerjee:2011pp,Sen:2012qq,Gupta:2014ns,Keeler:2014nn,Larsen:2015nx,Banerjee:2023quv}, which are restricted to the Bertotti-Robinson (AdS$_2\times S^2$) type extremal near-horizon background geometry featuring rotational symmetry. On the contrary, the Euclidean gravity setup considered in this study has achieved significant success over the last decade by computing logarithmic corrections for asymptotically flat black holes in various examples of the Einstein-Maxwell theory \cite{Bhattacharyya:2012ss,Sen:2013ns,Karan:2021teq} and ungauged $\mathcal{N}\geq 1$ supergravities \cite{Sen:2012rr,Karan:2019sk,Banerjee:2020wbr,Karan:2020sk,Charles:2015nn,Castro:2018tg,Banerjee:2021pdy}. More recently, the same investigation has also been extended to a few examples of asymptotically-AdS black holes in the four-dimensional gauged supergravity \cite{David:2021eoq,Bobev:2023dwx} and Einstein-Maxwell-dilaton theory with a negative cosmological constant \cite{Karan:2022dfy}. Notably, the logarithmic corrections for AdS black holes remain less explored till date, primarily due to substantial technical challenges. In this paper, we advance along this trendy direction and test the aforementioned Euclidean gravity setup on a specific class of AdS$_4$ black holes, which are inherent to string theory via the STU supergravity models.

The main technical contribution of this paper is twofold. Firstly, we derive the expression for the heat kernel coefficient $a_4(x)$ for two specific cases of $U(1)^2$-charged EMD-AdS theory, obtained as a consistent bosonic truncation of the STU supergravity. In this process, we expand the relevant bulk actions up to quadratic order to account for small quantum fluctuations of the entire STU field content. This quantization is performed around a generic class of four-dimensional classical backgrounds featuring a vanishing dilaton, yielding black hole solutions of a four-dimensional EM theory embedded into the STU-truncated systems. The considered heat kernel approach is entirely on-shell, where we utilize the EM-embedded background equations of motion to systematically manage the Seeley-DeWitt coefficients exclusively in terms of invariants induced by the background geometry and fields. In particular, we express the necessary $a_4(x)$ as a function of four-derivative background invariants involving trace anomalies such as the square of the four-dimensional Weyl tensor and the Euler density, which are derived from the two-derivative action of the truncated STU models. At any point, by imposing appropriate flat-space limits on the $U(1)^2$-charged EMD-AdS backgrounds, one can obtain similar heat kernel results in the counterpart of $U(1)^2$-charged EMD theory evolving with a vanishing cosmological constant. The computed formulas for $a_4(x)$, in simplified forms, are recorded in \cref{sdc30,sdc32}.

In the final part, the computed $a_4(x)$ relations are integrated over black hole geometries, encompassing both the full geometry and near-horizon configurations for non-extremal and extremal limits, respectively. These integrations yield the logarithmic corrections to the entropy of all asymptotically AdS$_4$ and flat$_4$ members of Kerr-Newman, Kerr, Reissner-Nordstr\"om and Schwarzschild black holes. The resulting formulas for the two cases of truncated STU supergravity models are presented in \cref{raI1,raI2,raI3,raI4,raI5,raI6,raI7,rfI1,rfI2,rfI3,rfI4,rfI5,raII1,raII2,raII3,raII4,raII5,raII6,raII7,rfII1,rfII2,rfII3,rfII4,rfII5}. Notably, our analysis reveals that the AdS$_4$ results are all non-topological, exhibiting a much richer and broader structure compared to the flat cases. We provide a consistent explanation for this observation, attributing the confirmed non-topological component in the logarithmic entropy corrections as a natural and generic contribution induced by the boundary of AdS black hole backgrounds. In contrast, when the AdS boundary disappears, as in the case of flat black hole backgrounds, a vanishing charge ensures confirmed topological logarithmic correction results. Furthermore, the extremal or zero temperature limit guarantees the same nature for the charged but non-rotating background, whereas the charged-rotating result remains non-topological. This observation might set a stringent criterion for 4D supergravity models to admit a UV completion, which is highly sensitive to the microscopic analysis of the relevant black holes within the framework of string theory counterparts.

The rest of this paper is outlined as follows. In \cref{setup}, we provide a concise and efficient guide for applying the heat kernel method to compute the logarithmic correction to black hole entropy, with special emphasis on the treatment of four-dimensional black hole backgrounds. In \cref{model}, we calculate the third-order Seeley-DeWitt coefficient $a_4(x)$ for the interested EMD-truncated STU supergravity models. \Cref{logresult} utilizes the computed heat kernel data to derive the logarithmic correction formulas for the entire Kerr-Newman-AdS and Kerr-Newman families of black holes embedded in the gauged and ungauged STU supergravity models, respectively. Finally, \cref{discuss} concludes this paper with a summary and relevant discussion, and provides an outlook for future research directions. Given the intricate nature of the current topic, we have included \cref{calcul,enhii} to provide comprehensive details on heat kernel computations and other relevant technical aspects.


\section{The setup}\label{setup}

This section aims to provide a comprehensive manual for calculating the logarithmic correction to the entropy of general black holes, considering both the extremal and near-extremal limits of their Hawking temperature. We explicitly evaluate the one-loop quantum effective action of the relevant gravitational theory fluctuated around the black hole backgrounds of interest. This manual is developed by revisiting the Euclidean quantum gravity frameworks proposed by Sen \cite{Sen:2008wa,Sen:2009wb,Sen:2009wc,Sen:2013ns} and employs the Seeley-DeWitt expansion of the heat kernel method \cite{Vassilevich:2003ll}.

\subsection{Euclidean quantum gravity and heat kernel expansion}  \label{nonext}

We consider charge and rotating black hole solutions in a generic class\footnote{This class incorporates an appropriate scaling symmetry \cite{Sen:2013ns}, ensuring that all bosonic terms possess two derivatives and fermionic terms have one derivative. It encompasses a wide range of theories, including Einstein's gravity coupled to scalar, gauge, and spinor fields, as well as various supergravities, while also accommodating the presence of a cosmological constant.} of $D$-dimensional Einstein's gravity characterized by the following path integral,
\begin{align}\label{set1}
	\begin{gathered}
		\mathcal{Z} = \int \mathscr{D}[g_{\mu\nu},\varphi]\exp\left(-\mathcal{S}_E[g_{\mu\nu},\varphi]\right).
	\end{gathered}
\end{align} 
Here, $\mathscr{D}[g_{\mu\nu},\varphi]$ is the measure of functional integration over the set of all massless fields $\varphi$ propagating through a spacetime geometry described by the metric $g_{\mu\nu}$, and $\mathcal{S}_E[g_{\mu\nu},\varphi]$ denotes the Wick-rotated action characterizing the Euclidean continuation of black hole solutions within the theory. To determine the entropy $S_{\text{bh}}(M, \vec{J}, \vec{Q})$ of a black hole with mass $M$, angular momenta $\vec{J}$, and charges $\vec{Q}$, we follow the saddle-point approximation in the Gibbons and Hawking prescription \cite{Gibbons:1977ta}, employing the following Legendre transformation,   
\begin{align}\label{set3}
	S_{\text{bh}}(M, \vec{J}, \vec{Q}) = \ln {\mathcal{Z}}(\beta,\vec{\omega},\vec{\mu}) + \beta M + \vec{\omega}\cdot\vec{J} + \vec{\mu}\cdot \vec{Q}.
\end{align}
Here, ${\mathcal{Z}}(\beta,\vec{\omega},\vec{\mu})$ represents the path integral measure defined in \eqref{set1}, evaluated at a stationary saddle point that satisfies the classical equations of motion of the theory, subject to appropriate asymptotic boundary conditions. These boundary conditions are controlled by fixing a set of three parameters: $\beta$, $\vec{\omega}$, and $\vec{\mu}$, which respectively denote the inverse temperature (or the period of Euclideanized time), the angular velocities, and the electromagnetic chemical potentials associated with the Euclidean saddle-point solutions. Notably, in the definition provided in \eqref{set3}, black holes are not directly referenced; rather, they are manifested as Euclideanized saddle points. Their relevant parameters $M$, $\vec{J}$, and $\vec{Q}$ are determined in terms of $\beta$, $\vec{\omega}$, and $\vec{\mu}$ through the relations,
\begin{align}
	M = - \frac{\partial \ln {\mathcal{Z}}}{\partial \beta},\enspace \vec{J} = - \frac{\partial \ln {\mathcal{Z}}}{\partial \vec{\omega}},\enspace \vec{Q} = - \frac{\partial \ln {\mathcal{Z}}}{\partial \vec{\mu}}.	
\end{align}
To evaluate the path integral partition function ${\mathcal{Z}}(\beta,\vec{\omega},\vec{\mu})$, we follow the procedure outlined below. We begin by considering fluctuations in the entire set of field content and the metric describing gravity, denoted by $\phi_m =\lbrace g_{\mu\nu}, \varphi \rbrace$, around their classical background or saddle-point values $(\bar{g}_{\mu\nu},\bar{\varphi})$, such that $\frac{\delta }{\delta \phi_m}\mathcal{S}_E[\bar{g}_{\mu\nu},\bar{\varphi}] =0$, i.e.,
\begin{equation}\label{set2}
	g_{\mu\nu} = \bar{g}_{\mu\nu} + h_{\mu\nu}, \enspace \varphi = \bar\varphi + \tilde{\varphi}.
\end{equation}
Here, $\tilde{\phi}_m =\lbrace h_{\mu\nu},\tilde{\varphi}\rbrace$ denotes the entire set of small quantum fluctuations that are in the Planck scale order. Consequently, the effective action, defined as $\mathpzc{W} = -\ln {\mathcal{Z}}(\beta,\vec{\omega},\vec{\mu})$, can be expanded in different-order quantum loop expansions, 
\begin{align}\label{setx1}
	\mathpzc{W} = \mathcal{S}_E[\bar{g}_{\mu\nu},\bar{\varphi}]-\ln \int \mathscr{D}[\tilde{\phi}_m]\exp\left(-\int \mathrm{d}^Dx\sqrt{\det \bar{g}}\thinspace\tilde{\phi}_m \mathcal{H}\tilde{\phi}^m\right)+ \cdots,
\end{align}
where $\mathcal{H}= \frac{\delta^2\mathcal{S}_E}{\delta {\phi}_m^2}$ is the kinetic operator controlling the quadratic field fluctuations appearing in the one-loop. The first term $\mathcal{S}_E[\bar{g}_{\mu\nu},\bar{\varphi}]$ corresponds to the classical or on-shell action, which dominates the transformation \eqref{set3} and leads to the Bekenstein-Hawking area law \cite{Gibbons:1977ta,Hawking:1978td}, or its Wald generalization \cite{Wald:1993rw} when higher-derivative terms are integrated into Einstein's framework,
\begin{align}\label{set4}
	S_{\text{BH}}(M, \vec{J}, \vec{Q}) = \frac{\mathcal{A}_{H}(M, \vec{J}, \vec{Q})}{4G_N} + \cdots.
\end{align}
On the other hand, the term next to the leading saddle-point contribution in the setup \eqref{setx1} is the one-loop effective action (OLEA). It involves a Gaussian integral over the quadratic fluctuations with the kinetic operator $\mathcal{H}$ and can be expressed as
\begin{align}\label{set5}
	\mathpzc{W}_{1\text{-loop}} = \frac{\chi}{2}\ln\det\mathcal{H} = \frac{\chi}{2}\mathrm{Tr}\ln \mathcal{H},
\end{align}       
where $\mathrm{Tr}$ denotes the trace operation performed over spacetime and all internal indices of the field fluctuations, and $\chi = \pm 1$ for bosonic and fermionic fluctuations, respectively. With this setup, the quantum-corrected black hole entropy up to one-loop is given by
\begin{align}\label{set6}
	S_{\text{bh}}(M, \vec{J}, \vec{Q}) = S_{\text{BH}}(M, \vec{J}, \vec{Q}) - \mathpzc{W}_{1\text{-loop}}  + \beta M + \vec{\omega}\cdot\vec{J} + \vec{\mu}\cdot \vec{Q}.
\end{align}   

The entire problem is now centered on the evaluation of $\mathpzc{W}_{1\text{-loop}}$, as it yields the desired one-loop quantum correction to the black hole entropy. To address this purpose, we employ the conventional heat kernel method \cite{Hawking:1977te,Denardo:1982tb,Avramidi:1994th,Barvinsky:2015,Vassilevich:2003ll} and rewrite the OLEA as 
\begin{align}\label{set9}
	\begin{gathered}
		\mathpzc{W}_{1\text{-loop}} = -\frac{\chi}{2}\int_\epsilon^\infty \frac{\mathrm{d}\tau}{\tau} K(\tau),\quad K(\tau) \equiv \mathrm{Tr}\left(e^{-\tau \mathcal{H}}\right),
	\end{gathered}
\end{align} 
where $\tau$ is an auxiliary proper time parameter with dimensions of $\left(\mathrm{length}\right)^2$. $K(\tau)$ is the heat kernel trace of the quadratic operator $\mathcal{H}$ with the orthonormal eigenfunctions $\lbrace f_i(x)\rbrace$ and eigenvalues $\lbrace h_i\rbrace$, which can be expressed in the following spectral decomposition form,
\begin{align}\label{set7}
	K(\tau) = \int \mathrm{d}^Dx\sqrt{\det \bar{g}}\thinspace \sum_i e^{-h_i\tau}f_i(x)f_i(x) = \sum_i e^{-h_i\tau}.
\end{align}
Notice that the heat kernel representation of the OLEA given above suffers from a UV divergence as $\tau \to 0$, which is regulated by the UV cut-off parameter $\epsilon$ introduced in the lower integration limit. In a UV-regulated theory, $\epsilon$ is typically constrained by the square of the Planck length, which is of the same order as the gravitational constant $G_N$ in the convention employed in this paper.

Further, the upper integration limit in \eqref{set7} suggests that the OLEA also suffers from an infrared divergence induced due to the infinite volume spacetime. However, Sen demonstrated in \cite{Sen:2013ns} that this infrared divergence could be regulated by excluding the component of the OLEA that accounts for the thermal gas contribution from all particles in our theory, which remains in equilibrium with the black hole saddle-point. This approach aids in isolating the precise portion of the OLEA exclusively associated with the quantum entropy of the black hole, as required in the formula \eqref{set6}. To achieve this, we confine a black hole of radius $\mathfrak{R}$ (such that the horizon area scales as $\mathcal{A}_H \sim \mathfrak{R}^{D-2}$) inside a box of size $L$. The dominant contribution to the OLEA from the thermal gas in equilibrium with the black hole, characterized by inverse temperature $\beta$, angular velocity $\vec{\omega}$, and chemical potential $\vec{\mu}$, is given by \cite{Sen:2013ns}
\begin{align}\label{set11}
	\mathpzc{W}_{1\text{-loop},\,\mathrm{gas}} \simeq L^{D-1}f(\beta, \vec{\omega}, \vec{\mu}),
\end{align}
where $f$ is a function that scales as $f(\lambda\beta, \vec{\omega}, \lambda\vec{\mu}) = \lambda^{-D+1} f(\beta, \vec{\omega}, \vec{\mu})$ under the same scaling of the metric $\bar{g}_{\mu\nu}$ and gauge fields $\bar{A}_\mu$, as well as the associated black hole parameters, i.e.,
\begin{align}\label{set10}
	\bar{g}_{\mu\nu} \to \lambda^2 \bar{g}_{\mu\nu}, \enspace \bar{A}_\mu \to \lambda \bar{A}_\mu, \enspace M \to \lambda^{D-3} M, \enspace \vec{Q} \to \lambda^{D-3}\vec{Q}, \enspace \vec{J} \to \lambda^{D-2}\vec{J}.
\end{align}
Here, $\lambda$ represents a common length scale crucial for logarithmic correction computations \cite{Sen:2013ns}. Moreover, the scalings in \eqref{set10} modify the black hole radius $\mathfrak{R}(M,\vec{J},\vec{Q})$ and its Bekenstein-Hawking entropy $S_{\text{BH}}(M,\vec{J},\vec{Q})$ by $\lambda$ and $\lambda^{D-2}$, respectively. 

Then the key idea involves considering a new reference black hole solution with a fixed radius $\mathfrak{R}_0$, confined within an identical box of size $L_0 = L\mathfrak{R}_0/\mathfrak{R}$. This new solution is related to the original black hole solution through parameter rescaling depicted in \eqref{set10} for $\lambda = \mathfrak{R}_0/\mathfrak{R}$. For this new black hole system, it is straightforward to verify that the dominant thermal gas contribution to the OLEA is given by $\left(L\lambda\right)^{D-1}f\left(\beta\lambda, \vec{\omega}, \vec{\mu}\lambda\right)$, which is identical to the contribution \eqref{set11} in the original black hole system (for details, refer to appendix B of \cite{Sen:2013ns}). Thus, the subtraction of the OLEAs between the original and rescaled black hole systems effectively cancels out the leading thermal gas contributions, yielding the difference solely related to the black holes. For simplicity, we will denote this thermal gas regulated difference in the OLEAs of the original and new black holes by $\Delta \mathpzc{W}$. Notably, the eigenvalues $h^{(0)}_i$ of the kinetic operator for the new black hole system are related to the eigenvalues $h_i$ of the original system via the same scaling \eqref{set10} with $\lambda = \mathfrak{R}_0/\mathfrak{R}$ and are given by
\begin{align}\label{set12}
	h^{(0)}_i = h_i/\lambda^2.
\end{align}
Therefore, the specific form \eqref{set9} of the OLEA, along with the relations \eqref{set7} and \eqref{set12}, allows us to express $\Delta \mathpzc{W}$ as
\begin{align}\label{set13}
	\Delta \mathpzc{W} &= -\frac{\chi}{2} \left[\int_\epsilon^\infty \frac{\mathrm{d}\tau}{\tau}\sum_i e^{-h_i\tau} - \int_\epsilon^\infty \frac{\mathrm{d}\tau}{\tau}\sum_i e^{-h_i\tau/\lambda^2}\right]\nonumber\\
	& = -\frac{\chi}{2}\int_\epsilon^{\epsilon/\lambda^2} \frac{\mathrm{d}\tau}{\tau} K(\tau),
\end{align}
where the final step is obtained by rescaling the second integration variable $\tau/\lambda^2 \to \tau$. The infrared cutoff at the upper limit of the OLEA integration \eqref{set13} results in a dominant contribution within the range $\epsilon < \tau < \epsilon/\lambda^2$ or $\epsilon/\mathfrak{R}^2 < \tau/\mathfrak{R}^2 < \epsilon/\mathfrak{R}_0^2$. Within this interval, we need to take the so-called {large charge limit} on both black holes by setting $\mathfrak{R} \gg \sqrt{\epsilon}$ and ${\mathfrak{R}_0} \gg \sqrt{\epsilon}$ or $\lambda \gg 1$, given that $\sqrt{\epsilon}$ is in the Planck length order. Consequently, we can utilize the short-time asymptotic expansion at $\tau \to 0$ of the heat kernel trace $K(\tau)$ as
\begin{align}\label{set14}
	K(\tau) \overset{\tau \to 0}{=} \int \mathrm{d}^Dx \sqrt{\text{det}\thinspace \bar{g}}\thinspace \sum_{n=0}^\infty \tau^{n-\frac{D}{2}}a_{2n}(x),
\end{align}
where the functions $a_{2n}(x)$ represent the well-known {Seeley-DeWitt coefficients} \cite{Seeley:1966tt,Seeley:1969uu,DeWitt:1965ff,DeWitt:1967gg,DeWitt:1967hh,DeWitt:1967ii}. 

It is noteworthy that opting for the large-charge limit $\lambda \gg 1$ allows us to carefully delineate {massless} fluctuations within the existing scaling configuration, which is a crucial fundamental in computing the logarithmic corrections. The appearance of massive fluctuations necessarily modulate the heat-trace expansion \eqref{set14} by an additional prefactor $e^{-\tau m^2}$ (e.g., see \cite{Barvinsky:2015}), where their mass $m$ must be scaled as $m \to m/\lambda$ to accommodate the rescaling $\tau/\lambda^2 \to \tau$ in the step \eqref{set13}. Following the same prescription as in \cite{Banerjee:2011oo,Banerjee:2011pp,Sen:2012rr,Sen:2012qq,Sen:2013ns,Jeon:2017ij}, we utilize the limit $\lambda \gg 1$ to define ``masslessness'' as any fluctuation whose mass $m$ is of the order of $\lambda^{-1}$ or less. This characteristic effectively suppresses the term $e^{-\tau m^2}$, as if setting $m=0$ from the outset, thereby excluding all massive contributions from the current heat kernel setup in computing logarithmic corrections for all generic black holes.\footnote{More precisely, within the ``massless'' limit, the introduction of the logarithmic term $\ln \lambda$ occurs in step \eqref{set15}, a term notably absent when employing expansion \eqref{set14} modulated by $e^{-\tau m^2}$ for the massive fluctuations, even at $n = \frac{D}{2}$.}     

We can now substitute the heat trace expansion \eqref{set14} into the OLEA form \eqref{set13} and integrate it over $\tau$, yielding\footnote{Note that the spin-signature parameter $\chi$ is absorbed by the Seeley-DeWitt coefficients in the typical OLEA form \eqref{set15}. However, it will be readjusted in the working formula \eqref{comp8}.} 
\begin{align}\label{set15}
	\Delta \mathpzc{W} = \int \mathrm{d}^Dx \sqrt{\text{det}\thinspace \bar{g}}\thinspace \bigg[& \frac{a_0(x)}{D\epsilon^{\frac{D}{2}}}\left(\lambda^D-1\right)+ \frac{a_2(x)}{(D-2)\epsilon^{\frac{D}{2}-1}}\left(\lambda^{D-2}-1\right) \nonumber \\
	& + \mathcal{O}(\epsilon^{-1}) + a_D(x)\ln \lambda + \mathcal{O}(\epsilon)\bigg].
\end{align}
It is evident that a logarithmic term involving the Seeley-DeWitt coefficient $a_{2n}(x)$ with $n=\frac{D}{2}$ emerges, which is a robust result stemming from the $\epsilon$ cut-off independent part of the bulk effective action and is therefore regularization scheme independent. Hence, we can disregard all divergent and vanishing terms as $\epsilon \to 0$ and extract the leading contribution in the explicit OLEA of the original black hole with radius $\mathfrak{R}$ as
\begin{align}\label{set16}
	\mathpzc{W}_{1\text{-loop}}^{\text{(log)}} = - \int \mathrm{d}^Dx \sqrt{\text{det}\thinspace \bar{g}}\thinspace a_D(x) \ln \mathfrak{R}.
\end{align} 
Putting all this together, we find that the leading one-loop quantum-gravitational effects on the black hole entropy formula in $D$-dimensional theory are given by (with $\mathcal{A}_H \sim \mathfrak{R}^{D-2}$)
\begin{align}\label{set17}
	S_{\text{bh}}(M, \vec{J}, \vec{Q}) &\simeq S_{\text{BH}}(M, \vec{J}, \vec{Q})  + \beta M + \vec{\omega}\cdot\vec{J} + \vec{\mu}\cdot \vec{Q} \nonumber\\
	&\quad + \frac{1}{(D-2)}\left[ \int_{\text{BH geometry}} \mathrm{d}^Dx \sqrt{\text{det}\thinspace \bar{g}}\thinspace a_D(x)+ \mathcal{C}_{\text{zm}}\right] \ln \left(\frac{\mathcal{A}_H}{G_N}\right).
\end{align}
Here, it is important to address the inclusion of $\mathcal{C}_{\text{zm}}$ as an additional prefactor of the logarithmic correction term $\ln \mathcal{A}_H$ in \eqref{set17}. Primarily, $\mathcal{C}_{\text{zm}}$ captures some global corrections that are not accounted for into the local correction term involving the integration of the Seeley-DeWitt coefficient $a_D(x)$ over all spacetimes. One significant source of such global contributions arises from the zero modes present in the theory, identified when eigenvalues satisfy $h_i = 0$ or $\mathcal{H} f_i(x) = 0$ in the setup \eqref{set7}. When studying quantum fluctuations around an asymptotic black hole solution, these zero modes represent typical symmetries (e.g., gauge transformations) that do not vanish even at infinity. In their presence, the OLEA deviates from its Gaussian integral form, as indicated in \eqref{setx1}. Consequently, the heat kernel methodology discussed cannot be used to evaluate the correction to the OLEA arising from zero modes. A systematic resolution of this issue involves removing all zero-mode contributions from the heat kernel \eqref{set7} and adding them back in terms of an overall volume factor associated with the asymptotic symmetry groups responsible for inducing the zero modes. This procedure contributes one part of the correction term $\mathcal{C}_{\text{zm}}$ in \eqref{set17}. 

Furthermore, the effective action in the current setup is initially defined through the Euclidean path integral under thermal boundary conditions and identified as the free energy within the canonical ensemble. In contrast, the quantum black hole entropy formulation occurs within the microcanonical ensemble, where the black hole mass and charges remain fixed. The transformation between these ensembles has been accomplished via the Legendre transform \eqref{set3}, which induces a logarithmic term and is eventually absorbed into $\mathcal{C}_{\text{zm}}$. 

For a more detailed discussion of the non-local or global corrections, readers are referred to \cite{Banerjee:2011oo,Banerjee:2011pp,Sen:2012rr,Sen:2012qq,Sen:2013ns,Bhattacharyya:2012ye,Charles:2015nn,David:2021eoq,Liu:2017vbl,Camporesi:1994ye,H:2023qko}. Additionally, the data regarding $\mathcal{C}_{\text{zm}}$ for the four-dimensional asymptotically flat and AdS black holes of interest in this paper are well-known and summarized in \cref{4dcomp,czero}.
{
	\renewcommand{\arraystretch}{1.3}
	\begin{table}[t]
		\centering
		\hspace{-0.2in}
		\begin{tabular}{|>{}p{3.0in}|>{\centering}p{0.5in}|}
			\hline
			\textbf{Black Hole Backgrounds} & \textbf{$\bm{\mathcal{C}_{\text{zm}}}$} \tabularnewline \hline
			Schwarzschild & $-3$ \tabularnewline 
			Non-extremal Kerr & $-1$ \tabularnewline 
			Non-extremal Reissner-Nordstr\"om & $-3$ \tabularnewline 
			Non-extremal Kerr-Newman & $-1$ \tabularnewline 
			Extremal Kerr near-horizon & $-4$ \tabularnewline 
			Extremal Reissner-Nordstr\"om near-horizon & $-6$ \tabularnewline 
			Extremal Kerr-Newman near-horizon & $-4$ \tabularnewline \hline
		\end{tabular}
		\caption{$\mathcal{C}_{\text{zm}}$ contributions to the logarithmic corrected black hole entropy. The results are same for both the asymptotically flat and AdS partners of each background geometry.}\label{czero}
	\end{table}
}


\subsection{Extremal black holes and quantum entropy function formalism}\label{ext}

We would like to highlight that the Euclidean quantum gravity framework outlined in \cref{nonext} does not directly accommodate extremal black holes. This limitation arises because naively applying the extremal limit $\beta \to \infty$ or $T_{\mathrm{bh}}= (\frac{\partial S_{\text{bh}}}{\partial M})^{-1} \to 0$ results in a divergent OLEA and quantum black hole entropy in the setup \eqref{set17}. However, we can interpret extremality as the non-radiating stable ``ground state'' of the non-extremal or finite temperature black hole setup, where the divergence stemming from extremality can be viewed as an infinite shift to the ground state energy. In the analysis of this paper, we primarily employ the well-established quantum entropy function (QEF) formalism \cite{Sen:2008wa,Sen:2009wb,Sen:2009wc} to regulate these divergences and adjust the quantum entropy formula \eqref{set17} to account for extremal black holes.

The near-horizon geometry (NHG) at extremality is interpreted as the ground state limit on the non-extremal horizon. This crucial feature is utilized in QEF formalism to define the quantum entropy of extremal black holes solely through the near-horizon analysis, thus bypassing the need for detailed knowledge of the entire spacetime. The extremal NHG is well-defined and gives rise to a new class of AdS$_2$ solutions \cite{Sen:2008wa}. Consequently, according to the AdS/CFT correspondence, the entropy of extremal black holes precisely corresponds to the entropy calculated from the full partition function in AdS$_2$, which is equivalent to the partition function of the boundary CFT$_1$, 
\begin{align}\label{ext1}
	\lim_{\beta \to \infty}\mathcal{Z}_{\text{AdS}_2} = \lim_{\beta \to \infty} \mathcal{Z}_{\text{CFT}_1}.
\end{align}
Here, $\beta$ serves as an infrared regulator on both sides, effectively regularizing the infinite volume of AdS$_2$ as well as the infinite length of the CFT$_1$ boundary. On the CFT$_1$ side, as $\beta$ tends to infinity, only the ground state (with energy $E_0$ and degeneracy $d_0$) contributes to the partition function,
\begin{align}\label{ext2}
	\lim_{\beta \to \infty} \mathcal{Z}_{\text{CFT}_1} = d_0 e^{-\beta E_0}.
\end{align}
Consequently, computing entropy from the partition function \eqref{ext2} yields $\ln d_0$, with $d_0$ interpreted as the microstate degeneracy underlying the entropy of extremal black holes. On the gravity or AdS$_2$ side, the trick is to choose appropriate coordinates such that the regularized path integral $\mathcal{Z}_{\text{AdS}_2}$ can be expressed in a similar form as \eqref{ext2},
\begin{align}\label{ext3}
	\lim_{\beta \to \infty}\mathcal{Z}_{\text{AdS}_2} = \mathcal{Z}^{\text{finite}}e^{-\beta \mathcal{C} + \mathcal{O}(\beta^{-1})}.
\end{align}
Here, the linear term involving a constant $\mathcal{C}$ in the $\beta$-dependent part corresponds to an infinite shift in the ground state energy. At any point, we can disregard this divergent term, along with all other vanishing $\mathcal{O}(\beta^{-1})$ terms, in the limit $\beta \to \infty$ and identify the cut-off independent quantity $\mathcal{Z}^{\text{finite}}$ with $d_0$. This $\mathcal{Z}^{\text{finite}}$ piece is known as the quantum entropy function \cite{Sen:2008wa,Sen:2009wb,Sen:2009wc}, which describes a macroscopic definition of the horizon degeneracy of extremal black holes. By computing the logarithm of $\mathcal{Z}^{\text{finite}}$ provides a finite and unambiguous result for the extremal black hole entropy, which is independent of any regularization procedure.

Based on the quantum entropy function formalism described above, we can refine the formulation of the OLEA part in formula \eqref{set17} as follows
\begin{align}\label{set20}
	\int_{\text{BH geometry}} \mathrm{d}^Dx \sqrt{\text{det}\thinspace \bar{g}}\thinspace a_D(x)\Big\rvert_{\beta \to \infty} \equiv {\left\langle \int_{\text{near-horizon}} \mathrm{d}^Dx \sqrt{\text{det}\thinspace \bar{g}}\thinspace a_D(x) \right\rangle}_{\text{AdS}_2}^{\text{finite}}.
\end{align}
Here, the notation $\langle \rangle$ represents the integration of the coefficient $a_D(x)$ over the finite extremal NHG. This integration explicitly excludes any boundary independent terms of the regulated $\text{AdS}_2$ after structuring the Euclideanized extremal NHG into the form $\text{AdS}_2 \times K^{D-2}$, where $K$ denotes a $(D-2)$-dimensional space fibered over the $\text{AdS}_2$ part and encompasses all the compact or angular coordinates. For more technical details, interested readers can refer to \cite{Banerjee:2011pp,Sen:2012rr,Sen:2012qq,Bhattacharyya:2012ss,Karan:2019sk,Karan:2020sk,Banerjee:2020wbr,Karan:2021teq}.


\subsection{Working formulas and computations for 4D black holes}\label{4dcomp}
In this paper, the working formula for computing the logarithmic correction to the entropy of black holes in a four-dimensional theory is given by
\begin{subequations}\label{comp1}
	\begin{align}\label{comp1a}
		\Delta S_{\text{BH}} &= \frac{1}{2}\left(\mathcal{C}_{\text{local}}+\mathcal{C}_{\text{zm}}\right)\ln\left(\frac{\mathcal{A}_{H}}{G_N}\right).
	\end{align}
	The local contribution, denoted as $\mathcal{C}_{\text{local}}$, is identified as the density of the Seeley-DeWitt coefficient $a_4(x)$ integrated over the finite near-horizon and full geometries respectively for extremal and non-extremal black holes,
	\begin{align}\label{comp1b}
		\mathcal{C}_{\text{local}} &= \int_{\text{BH geometry}} \mathrm{d}^4x \sqrt{\text{det}\thinspace \bar{g}}\thinspace a_4(x).
	\end{align}
	The global contribution $\mathcal{C}{_\text{zm}}$, which captures corrections from the available zero-modes and the changes in ensemble, has been extensively computed and analyzed in previous works \cite{Banerjee:2011oo,Banerjee:2011pp,Sen:2012rr,Sen:2012qq,Sen:2013ns,Bhattacharyya:2012ye,Charles:2015nn,David:2021eoq,Liu:2017vbl,Camporesi:1994ye}. We can consolidate the available data and write a common but compact $\mathcal{C}_{\text{zm}}$ formula applicable for all 4D asymptotically flat and AdS black holes as 
	\begin{align}\label{comp1c}
		\mathcal{C}_{\text{zm}} = -(3+\mathbb{K})+ 3\delta_{\text{non-ext}} +  2\mathbb{N}_{\text{BPS}}.
	\end{align}
\end{subequations}
Here, $\mathbb{K}$ takes the value 3 for spherically symmetric non-rotating black holes and 1 for other cases. $\delta_{\text{non-ext}}$ is 0 for extremal black holes and 1 otherwise, while $\mathbb{N}_{\text{BPS}}$ is 4 for black holes preserving supercharges (i.e., BPS black holes) and 0 otherwise.

In the formula for $\mathcal{C}_{\text{zm}}$ \eqref{comp1c}, the $-\mathbb{K}$ contribution arises from the change in the ensemble from canonical to microcanonical via the transition \eqref{set3} and is related to the number of unbroken rotational symmetries of the black hole. The $-3$ contribution is associated with the SL($2,\mathbb{R}$) symmetry of AdS$_2$ spaces characterizing the near-horizon geometry of extremal black holes. This contribution is countered by the additional $3\delta_{\text{non-ext}}$ contribution when transitioning from the near-horizon to the full geometry for non-extremal black holes. Furthermore, the contribution $2\mathbb{N}_{\text{BPS}}$ arises from the fermionic generators of the PSU(1,1|2) near-horizon symmetry for BPS black holes in supergravity theories.

For the local piece \eqref{comp1b} in logarithmic correction, this paper aims to follow the methodology outlined in \cite{Vassilevich:2003ll} and evaluate the heat kernel coefficient $a_4(x)$ solely in terms of the background fields and covariant derivatives appearing in the kinetic operator $\mathcal{H}$ characterizing the one-loop fluctuations. The entire strategy is summarized as follows. First, the quadratic fluctuated action needs to be adjusted as
\begin{align}\label{comp2}
	\delta^2 \mathcal{S}[\tilde{\phi}_m]= \int \mathrm{d}^4x \sqrt{\text{det}\thinspace \bar{g}}\thinspace \tilde{\phi}_m \mathcal{H}^m_n\tilde{\phi}^n,
\end{align}
for the quantum fluctuations $\lbrace \tilde{\phi}_m\rbrace$ so that the differential operator $\mathcal{H}$ takes the following Hermitian and Laplace-type form,
\begin{align}\label{comp3}
	-\mathcal{H}^m_n = \left(D_\rho D^\rho\right) \mathcal{I}^m_n + 2(\omega_\rho D^\rho)^m_n + P^m_n.
\end{align}
Here, $m$ and $n$ simultaneously label the types of fluctuations as well as their tensor indices. $D_\rho$ represents the covariant derivative with connections controlled by the background metric. $\mathcal{I}$ serves as the identity operator in the space of field fluctuations, which functions as an effective metric or a projection operator for raising and lowering the indices of relevant matrices, and defines the trace operation in the central formula \eqref{comp8}.\footnote{In general, $\mathcal{I}$ is $1$ for scalars, $\bar{g}^{\mu\nu}$ for vectors, and takes the DeWitt metric form \eqref{dewitt} for spin-2 metric fluctuations. If the fluctuated theory incorporates fermions, $\mathcal{I}$ is identified as $\mathbb{I}_4$ for spin-$\frac{1}{2}$ Dirac spinors and $\mathbb{I}_4\bar{g}^{\mu\nu}$ for spin-$\frac{3}{2}$ Rarita-Schwinger spinors, where $\mathbb{I}_4$ denotes the identity matrix of the Clifford algebra describing the spinor form of the four-dimensional fermionic fluctuations.} Also, $\omega_\rho$ and $P$ are matrices induced from the background metric and fields. More precisely, the matrix $\omega_\rho$ in the operator form \eqref{comp3} is identified as a background gauged connection determining the non-minimal coupling between fluctuations. Consequently, we can reformulate the quadratic operator $\mathcal{H}$ into a more generic form,
\begin{align}\label{comp4}
	-\mathcal{H}^m_n = \left(\mathcal{D}_\rho\mathcal{D}^\rho\right) \mathcal{I}^m_n + E^m_n,
\end{align}
where $\mathcal{D}_\rho$ is a modified covariant derivative with the gauge connection $\omega_\rho$, defined as
\begin{align}\label{comp5}
	\mathcal{D}_\rho \tilde{\phi}_m= D_\rho \tilde{\phi}_m + (\omega_\rho)^m_n {\tilde{\phi}}^n \quad\forall\thinspace m\neq n.
\end{align}
Note that by definition, $\omega_\rho$ vanishes for any fluctuation that is minimally coupled to the background gravity via $\sqrt{\text{det}\thinspace \bar{g}}$ in the quadratic action form \eqref{comp2}.  Furthermore, there exists an effective curvature $\Omega_{\rho\sigma}$ associated with $\mathcal{D}_\rho$, given by the following relation,
\begin{align}\label{comp6}
	\tilde{\phi}_m\left(\Omega_{\rho\sigma}\right)^m_n \tilde{\phi}^m =  \tilde{\phi}_m[\mathcal{D}_\rho,\mathcal{D}_\sigma] \tilde{\phi}^m =  \tilde{\phi}_m[D_\rho,D_\sigma] \tilde{\phi}^m +  \tilde{\phi}_m{D_{[\rho}\omega_{\sigma]}} ^m_n \tilde{\phi}^n +  \tilde{\phi}_m[\omega_\rho,\omega_\sigma]^m_n \tilde{\phi}^n,
\end{align}
where the brackets indicate commutation operations. In the first part of \eqref{comp6}, the covariant derivative commutation $[D_\rho,D_\sigma]$ operating on scalars $\Phi$, vectors or gauge fields $a_\mu$, spin-1/2 Dirac fields $\psi$, spin-3/2 Rarita-Schwinger fields $\psi_\mu$, and metric or graviton fields $h_{\mu\nu}$ is defined by the following standard relations,
{	\allowdisplaybreaks
	\begin{subequations}\label{commutations}
		\begin{align}
			[D_\rho,D_\sigma]\Phi &= 0, \label{cs}\\
			[D_\rho,D_\sigma]a_\mu &= R\indices{_\mu^\nu_{\rho\sigma}} a_\nu,\label{cv}\\
			[D_\rho,D_\sigma]\psi &= \frac{1}{8}[\gamma^\alpha,\gamma^\beta] R_{\alpha\beta\rho\sigma}\lambda,\label{cd}\\
			[D_\rho,D_\sigma]\psi_\mu &=R\indices{_\mu^\nu_{\rho\sigma}}\psi_\nu + \frac{1}{8}[\gamma^\alpha,\gamma^\beta] R_{\alpha\beta\rho\sigma}\psi_\mu,\label{crs}\\
			[D_\rho,D_\sigma]h_{\mu\nu} &= R\indices{_\mu^\alpha_{\rho\sigma}}h_{\alpha\nu} + R\indices{_\nu^\alpha_{\rho\sigma}}h_{\mu\alpha},\label{ct}
		\end{align}
\end{subequations}}
where the gamma matrices $\gamma^\alpha$ are associated with the Clifford algebra describing the spinors. Next, the matrix $E$ introduced in the operator form \eqref{comp4} accounts for the effective potential for the operator $\mathcal{H}$ via the following expression,
\begin{align}\label{comp7}
	\tilde{\phi}_mE^m_n \tilde{\phi}^n &= \tilde{\phi}_mP^m_n\tilde{\phi}^n - \tilde{\phi}_m(D_\rho\omega^\rho)^m_n\tilde{\phi}^n - \tilde{\phi}_m(\omega_\rho)^{mp}(\omega^\rho)_{pn}\tilde{\phi}^n.
\end{align}
Finally, incorporating all the aforementioned matrix-valued data, the Seeley-DeWitt coefficient $a_4(x)$ can be determined using the formula \cite{Vassilevich:2003ll}
\begin{align}\label{comp8}
	a_4(x) &= \frac{\chi}{16\pi^2} \mathrm{Tr}\bigg[ \frac{1}{2}E^2 +\frac{1}{6}RE + \frac{1}{12}\Omega_{\rho\sigma}\Omega^{\rho\sigma} +\frac{1}{180}\left(R_{\mu\nu\rho\sigma}R^{\mu\nu\rho\sigma}-R_{\mu\nu}R^{\mu\nu}+ \frac{5}{2} R^2\right)\mathcal{I}\bigg],
\end{align}
where $R_{\mu\nu\rho\sigma}$, $R_{\mu\nu}$, and $R$ are the background Riemann and Ricci curvature tensors and Ricci scalar, respectively. Here, the $\mathrm{Tr}$ operation over a specific matrix is defined by contracting them using the appropriate $\mathcal{I}$. In the analysis of this paper, we shall ignore all the total derivative terms while computing $a_4(x)$ since they appear as vanishing boundary contributions to the integral \eqref{comp1b}. 

Furthermore, in order to incorporate fermionic fluctuations into the current heat kernel setup, a specific adjustment is necessary, as demonstrated in \cite{Sen:2012qq}. The quadratic action of fermions $\Psi$ is always characterized by a first-order operator $\slashed{D}$,
\begin{align}\label{comp9}
	\delta^2 \mathcal{S}[{\Psi}_m,\bar{\Psi}_m]= \int \mathrm{d}^4x \sqrt{\text{det}\thinspace \bar{g}}\thinspace \bar{\Psi}_m \slashed{D}^m_n{\Psi}^n, \enspace \bar{\Psi}_m = \left(\Psi_m\right)^\dagger.
\end{align}
However, the fermionic operator $\slashed{D}$ can be bosonized into the desired second-order Laplace-type form \eqref{comp3} by adjusting the one-loop determinant form \eqref{set5} as follows
\begin{align}\label{comp10}
	\text{ln det}\thinspace \slashed{D}=\text{ln det}\thinspace \slashed{D}^\dagger = \frac{1}{2}\ln\det \mathcal{H},\quad \mathcal{H} = \slashed{D}^\dagger\slashed{D},
\end{align}
where the appearance of an additional factor of $1/2$ needs to be compensated by setting the $\chi$ value as $-1$ and $-1/2$ in the formula \eqref{comp8} for the case of complex Dirac and real Majorana spinors, respectively. $\chi$ is always set to $+1$ for all bosons and scalars. Additionally, the signature of $\chi$ must be reversed for the ghost fields induced during the process of gauge-fixing the theory. It is important to note that the treatment outlined in \eqref{comp10} cannot accommodate Weyl spinors with both left and right chirality states. For further elaboration and examples regarding the heat kernel computation of various elementary fermionic cases, readers are encouraged to review \cite{Karan:2018ac}.

The heat kernel approach offers a significant advantage: after fluctuating the action around an arbitrary classical background up to a quadratic order, we have systematic steps and a straightforward formula to compute $a_4(x)$ only in terms of background curvature invariants. This can be particularly useful for determining the quantum entropy of all black holes in the theory under consideration. For example, please refer to \cref{model,logresult}.


\section{\boldmath Seeley-DeWitt coefficient $a_4(x)$ in STU supergravity}\label{model}

This section aims to revisit the model setup of STU supergravity in terms of Einstein-Maxwell-dilaton (EMD) systems, and subsequently embed the EM-AdS and EM backgrounds \cite{Carter:1968ks,Plebanski:1976gy,Caldarelli:1999x,Adamo:2014lk} within it. We then demonstrate systematic computations of the Seeley-DeWitt coefficient $a_4(x)$ for the fluctuations of the STU content around the embedded black hole backgrounds. This computation is essential for deriving the logarithmic correction relation presented in \cref{logresult}.

\subsection{The model setup}\label{setupmodel}
Four-dimensional maximal $\mathcal{N} = 8$ supergravity theory originates from the $T^7$ reduction of eleven-dimensional supergravity, through ten-dimensional type IIA supergravity, where the bosonic content comprises the metric and a large number of $U(1)$ gauge fields as well as scalar fields \cite{Cremmer:1978ds,Cremmer:1979up}. To generate a solution encompassing the most general black hole in $\mathcal{N} = 8$ supergravity, the global symmetries of the field equations (i.e., classical U-dualities)  indicate that reducing the theory to only four gauge fields is sufficient \cite{Cvetic:1996zq}. This pertinent supergravity theory, commonly referred to as the STU model, represents $\mathcal{N} = 2$ supergravity coupled with three vector multiplets \cite{Cremmer:1984hj,Duff:1995sm}. In each vector multiplet, there exists a gauge field, a dilaton, and an axion, while the fourth gauge field is part of the $\mathcal{N} = 2$ supergravity multiplet. 
The scope of this study is to nullify the influence of the three axionic scalars and ensure the vanishing of their sources. To achieve this, we explore field configurations where the four $U(1)$ or Maxwell field strengths exclusively exhibit electric components without any magnetic components. Our focus is solely on the bosonic sector, which provides black hole solutions for STU supergravity and often viewed as a $U(1)^4$-charged EMD model, with the Einstein gravity sector being non-minimally coupled to the three Maxwell sectors through three dilaton fields \cite{Cvetic:1999xp,Clement:2013fc,Chow:2014cca,Cvetic:2014vsa,Cvetic:2021lss,Anabalon:2022aig}. The pertinent field content comprises the metric $g_{\mu\nu}$, four $U(1)$ gauge or Maxwell fields $\mathcal{A}_{I\mu}$, and three dilatons $\Phi_i$, where $I = 1, 2, 3, 4$ and $i= 1, 2, 3$. Their dynamics and equations of motion are described by the following action\footnote{The action structure \eqref{mod1} interprets only the massless sector of the STU model and necessarily excludes all the relevant scalar potential or effective dilaton mass terms, as they do not contribute to the logarithmic correction results computed in \cref{logresult}.}
{
	\allowdisplaybreaks
	\begin{subequations}\label{mod1}
		\begin{align}\label{mod1a}
			\mathcal{S}[g_{\mu\nu}, \mathcal{A}_{I\mu}, \Phi_i] &= \int \mathrm{d}^4x \sqrt{\det g} \Big(\mathcal{R}-2\Lambda- 2\sum_{i=1}^{3}D^\mu\Phi_iD_\mu\Phi_i - \sum_{I=1}^{4} \mathfrak{f}_I(\vec{\Phi})\mathcal{F}_{I\mu\nu}\mathcal{F}_{I}^{\mu\nu} \Big),
		\end{align}
		where we have defined, 
		\begin{align}\label{mod1b}
			\mathfrak{f}_I(\vec{\Phi}) = e^{-2\vec{\mathfrak{a}}_I\cdot \vec{\Phi}}, \quad \vec{{\Phi}} = (\Phi_1, \Phi_2, \Phi_3), \quad \vec{\mathfrak{a}}_1 = (1, 1, 1),\\
			\vec{\mathfrak{a}}_2 = (1, -1, -1), \quad \vec{\mathfrak{a}}_3 = (-1, 1, -1), \quad \vec{\mathfrak{a}}_4 = (-1, -1, 1),
		\end{align}
\end{subequations}}
where $\mathcal{R} = g^{\mu\nu}\mathcal{R}_{\mu\nu}$ corresponds to the Ricci scalar characterizing the Einstein sector and $\mathcal{F}_{I\mu\nu} = D_{[\mu}\mathcal{A}_{I\nu]}$ represents the 2-form gauge field strengths that govern the four distinct Maxwell sectors. The presence of negative cosmological constant $\Lambda$ induces the AdS$_4$ backgrounds of boundary $\ell$ (such that $\ell^2 = -3/\Lambda$) within the gauged version of STU supergravity. At any point, we can set the limit $\ell \to \infty$ or $\Lambda = 0$ in order to transition into the ungauged STU supergravity admitting the flat$_4$ black hole backgrounds.

The STU supergravity model \eqref{mod1} has the scope for additional simplification, leading to irreducible $U(1)^2$-charged EMD systems including just a single dilaton and two Maxwell or $U(1)$ gauged fields \cite{Clement:2013fc,Chow:2014cca,Cvetic:2014vsa,Lu:2013eoa,Mai:2021yny}. This truncation gives rise to two distinct scenarios. The first involves setting $\mathcal{A}_{1\mu} = \mathcal{A}_{2\mu}$ and $\mathcal{A}_{3\mu} = \mathcal{A}_{4\mu}$, while the second scenario entails setting $\mathcal{A}_{2\mu} = \mathcal{A}_{3\mu} = \mathcal{A}_{4\mu}$. In both cases, the relevant action form is formulated as
\begin{subequations}\label{mod2}
	\begin{align}\label{mod2a}
		\mathcal{S}[g_{\mu\nu}, {A}_{1\mu}, {A}_{2\mu}, \Phi] &= \int \mathrm{d}^4x \sqrt{g} \left(\mathcal{R}-2\Lambda- 2 D^\mu\Phi D_\mu\Phi -  {f}_1({\Phi}){F}_{1\mu\nu}{F}\indices{_1^\mu^\nu} -  {f}_2({\Phi}){F}_{2\mu\nu}{F}\indices{_2^\mu^\nu} \right),
	\end{align}
	where the two Maxwell field strengths and dilaton coupling functions are defined as
	\begin{align}\label{mod2b}
		{F}_{1\mu\nu} = D_{[\mu}{A}_{1\nu]}, \quad {F}_{2\mu\nu} = D_{[\mu}{A}_{2\nu]}, \quad f_1(\Phi)=e^{-2\kappa_1\Phi}, \quad f_2(\Phi)=e^{-2\kappa_2\Phi}. 
	\end{align}
	It is important to note that the dilaton coupling constants $(\kappa_1, \kappa_2)$ must satisfy,
	\begin{align}\label{mod2c}
		\kappa_1 \kappa_2 = -1, \quad N_1 \kappa_1 + N_2 \kappa_2 = 0, \quad N_1 + N_2 = 4,	
	\end{align}
	which ensure the consistency of the $U(1)^2$-charged EMD model \eqref{mod2} as a consistent truncation of STU supergravity \eqref{mod1}. There are only two special cases of the STU truncations:  $\left(N_1, N_2\right) = \left(2, 2\right)$ and $(1, 3)$, which in turn correspond to $\left(\kappa_1, \kappa_2\right) = \left(1, -1\right)$ and $(\sqrt{3}, -1/\sqrt{3})$, respectively \cite{Cvetic:2014vsa}. Notably, the former type directly intersects with the exact bosonic sector of $\mathcal{N}=4$ supergravity \cite{Kallosh:1992ii,Kol:1996hf,Krasnitz:1997gn}. These two $U(1)^2$-charged systems hold particular significance within this paper, as they exhibit considerably lower complexity compared to the complete STU model, while still preserving all fundamental characteristics and offering an irreducible form of supergravity. It is essential to note that the relevant $U(1)$ gauge or Maxwell fields, $A_{1\mu}$ and $A_{2\mu}$, exclusively carry electric charges.
\end{subequations}

We are now interested in exploring the nature of the equations of motion that govern STU supergravity. These equations stem from the evolution of the action \eqref{mod2a} with respect to all STU components in the context of an arbitrary background solution denoted as $(\bar{g}_{\mu\nu},\, \bar{A}_{1\mu},\, \bar{A}_{2\mu}, \bar{\Phi})$. The gravitational field equations are derived as follows
\begin{subequations}\label{mod3}
	\begin{align}
		R_{\mu\nu}-\frac{1}{2}\bar{g}_{\mu\nu}R + \Lambda \bar{g}_{\mu\nu} = T_{\mu\nu}^{\mathrm{(dilaton)}} + T_{\mu\nu}^{U(1)^2},\label{mod3a}
	\end{align}
	where $R_{\mu\nu}$ and $R$ represent the background Ricci curvature tensor and scalar, respectively. The total stress-energy tensor consists of distinct dilaton and $U(1)^2$ or Maxwell components:
	\begin{align}
		T_{\mu\nu}^{\mathrm{(dilaton)}} &= 2 \Big( D_\mu\bar{\Phi}D_\nu \bar{\Phi} - \frac{1}{2} \bar{g}_{\mu\nu}D_\rho\bar{\Phi}D^\rho \bar{\Phi}\Big),\label{mod3b}\\[5pt]
		T_{\mu\nu}^{U(1)^2}  &= e^{-2\kappa_1 \bar{\Phi}}\Big( 2\bar{F}_{1\mu\rho}\bar{F}\indices{_{1\nu}^\rho}-\frac{1}{2}\bar{g}_{\mu\nu}\bar{F}_{1\rho\sigma}\bar{F}\indices{_1^\rho^\sigma}\Big)\nonumber\\
		&\quad + e^{-2\kappa_2 \bar{\Phi}}\Big( 2\bar{F}_{2\mu\rho}\bar{F}\indices{_{2\nu}^\rho}-\frac{1}{2}\bar{g}_{\mu\nu}\bar{F}_{2\rho\sigma}\bar{F}\indices{_2^\rho^\sigma}\Big),\label{mod3c}
	\end{align}
\end{subequations}
where $\bar{F}_{1\mu\nu}= {D_{[\mu}\bar{A}_{1\nu]}}$ and $\bar{F}_{2\mu\nu}= {D_{[\mu}\bar{A}_{2\nu]}}$ symbolize the background Maxwell field strengths. The two sets of Maxwell and Maxwell-Bianchi equations take the form,
\begin{align}\label{mod4}
	\begin{split}
		D_\mu \left(e^{-2\kappa_1 \bar{\Phi}}\bar{F}\indices{_1^\mu^\nu}\right) & = 0, \enspace D_{[\mu}\bar{F}_{1\rho\sigma]}=0, \\
		D_\mu \left(e^{-2\kappa_2 \bar{\Phi}}\bar{F}\indices{_2^\mu^\nu}\right) & = 0, \enspace D_{[\mu}\bar{F}_{2\rho\sigma]}=0.
	\end{split}
\end{align}
Finally, the dilaton evolution equation is expressed as:
\begin{align}\label{mod5}
	D_\mu D^\mu \bar{\Phi} + \frac{1}{2}\left(\kappa_1e^{-2\kappa_1\bar{\Phi}}\bar{F}_{1\mu\nu}\bar{F}\indices{_1^\mu^\nu} + \kappa_2e^{-2\kappa_2\bar{\Phi}}\bar{F}_{2\mu\nu}\bar{F}\indices{_2^\mu^\nu}\right) =0.
\end{align}
Notably, the aforementioned STU equations of motion, together with their corresponding background solutions, always remain unchanged under the following transformations 
\begin{align}\label{mod6}
	\begin{gathered}
		\left\lbrace Q_1 \longrightarrow Q_2,\, Q_2 \longrightarrow Q_1 \right\rbrace, \enspace (\kappa_1, \kappa_2, \bar{\Phi}) \longrightarrow (-\kappa_1, -\kappa_2, -\bar{\Phi}), \\[3pt]
		f_1(\bar{\Phi}) \longrightarrow \frac{1}{f_1(\bar{\Phi})},\enspace f_2(\bar{\Phi}) \longrightarrow \frac{1}{f_2(\bar{\Phi})},
	\end{gathered}
\end{align}
where any change in the signature of dilaton coupling constants necessitates a corresponding reversal in the signature of the background dilaton as well as its coupling functions, and vice versa. This further suggests that the STU background solutions remain invariant when electric charges $(Q_1, Q_2)$ associated with the background Maxwell fields $(A_{1\mu}, A_{2\mu})$ are interchanged. It is important to note that all the above-mentioned equations of motion and identities hold appropriately for the two cases of $U(1)^2$-charged STU truncations when $(\kappa_1, \kappa_2) = (1, -1)$ and $(\sqrt{3}, -1/\sqrt{3})$, satisfying $\kappa_1\kappa_2 = -1$.


\subsection{Embedding of the Einstein-Maxwell backgrounds}\label{embeddingEM}
The $U(1)^2$-charged STU supergravity model \eqref{mod2} emerges as a natural extension of the Einstein-Maxwell (EM) theory where the Einstein gravity sector is minimally-coupled to a single Maxwell sector described by the action, 
\begin{align}\label{mod7}
	\mathcal{S}[g_{\mu\nu}, A_\mu]= \int \mathrm{d}^4x \sqrt{\det g} \left(\mathcal{R}-2\Lambda- {F}_{\mu\nu}{F}^{\mu\nu} \right).
\end{align}
Within this EM framework, the well-known Kerr-Newman family of black holes represents the general background solutions, satisfying the following field equations
\begin{align}\label{mod8}
	\begin{gathered}
		R_{\mu\nu}- \bar{g}_{\mu\nu}\Lambda = 2 \bar{F}_{\mu\rho}\bar{F}\indices{_\nu^\rho}- \frac{1}{2}\bar{g}_{\mu\nu}\bar{F}_{\rho\sigma}\bar{F}^{\rho\sigma}, \quad R = 4\Lambda, \\
		D_\mu \bar{F}^{\mu\nu} = 0, \quad D_{[\mu}\bar{F}_{\rho\sigma]}=0. 
	\end{gathered}
\end{align}
This family of EM background solutions encompasses all the four-dimensional asymptotically-AdS and asymptotically-flat counterparts of Schwarzschild (both stationary and static), Reissner-Nordström (stationary with charge), Kerr (static with rotation), and Kerr-Newman (both rotating and charged) black holes. Our current goal is to systematically recover all of these EM backgrounds within the STU supergravity model \eqref{mod2}. This motivation stems from the intriguing properties and broader implications of the EM solutions, which already found a robust microscopic foundation within string theory. The four-dimensional EM theory \eqref{mod7} features a single $U(1)$ Maxwell field coupled minimally to the metric field and is known to admit supersymmetric black hole solutions by structuring the bosonic sector of a pure $\mathcal{N}=2$ supergravity \cite{Freedman:2012xp}. However, while the STU supergravity we focus on is a more generalized theory, there is no direct and consistent truncation that reduces STU models to the EM theory. 
We now elucidate how the systematic recovery of the Kerr-Newman family of black holes within STU supergravity models is achievable through special embedding choices.
{
	\renewcommand{\arraystretch}{2.1}
	\begin{table}[t]
		\centering
		\hspace{-0.2in}
		\begin{tabular}{|>{\centering}p{2.5in}|>{\centering}p{3.0in}|}
			\hline
			\textbf{Types of $U(1)^2$-charged STU Supergravity Models} & \textbf{Constraints on Background $U(1)$ Charges and Field Strengths} \tabularnewline \hline
			Case I ($\kappa_1 =1, \kappa_2=-1$) & $Q_1 = Q_2,\enspace \bar{F}_{1\mu\nu}\bar{F}\indices{_1^\mu^\nu} = \bar{F}_{2\mu\nu}\bar{F}\indices{_2^\mu^\nu}$ \tabularnewline \hline
			Case II ($\kappa_1 =\sqrt{3}, \kappa_2=-\frac{1}{\sqrt{3}}$)  & $Q_1=\frac{1}{\sqrt{3}}Q_2,\enspace \bar{F}_{1\mu\nu}\bar{F}\indices{_1^\mu^\nu} = \frac{1}{3}\bar{F}_{2\mu\nu}\bar{F}\indices{_2^\mu^\nu}$ \tabularnewline \hline
		\end{tabular}
		\caption{The choice of constraints on two cases of $U(1)^2$-charged STU truncations \eqref{mod2} for embedding Einstein-Maxwell (EM) black hole backgrounds.}\label{embeddings}
	\end{table}
} 

While the STU model \eqref{mod2} is indeed a natural extension of the EM theory \eqref{mod7}, the recovery of the EM theory through a trivial truncation is not feasible due to the presence of non-minimal dilaton coupling functions $f_1(\Phi) = e^{-2\kappa_1\Phi}$ and $f_2(\Phi) = e^{-2\kappa_2\Phi}$. Even when considering a vanishing dilaton background ($\bar{\Phi} = 0$), the EM backgrounds do not conform to the STU evolution equations \eqref{mod3} to \eqref{mod5}. However, a special scenario exists where the EM backgrounds can be revived by constraining the STU equations of motion with $\kappa_1 \bar{F}_{1\mu\nu}\bar{F}\indices{_1^\mu^\nu} + \kappa_2 \bar{F}_{2\mu\nu}\bar{F}\indices{_2^\mu^\nu} = 0$ for any non-vanishing dilaton coupling constants $\kappa_1$ and $\kappa_2$. This particular EM embedding is achievable by appropriately scaling the charges of two background $U(1)$ or Maxwell fields in STU supergravity. In the context of two specific $U(1)^2$-truncated STU cases, the EM embedding conditions can be summarized as follows (also see \Cref{embeddings})
\begin{align}\label{mod9}
	\begin{gathered}
		{Q_2} = {\kappa_1}{Q_1} = -\frac{{Q_1}}{{\kappa_2}}\, \xrightarrow[\text{STU supergravity}]{\text{EM embedding}}\, \bar{F}_{2\mu\nu}\bar{F}\indices{_2^\mu^\nu} = -\frac{\kappa_1}{\kappa_2}\bar{F}_{1\mu\nu}\bar{F}\indices{_1^\mu^\nu},\enspace \bar{\Phi} =0,
	\end{gathered}  
\end{align}
where $(Q_1, Q_2)$ represent the electric charges for the background $U(1)$ or Maxwell fields $(\bar{A}_{1\mu}, \bar{A}_{2\mu})$. Technically, the aforementioned embedding choice effectively decouples all the non-minimal `Maxwell-dilaton' background terms and interprets $\bar{\Phi}=0$ as a non-trivial solution of the STU field equations. As a result, the dilaton contribution \eqref{mod3b} of the stress-energy tensor disappears, while the Maxwell component \eqref{mod3c} undergoes modification to represent an effective background $U(1)$ field with a net charge of $\sqrt{Q_1^2 + Q_2^2}$. This further enforces all STU evolution equations \eqref{mod3} to \eqref{mod5} to truncate into the exact Einstein-Maxwell background equations \eqref{mod8}, subject to the following definitions (also refer to \cref{kna})
\begin{align}\label{mod10}
	\begin{gathered}
		\bar{F}_{\mu\rho}\bar{F}\indices{_\nu^\rho} = \bar{F}_{1\mu\rho}\bar{F}\indices{_1^\rho_\nu} + \bar{F}_{2\mu\rho}\bar{F}\indices{_2^\rho_\nu}, \\[5pt]
		\bar{F}_{\mu\nu}\bar{F}^{\mu\nu} = \bar{F}_{1\mu\nu}\bar{F}\indices{_1^\mu^\nu} + \bar{F}_{2\mu\nu}\bar{F}\indices{_2^\mu^\nu}= \kappa_1^2 \bar{F}_{1\mu\nu}\bar{F}\indices{_1^\mu^\nu} + \kappa_2^2\bar{F}_{2\mu\nu}\bar{F}\indices{_2^\mu^\nu}.
	\end{gathered}
\end{align}  
The above analysis further substantiates the concept of `EM embedding' and ensures the emergence of dilaton-free EM backgrounds as solutions within the framework of the $U(1)^2$-charged STU model in supergravity.\footnote{At any point, one can always employ the well-known EM embedding directly into the $U(1)^4$-charged STU model \eqref{mod1} by setting all related Maxwell gauged field charges equal (e.g., see \cite{Mai:2021yny}). However, this choice leads to insurmountable heat-kernel computations for the same embedded EM backgrounds due to the presence of four Maxwell and three dilaton fluctuations. Notably, these difficulties are systematically bypassed in \cref{SDCcomputation} by switching to the $U(1)^2$-charged STU models \eqref{mod2a} along with the EM embedding choice depicted in \cref{embeddingEM}.} In summary, this approach employs the embedding constraint \eqref{mod9} to transform the general two-charged STU background solutions $(\bar{g}_{\mu\nu}, \bar{A}_{1\mu}, \bar{A}_{2\mu}, \bar{\Phi})$ into the single-charged Kerr-Newman family of black holes (both asymptotically flat and AdS) satisfying the EM equations of motion \eqref{mod8}. Notably, while the Kerr and Schwarzschild black holes do not necessitate this specific embedding due to being charge neutral, the charged Kerr-Newman and Reissner-Nordstr\"om black holes strictly require the prescribed scalings as detailed in \Cref{embeddings} in order to be uplifted within the STU models.

Before delving into the computation of the Seeley-DeWitt coefficient $a_4(x)$ and the logarithmic correction to the entropy of black holes within EM-embedded STU supergravity models, it is important to clarify a crucial point. There is no need for concern when implementing a vanishing dilaton background, $\bar{\Phi}=0$, within the STU framework using the embedding choice \eqref{mod9}. This choice simply entails a modification of the STU equations of motion to accommodate dilaton-free EM black hole backgrounds. However, throughout this study, we consistently work with the complete STU supergravity model \eqref{mod2}, where the non-minimal Maxwell-dilaton couplings are always active that are expressed through exponential functions $f_1(\Phi)$ and $f_2(\Phi)$. In fact, the subsequent section delves into the quantization of the STU model by expanding the dilaton coupling functions through a perturbative expansion for a small quantum fluctuation denoted as $\tilde{\Phi}$ around the embedded EM background with $\bar{\Phi}=0$. This expansion takes the form,
\begin{align}\label{mod11}
	f_I(\bar{\Phi}+ \tilde{\Phi})\big\vert_{\bar{\Phi}=0} = f_I(\bar{\Phi})\bigg\vert_{\bar{\Phi}=0} + \tilde{\Phi}\frac{\mathrm{d}f_I(\Phi)}{\mathrm{d}\Phi}\bigg\vert_{\bar{\Phi}=0} + \frac{\tilde{\Phi}^2}{2}\frac{\mathrm{d}^2f_I(\Phi)}{\mathrm{d}\Phi^2}\bigg\vert_{\bar{\Phi}=0}+ \cdots, \quad I = 1, 2.
\end{align}
Our objective is to explore the quadratic fluctuations of the complete STU content, including the dilaton, with the aim of evaluating the $a_4(x)$ coefficient and determining the contributions of logarithmic entropy corrections.


\subsection{Quadratic fluctuations: background matrices and $a_4(x)$ computation}\label{SDCcomputation}

As discussed in \cref{setupmodel}, the bosonic sector of STU supergravity directly intersects with the $U(1)^2$-charged Einstein-Maxwell-dilaton (EMD) theory, encompassing the graviton $g_{\mu\nu}$, two Maxwell fields $(A_{1\mu}, A_{2\mu})$, and a dilaton $\Phi$. This alignment is established for two specific choices of the dilaton coupling constants $(\kappa_1, \kappa_2)\equiv (1,-1)$ and $(\sqrt{3},-1/\sqrt{3})$. The corresponding interactions are detailed within the action form \eqref{mod2}. Our present focus lies in exploring the quadratic fluctuations around the embedded Einstein-Maxwell (EM) backgrounds and then follow the heat kernel treatment elucidated in \cref{4dcomp} to compute the Seeley-DeWitt coefficient $a_4(x)$. Throughout, we aim to provide systematic details of the relevant background matrices and their traces, expressed in terms of the arbitrary dilaton couplings $\kappa_1$ and $\kappa_2$. Finally, we present simplified $a_4(x)$ results for the two distinct STU cases.

We consider the following perturbations around the EM background $(\bar{g}_{\mu\nu},\, \bar{A}_{1\mu},\, \bar{A}_{2\mu})$ embedded within the STU supergravity models for small quantum fluctuations of metric or graviton $h_{\mu\nu}$, two Maxwell fields ${a}_{1\mu}$ and ${a}_{2\mu}$, and dilaton $\tilde{\Phi}$
\begin{subequations}\label{sdc1}
	\begin{align}\label{sdc1a}
		g_{\mu\nu} &= \bar{g}_{\mu\nu} + \sqrt{2}h_{\mu\nu}, \enspace A_{1\mu} = \bar{A}_{1\mu} + \frac{1}{2}a_{I\mu}, \enspace A_{2\mu} = \bar{A}_{2\mu} + \frac{1}{2}a_{2\mu},\enspace \Phi = 0 + \frac{1}{2}\tilde{\Phi},
	\end{align} 
	where 
	\begin{align}\label{sdc1b}
		F_{I\mu\nu} &= \bar{F}_{I\mu\nu}+\frac{1}{2}f_{I\mu\nu}, \quad {f}_{I\mu\nu}={D_{[\mu}a_{I\nu]}}, \quad I = 1,2.
	\end{align}
\end{subequations}
The graviton and Maxwell fluctuations are adjusted using a specific normalization factor, following the convention employed by Sen \textit{et al.} in \cite{Bhattacharyya:2012ss}. Additionally, the dilaton fluctuation has been scaled by a $\frac{1}{2}$ factor. This collective normalization choice will be advantageous in ensuring that all the effective kinetic components are in the same state of normalization within the Laplace-type operator form \eqref{sdc10}. It is important to note that the dilaton acts as its own fluctuation, while simultaneously sharing the common EM background with the graviton and Maxwell fluctuations. We then proceed with expanding the STU action \eqref{mod2} in terms of the fluctuations \eqref{sdc1}. We specifically focus on the quadratic part of the contribution, as required by the one-loop quantum correction setup in \cref{setup}. We ascertained that this quadratic contribution entails several terms with complicated non-minimal couplings via the background Maxwell field strengths $\bar{F}_{1\mu\nu}$ and $\bar{F}_{2\mu\nu}$. By considering terms up to total derivatives and making use of the expansion \eqref{mod11}, we systematically decouple the quadratic fluctuated STU action into the following ``Einstein-dilaton'' and ``Maxwell-dilaton'' sectors: 
{
	\allowdisplaybreaks
	\begin{subequations}\label{sdc2}
		\begin{align}
			&\delta^2 \left(\sqrt{\det g}\,\big(\mathcal{R}-2\Lambda- 2 D^\mu\Phi D_\mu\Phi\big)\right) = \frac{1}{2}\sqrt{\det \bar{g}} \bigg[h_{\mu\nu}D_\rho D^\rho h^{\mu\nu}- h\indices{^\mu_\mu}D_\rho D^\rho h\indices{^\nu_\nu}  \nonumber\\
			&\hspace{0.7in}  -2 h^{\nu\rho}D_\mu D_\nu h\indices{^\mu_\rho}  + 2h^{\mu\nu}D_\mu D_\nu h\indices{^\alpha_\alpha} + \tilde{\Phi} D_\rho D^\rho \tilde{\Phi} -\left(R-2\Lambda\right)h_{\mu\nu}h^{\mu\nu}   \nonumber\\
			&\hspace{0.7in} +2 {R}_{\mu\nu}\big(2h^{\mu\rho}h\indices{^\nu_\rho}-h\indices{^\alpha_\alpha} h^{\mu\nu}\big)   + \frac{1}{2}\left(R -2\Lambda\right)(h\indices{^\alpha_\alpha})^2\bigg],\label{sdc2a}
		\end{align}
		\begin{align}
			&\delta^2 \left(-\sqrt{\det g}\big({f}_1({\Phi}){F}_{1\mu\nu}{F}\indices{_1^\mu^\nu} +  {f}_2({\Phi}){F}_{2\mu\nu}{F}\indices{_2^\mu^\nu}\big)\right) \nonumber\\
			&\hspace{0.4in} = -\sqrt{\det \bar{g}}  \bigg[ \frac{1}{4}\left(f_{1\mu\nu}f\indices{_1^\mu^\nu}+ f_{2\mu\nu}f\indices{_2^\mu^\nu}\right) + 2\left( \bar{F}_{1\mu\nu}\bar{F}_{1\alpha\beta} + \bar{F}_{2\mu\nu}\bar{F}_{2\alpha\beta}\right)h^{\mu\alpha}h^{\nu\beta} \nonumber \\
			&\hspace{0.4in}  + 4 \left(\bar{F}_{1\mu\nu}\bar{F}\indices{_1^\mu^\alpha} + \bar{F}_{2\mu\nu}\bar{F}\indices{_2^\mu^\alpha}\right)h^{\nu\beta}h_{\alpha\beta}
			- 2\left(\bar{F}_{1\mu\nu}\bar{F}\indices{_1_\alpha^\nu}+ \bar{F}_{2\mu\nu}\bar{F}\indices{_2_\alpha^\nu}\right)h\indices{^\rho_\rho}h^{\mu\alpha} \nonumber \\ 
			&\hspace{0.4in} - \frac{1}{2}\left(\bar{F}_{1\mu\nu}\bar{F}\indices{_1^\mu^\nu} + \bar{F}_{2\mu\nu}\bar{F}\indices{_2^\mu^\nu}\right)\Big(h_{\alpha\beta}h^{\alpha\beta}-\frac{1}{2}(h\indices{^\rho_\rho})^2\Big) \nonumber\\
			&\hspace{0.4in} - 2\sqrt{2}\left(\bar{F}_{1\mu\nu}f\indices{_1_\alpha^\nu} + \bar{F}_{2\mu\nu}f\indices{_2_\alpha^\nu}\right)h^{\mu\alpha} + \frac{\sqrt{2}}{2}\left(\bar{F}_{1\mu\nu} f\indices{_1^\mu^\nu} + \bar{F}_{2\mu\nu} f\indices{_2^\mu^\nu} \right)h\indices{^\rho_\rho} \nonumber\\[3pt]
			&\hspace{0.4in} + \frac{1}{2} \left({\kappa_1}^2 \bar{F}_{1\mu\nu}\bar{F}\indices{_1^\mu^\nu}+ {\kappa_2}^2 \bar{F}_{2\mu\nu}\bar{F}\indices{_2^\mu^\nu}\right) \tilde{\Phi}^2   - \left(\kappa_1\bar{F}_{1\mu\nu} f\indices{_1^\mu^\nu} + \kappa_2\bar{F}_{2\mu\nu} f\indices{_2^\mu^\nu}\right) \tilde{\Phi} \nonumber \\[3pt]
			&\hspace{0.4in} + 2\sqrt{2}\left(\kappa_1 \bar{F}_{1\mu\nu}\bar{F}\indices{_1_\alpha^\nu} + \kappa_2 \bar{F}_{2\mu\nu}\bar{F}\indices{_2_\alpha^\nu}\right)\tilde{\Phi} h^{\mu\alpha} \nonumber \\
			&\hspace{0.4in} - \frac{\sqrt{2}}{2}\left(\kappa_1 \bar{F}_{1\mu\nu}\bar{F}\indices{_1^\mu^\nu} + \kappa_2 \bar{F}_{2\mu\nu}\bar{F}\indices{_2^\mu^\nu}\right) \tilde{\Phi} h\indices{^\alpha_\alpha} \bigg],\label{sdc2b}
		\end{align}
		where the respective kinetic parts of Maxwell fluctuations need to be identified from the terms $f_{1\mu\nu}f\indices{_1^\mu^\nu}$ and $f_{2\mu\nu}f\indices{_2^\mu^\nu}$ using the following relations
		\begin{align}\label{sdc2c}
			-\frac{1}{2}f_{I\mu\nu}f\indices{_I^\mu^\nu} = a_{I\mu} \left(\bar{g}^{\mu\nu}D_\rho D^\rho- R^{\mu\nu} \right) a_{I\nu} + (D_\mu {a_I}^\mu)^2, \quad I= 1,2. 
		\end{align}
	\end{subequations}
}
We continue with the objective of formulating the desired Laplace-type structure \eqref{comp3} for the differential operator that governs the quadratic STU fluctuations. To achieve this, we carried out a series of adjustments on the fluctuated action form \eqref{sdc2} that are in the following order.
\begin{enumerate}
	
	\item
	The heat kernel treatment \ref{4dcomp} necessitates that there be a distinct kinetic term corresponding to each off-shell degree of freedom or fluctuation present in the theory of interest. However, the quadratic action form \eqref{sdc2} includes a few redundant components in the kinetic contributions of both the graviton and Maxwell fluctuations. To address this, we adopt the conventional practice of gauge-fixing the fluctuated theory by incorporating the gauge-fixing term,
	\begin{align}\label{sdc3}
		\hspace{-0.15in}-\int \mathrm{d}^4x \sqrt{\det \bar{g}}\Big[\Big(D^\mu h_{\mu\rho}-\frac{1}{2}D_\rho h\indices{^\alpha_\alpha}\Big)\Big(D_\nu h^{\nu\rho}-\frac{1}{2}D^\rho h\indices{^\beta_\beta}\Big) + \frac{1}{2}\Big((D^\mu a_{1\mu})^2+ (D^\mu a_{2\mu})^2\Big) \Big],
	\end{align}
	where the graviton and Maxwell modes are accommodated through the choice of harmonic gauge $D^\mu h_{\mu\rho}-\frac{1}{2}D_\rho h\indices{^\mu_\mu}=0$ and Lorentz gauge $D^\mu a\indices{_1_\mu}=0, D^\mu a\indices{_2_\mu}=0$, respectively. The gauge-fixing \eqref{sdc3} is applicable to both the asymptotically flat and AdS cases, as the presence of the cosmological term always preserves the gauge invariance under the considered background field transformations. The introduction of this gauge-fixing procedure gives rise to an additional ghost term \cite{Banerjee:2011oo},
	\begin{align}\label{sdc4}
		&\int \mathrm{d}^4x \sqrt{\det \bar{g}}\Big[ 2b_{1\mu}\left(\bar{g}^{\mu\nu}D_\rho D^\rho + R^{\mu\nu}\right) c_{1\nu} + 2b_{2\mu}\left(\bar{g}^{\mu\nu}D_\rho D^\rho + R^{\mu\nu}\right) c_{2\nu} \nonumber \\
		&\qquad\qquad\qquad + 2b_1 D_\rho D^\rho c_1 + 2b_2 D_\rho D^\rho c_2  -4 b_1 \bar{F}^{\rho\nu} D_\rho c_{1\nu} -4 b_2 \bar{F}^{\rho\nu} D_\rho c_{2\nu} \Big].
	\end{align}
	This term involves two sets of vector ghosts $(b_{1\mu}, b_{2\mu}, c_{1\mu}, c_{2\mu})$ and scalar ghosts $(b_1, b_2, c_1, c_2)$, arising respectively from the diffeomorphism and gauge invariance of the graviton and Maxwell fluctuations. This ghost term is minimal and remains non-interacting with the gauge-fixed component. Consequently, we have the flexibility to treat and evaluate the heat kernel contribution of the ghost part \eqref{sdc4} separately.
	
	\item The preceding gauge-fixing procedure needs to be complemented by an appropriate treatment of the trace mode of the graviton for the remaining kinetic component $h\indices{^\mu_\mu}D_\rho D^\rho h\indices{^\nu_\nu}$. Readers familiar with the literature (e.g., see \cite{Christensen:1979iy,Christensen:1978md,Gibbons:1978ji}) may recognize the standard treatment of decomposing the graviton into its trace $h = h\indices{^\mu_\mu} = \bar{g}^{\mu\nu}h_{\mu\nu}$ and a symmetric traceless part $\hat{h}_{\mu\nu} = h_{\mu\nu} - \frac{1}{4}\bar{g}_{\mu\nu}h$. This decomposition enables to effectively treat the trace $h$ and traceless $\hat{h}_{\mu\nu}$ graviton components as distinct fluctuations that transform under the irreducible representations of $SL(2, \mathbb{C})$. As a consequence, the gauge-fixed kinetic portion of the graviton can be reformulated as
	\begin{align}\label{sdc5}
		h_{\mu\nu}D_\rho D^\rho h^{\mu\nu}- \frac{1}{2}h\indices{^\mu_\mu}D_\rho D^\rho h\indices{^\nu_\nu} = \mathcal{G}^{\hat{h}_{\mu\nu}\hat{h}_{\alpha\beta}}	\hat{h}_{\mu\nu}D_\rho D^\rho \hat{h}_{\alpha\beta}- \frac{1}{4} h D_\rho D^\rho h,
	\end{align}   
	where the inner product between traceless graviton modes is defined by the following DeWitt metric,\footnote{
		In general, there exists a family of DeWitt metrics, parameterized by $\lambda_{\text{DW}}$ (referred to as the DeWitt parameter), which takes the following form \cite{DeWitt:1967gg}
		\begin{align}\label{dewitt}
			\mathcal{G}^{\hat{h}_{\mu\nu}\hat{h}_{\alpha\beta}} = \frac{1}{2}\Big(\bar{g}^{\mu\alpha} \bar{g}^{\nu\beta} + \bar{g}^{\mu\beta} \bar{g}^{\nu\alpha}+ \lambda_{\text{DW}}\bar{g}^{\mu\nu}\bar{g}^{\alpha\beta}\Big),
		\end{align}
		where the covariant or inverse operation is also represented by a similar relation. The specific relation \eqref{sdc6}, employed in this paper as well as in the works \cite{Charles:2015nn,Karan:2022dfy,Castro:2018tg,Banerjee:2021pdy}, corresponds to the choice $\lambda_{\text{DW}}= -1/2$. However, interested readers are also referred to an alternative choice $\lambda_{\text{DW}}= -1$ considered in \cite{Bhattacharyya:2012ss,Sen:2012qq,Karan:2019sk,Karan:2021teq,David:2021eoq}, where the approach does not involve decomposing the graviton fluctuation and treating its trace mode separately. 
	}
	\begin{align}\label{sdc6}
		\mathcal{G}^{\hat{h}_{\mu\nu}\hat{h}_{\alpha\beta}} = \frac{1}{2}\Big(\bar{g}^{\mu\alpha} \bar{g}^{\nu\beta} + \bar{g}^{\mu\beta} \bar{g}^{\nu\alpha}-\frac{1}{2}\bar{g}^{\mu\nu}\bar{g}^{\alpha\beta}\Big).
	\end{align}
	The operator $\mathcal{G}^{\hat{h}_{\mu\nu}\hat{h}_{\alpha\beta}}$ serves as an effective metric that projects onto the symmetric and traceless piece of the graviton, containing $10-1=9$ off-shell degrees of freedom after gauge-fixing the STU theory. Consequently, in the subsequent heat kernel treatment of this study, $\mathcal{G}^{\hat{h}_{\mu\nu}\hat{h}_{\alpha\beta}}$ must be utilized as a projection operator to contract the pairs of indices for any matrix acting on $\hat{h}_{\mu\nu}$.
	For instance, if ${M}$ is an arbitrary matrix associated with the fluctuations of the traceless graviton component, i.e.,
	\begin{align}\label{sdc7a}
		\tilde{\phi}_m {M}^m_n\tilde{\phi}^n = \hat{h}_{\mu\nu} {{M}}^{\hat{h}_{\mu\nu}\hat{h}_{\alpha\beta}}\hat{h}_{\alpha\beta},
	\end{align}
	then the expression for ${M}^2$ is given by,
	\begin{align}\label{sdc7b}
		{({M}^2)}^{\hat{h}_{\mu\nu}\hat{h}_{\alpha\beta}} = \mathcal{G}_{\hat{h}_{\rho\sigma}\hat{h}_{\gamma\delta}} {{M}}^{\hat{h}_{\mu\nu}\hat{h}_{\rho\sigma}}{{M}}^{\hat{h}_{\gamma\delta}\hat{h}_{\alpha\beta}}.
	\end{align}
	Moreover, the trace operations on $M$ and $M^2$ are defined as
	\begin{align}\label{sdc7c}
		\begin{split}
			\text{Tr}({M})  = \mathcal{G}_{\hat{h}_{\mu\nu}\hat{h}_{\alpha\beta}}{{M}}^{\hat{h}_{\mu\nu}\hat{h}_{\alpha\beta}}, \quad \text{Tr}({M}^2) = \mathcal{G}_{\hat{h}_{\alpha\beta}\hat{h}_{\rho\sigma}}\mathcal{G}_{\hat{h}_{\mu\nu}\hat{h}_{\gamma\delta}}{{M}}^{\hat{h}_{\mu\nu}\hat{h}_{\rho\sigma}}{{M}}^{\hat{h}_{\alpha\beta}\hat{h}_{\gamma\delta}}. 
		\end{split}
	\end{align} 
	
	\item The revised kinetic contribution of the graviton, as presented in \cref{sdc5}, still requires further refinement. Notably, the graviton trace component involves a negative signature, treating $h$ as a ghost field. This scenario highlights the familiar conformal factor problem in gravity, resulting in a divergent and ill-defined contribution to the Euclideanized one-loop path integral \eqref{setx1}. To address this issue, we adopt the conventional procedure outlined by Gibbons \textit{et al.} \cite{Gibbons:1978ac,Mazur:1989by} by implementing a conformal rotation along the imaginary axis with a new real graviton trace component $\hat{h}$, i.e.,        
	\begin{align}\label{sdc8}
		\hat{h} = -\frac{i}{2}{h}.
	\end{align}
	It is important to note that the introduction of an additional $1/2$ factor serves the purpose of aligning the kinetic contribution of the graviton trace with the same normalization state as that of the traceless component and Maxwell fluctuations (please refer to \cref{sdc10}). 
	
	\item 
	As a subsequent step, we need to modify the quadratic fluctuated STU action to ensure its self-adjoint properties in the desired operator form \eqref{sdc5}. This technical requirement necessitates adjusting the quadratic action to incorporate the Hermitian counterpart of each term. To achieve this, we identify appropriate terms with fluctuations $\tilde{\phi}_m = \big\lbrace \hat{h}_{\mu\nu}, \hat{h}, a_{1\mu}, a_{2\mu}, \tilde{\Phi} \big\rbrace$ and then extract their Hermitian partners using the following schematic approach involving arbitrary matrices $\mathbb{N}$ and $\mathbb{K}^\rho$, 
	\begin{align}\label{sdc9}
		\tilde{\phi}_m \mathbb{N}^{mn}\tilde{\phi}_n  +\tilde{\phi}_m (\mathbb{K}^\rho)^{mn} D_\rho\tilde{\phi}_n &= \frac{1}{2} \tilde{\phi}_m \Big(\mathbb{N}^{mn} - \frac{1}{2} (D_\rho \mathbb{K}^\rho)^{mn} + (\mathbb{K}^\rho)^{mn} D_\rho  \Big) \tilde{\phi}_n \nonumber \\
		&\enspace + \frac{1}{2} \tilde{\phi}_n \Big(\mathbb{N}^{mn} - \frac{1}{2} (D_\rho \mathbb{K}^\rho)^{mn} - (\mathbb{K}^\rho)^{mn} D_\rho  \Big) \tilde{\phi}_m.
	\end{align}
	It is important to note that the above relations are derived by utilizing the commutative properties of the bosonic fluctuations and disregarding all the total derivative terms. This adjustment up to total derivatives is crucial since they lead to boundary terms that do not contribute to the integrations \eqref{sdc2} around asymptotically flat and AdS backgrounds. 
	
\end{enumerate}
\subsubsection{Gauge-fixed contribution}
With incorporating the above rectifications, we finally arrive at the desired form of quadratic fluctuated STU theory (without including the ghost contribution \eqref{sdc4}),
{
	\allowdisplaybreaks
	\begin{subequations}\label{sdc10}
		\begin{align}
			\delta^2 \mathcal{S}[\hat{h}_{\mu\nu}, \hat{h}, a_{1\mu}, a_{2\mu}, \tilde{\Phi}] = \frac{1}{2}\int \mathrm{d}^4x \sqrt{\det \bar{g}}\, \tilde{\phi}_m \mathcal{H}^m_n\tilde{\phi}^n,
		\end{align}
		where the simplified Laplace-type operator $\mathcal{H}$ is expressed into the following components
		\begin{align}
			\tilde{\phi}_m \mathcal{H}^m_n\tilde{\phi}^n &= \hat{h}_{\mu\nu}\bigg\lbrace \mathcal{G}^{\hat{h}_{\mu\nu}\hat{h}_{\alpha\beta}} D_\rho D^\rho \hat{h}_{\alpha\beta} + 2\Big( R^{\mu\alpha\nu\beta} - \bar{g}^{\mu\alpha}R^{\nu\beta} -2\bar{F}\indices{_1^{\mu\alpha}}\bar{F}\indices{_1^{\nu\beta}}     \nonumber \\ 
			& -2 \bar{F}\indices{_2^{\mu\alpha}}\bar{F}\indices{_2^{\nu\beta}} - \frac{1}{2}\bar{g}^{\mu\alpha} \bar{g}^{\nu\beta}\left(\bar{F}\indices{_1_\rho_\sigma}\bar{F}\indices{_1^{\rho\sigma}} + \bar{F}\indices{_2_\rho_\sigma}\bar{F}\indices{_2^{\rho\sigma}} -2\Lambda \right)\Big) \hat{h}_{\alpha\beta} \nonumber \\
			& - 2i \Big(\bar{F}\indices{_1^{\mu\alpha}}\bar{F}\indices{_1^\nu_\alpha} + \bar{F}\indices{_2^{\mu\alpha}}\bar{F}\indices{_2^\nu_\alpha}\Big)\hat{h} - \sqrt{2} \left(D^\nu \bar{F}\indices{_1^{\mu\alpha}}\right)a_{1\alpha}    \nonumber \\[3pt]
			&  - \sqrt{2} \left(D^\nu \bar{F}\indices{_2^{\mu\alpha}}\right)a_{2\alpha} - 2\sqrt{2} \Big(\kappa_1\bar{F}\indices{_1^{\mu\alpha}}\bar{F}\indices{_1^\nu_\alpha} + \kappa_2\bar{F}\indices{_2^{\mu\alpha}}\bar{F}\indices{_2^\nu_\alpha}\Big)\tilde{\Phi} \nonumber \\
			& + 2\sqrt{2} \Big(\bar{g}^{\nu\rho}\bar{F}\indices{_1^{\mu\alpha}} -\bar{g}^{\nu\alpha}\bar{F}\indices{_1^{\mu\rho}}\Big) (D_\rho a_{1\alpha}) + 2\sqrt{2} \Big(\bar{g}^{\nu\rho}\bar{F}\indices{_2^{\mu\alpha}} -\bar{g}^{\nu\alpha}\bar{F}\indices{_2^{\mu\rho}}\Big) (D_\rho a_{2\alpha}) \bigg\rbrace \nonumber \\
			&   +  \hat{h} \bigg\lbrace D_\rho D^\rho \hat{h} + 2\Lambda\hat{h} - 2i \Big(\bar{F}\indices{_1^{\mu\alpha}}\bar{F}\indices{_1^\nu_\alpha} + \bar{F}\indices{_2^{\mu\alpha}}\bar{F}\indices{_2^\nu_\alpha}\Big)\hat{h}_{\mu\nu} \bigg\rbrace \nonumber \\
			&    + \tilde{\Phi} \bigg\lbrace D_\rho D^\rho \tilde{\Phi} - \Big({\kappa_1}^2 \bar{F}\indices{_1_\mu_\nu}\bar{F}\indices{_1^{\mu\nu}}  + {\kappa_2}^2 \bar{F}\indices{_2_\mu_\nu}\bar{F}\indices{_2^{\mu\nu}}\Big)\tilde{\Phi}  \nonumber \\
			&  - 2\sqrt{2} \Big(\kappa_1\bar{F}\indices{_1^{\mu\alpha}}\bar{F}\indices{_1^\nu_\alpha} + \kappa_2\bar{F}\indices{_2^{\mu\alpha}}\bar{F}\indices{_2^\nu_\alpha}\Big)\hat{h}_{\mu\nu} + 2\kappa_1\bar{F}\indices{_1^\rho^\alpha}\left(D_\rho a_{1\alpha}\right) \nonumber \\
			&   + 2\kappa_2\bar{F}\indices{_2^\rho^\alpha}\left(D_\rho a_{2\alpha}\right) \bigg\rbrace + a_{1\alpha}\bigg\lbrace \bar{g}^{\alpha\beta}D_\rho D^\rho a_{1\beta} - {R}^{\alpha\beta}a_{1\beta} - \sqrt{2} \left(D^\nu \bar{F}\indices{_1^{\mu\alpha}}\right)\hat{h}_{\mu\nu} \nonumber \\
			&  - 2\kappa_1\bar{F}\indices{_1^\rho^\alpha}\big(D_\rho \tilde{\Phi} \big) - 2\sqrt{2} \Big(\bar{g}^{\nu\rho}\bar{F}\indices{_1^{\mu\alpha}} -\bar{g}^{\nu\alpha}\bar{F}\indices{_1^{\mu\rho}}\Big) (D_\rho \hat{h}_{\mu\nu}) \bigg\rbrace \nonumber \\
			&  + a_{2\alpha}\bigg\lbrace \bar{g}^{\alpha\beta}D_\rho D^\rho a_{2\beta} - {R}^{\alpha\beta}a_{2\beta} - \sqrt{2} \left(D^\nu \bar{F}\indices{_2^{\mu\alpha}}\right)\hat{h}_{\mu\nu}  \nonumber \\
			& - 2\kappa_2\bar{F}\indices{_2^\rho^\alpha}\big(D_\rho \tilde{\Phi} \big) - 2\sqrt{2} \Big(\bar{g}^{\nu\rho}\bar{F}\indices{_2^{\mu\alpha}} -\bar{g}^{\nu\alpha}\bar{F}\indices{_2^{\mu\rho}}\Big) (D_\rho \hat{h}_{\mu\nu})  \bigg\rbrace.
		\end{align}
\end{subequations}}
The above operator form represents different intricate interactions between the quadratic fluctuations of the STU model through the Einstein-Maxwell backgrounds. These interactions are noticed to be distributed among various non-minimal blocks that are determined by the background metric, curvature, and Maxwell strengths. To better organize these blocks, we have further restructured them in \eqref{sdc10} to pair each quadratic term symmetrically by interchanging indices within each traceless graviton $\hat{h}_{\mu\nu}$ and also with its quadratic counterpart ($\hat{h}_{\mu\nu}, \hat{h}_{\alpha\beta}$). During this process, one may make use of the background Einstein equation \eqref{mod3}, the Maxwell equation \eqref{mod4}, and the embedding condition \eqref{mod9} to facilitate necessary simplifications at appropriate steps. This adjustment is crucial to ensure correct trace results using the relations in \eqref{sdc7c}. At this stage, we are well-prepared to conduct the heat kernel treatment as outlined in \cref{4dcomp}, where we compare the Laplace-type operator $\mathcal{H}$ for STU theory with the schematic representation in \eqref{comp3} and proceed to extract the required background heat kernel matrices.

The effective projection or identity matrices for individual off-shell STU fluctuations are read off as      
\begin{align}\label{sdc11}
	\tilde{\phi}_m \mathcal{I}^{mn}\tilde{\phi}_n &= \hat{h}_{\mu\nu} \mathcal{G}^{\hat{h}_{\mu\nu}\hat{h}_{\alpha\beta}} \hat{h}_{\alpha\beta} + \hat{h}\hat{h} + a_{1\mu }\bar{g}^{\mu\nu} a_{1\nu} + a_{2\mu }\bar{g}^{\mu\nu} a_{2\nu} + \tilde{\Phi}\tilde{\Phi},
\end{align}
Notice that the traceless graviton employs the DeWitt metric $\mathcal{G}^{\hat{h}_{\mu\nu}\hat{h}_{\alpha\beta}}$ as an identity operator for the specific form given in \cref{sdc6}, whereas the same is replaced solely by the four-dimensional background metric $\bar{g}^{\mu\nu}$ for the $U(1)$ or Maxwell fluctuations. When we calculate the trace of \eqref{sdc11}, we find a total of nineteen off-shell degrees of freedom for the current gauged-fixed STU fluctuations $\lbrace \tilde{\phi}_m \rbrace$: nine stemming from the traceless graviton $\hat{h}_{\mu\nu}$, four from each Maxwell mode $a_{1\mu}$ and $a_{2\mu}$, and one from each scalar field $\hat{h}$ and $\tilde{\Phi}$, providing
\begin{align}\label{sdc12}
	\text{Tr}(\mathcal{I}) = 9 + 1 + 4 + 4 + 1 = 19.
\end{align}

Additionally, we derived the gauge-connection matrix $\omega^\rho$ by analyzing the non-minimal (i.e., linear-derivative) terms in the operator form \eqref{sdc10} as
{
	\allowdisplaybreaks
	\begin{align}\label{sdc13}
		\tilde{\phi}_m (\omega^\rho)^{mn}\tilde{\phi}_n &= \frac{\sqrt{2}}{2} \hat{h}_{\mu\nu}\Big(\bar{g}^{\mu\rho}\bar{F_1}^{\nu\alpha}+\bar{g}^{\nu\rho}\bar{F_1}^{\mu\alpha}-\bar{g}^{\mu\alpha}\bar{F_1}^{\nu\rho}-\bar{g}^{\nu\alpha}\bar{F_1}^{\mu\rho}\Big) a_{1\alpha} \nonumber \\
		& \quad +\frac{\sqrt{2}}{2}\hat{h}_{\mu\nu}\Big(\bar{g}^{\mu\rho}\bar{F_2}^{\nu\alpha}+\bar{g}^{\nu\rho}\bar{F_2}^{\mu\alpha}-\bar{g}^{\mu\alpha}\bar{F_2}^{\nu\rho}-\bar{g}^{\nu\alpha}\bar{F_2}^{\mu\rho}\Big) a_{2\alpha} \nonumber \\
		&\quad - \frac{\sqrt{2}}{2}  a_{1\alpha} \Big(\bar{g}^{\mu\rho}\bar{F_1}^{\nu\alpha}+\bar{g}^{\nu\rho}\bar{F_1}^{\mu\alpha}-\bar{g}^{\mu\alpha}\bar{F_1}^{\nu\rho}-\bar{g}^{\nu\alpha}\bar{F_1}^{\mu\rho}\Big)\hat{h}_{\mu\nu} \nonumber \\
		&\quad - \frac{\sqrt{2}}{2}  a_{2\alpha} \Big(\bar{g}^{\mu\rho}\bar{F_2}^{\nu\alpha}+\bar{g}^{\nu\rho}\bar{F_2}^{\mu\alpha}-\bar{g}^{\mu\alpha}\bar{F_2}^{\nu\rho}-\bar{g}^{\nu\alpha}\bar{F_2}^{\mu\rho}\Big)\hat{h}_{\mu\nu} \nonumber \\
		&\quad + \kappa_1 \Big(a_{1\alpha} \bar{F_1}^{\alpha\rho} \tilde{\Phi} - \tilde{\Phi}\bar{F_1}^{\alpha\rho} a_{1\alpha} \Big) + \kappa_2 \Big(a_{2\alpha} \bar{F_2}^{\alpha\rho} \tilde{\Phi} - \tilde{\Phi}\bar{F_2}^{\alpha\rho} a_{2\alpha} \Big).
	\end{align}
}
The relation above confirms the existence of only two primary sources of non-minimal couplings within the current quadratic fluctuation framework of the STU model. These couplings  directly govern the interactions of two Maxwell fluctuations $\left(a_{1\alpha}, a_{2\alpha}\right)$ with the traceless graviton $\hat{h}_{\mu\nu}$ and the dilaton $\tilde{\Phi}$. From a technical perspective, this interaction pattern gives rise to an indirect non-minimal coupling between $a_{1\alpha}$ and $a_{2\alpha}$, subsequently leading to the emergence of relevant components within the $E$ and $\Omega_{\rho\sigma}$ matrices when derived using the relation \eqref{sdc13} (for details, please refer to \cref{calcul}). Furthermore, it is worth noting that the components of $\omega^\rho$ are antisymmetric with respect to the relevant fluctuations. This particular characteristic is consistent with findings from previously reported similar heat kernel treatments, offering primary validation for the Laplace-type operator form derived in \eqref{sdc10}.

Similarly, we read off the matrix $P$ from the derivative-free or minimal part of the operator form \eqref{sdc10}. The relevant components are expressed as follows
{
	\allowdisplaybreaks
	\begin{align}\label{sdc14}
		\tilde{\phi}_m P^{mn}\tilde{\phi}_n &= \hat{h}_{\mu\nu}\bigg( R^{\mu\alpha\nu\beta}+R^{\mu\beta\nu\alpha} -\frac{1}{2}\left(\bar{g}^{\mu\alpha}R^{\nu\beta}+\bar{g}^{\nu\alpha}R^{\mu\beta}+ \bar{g}^{\mu\beta}R^{\nu\alpha}+\bar{g}^{\nu\beta}R^{\mu\alpha}\right) \nonumber\\
		&\quad\qquad -2\Big(\bar{F}\indices{_1^{\mu\alpha}}\bar{F}\indices{_1^{\nu\beta}}+\bar{F}\indices{_1^{\mu\beta}}\bar{F}\indices{_1^{\nu\alpha}}+ \bar{F}\indices{_2^{\mu\alpha}}\bar{F}\indices{_2^{\nu\beta}}+\bar{F}\indices{_2^{\mu\beta}}\bar{F}\indices{_2^{\nu\alpha}}\Big) \nonumber\\
		&\quad\qquad - \frac{1}{2}\Big(\bar{F}\indices{_1_\rho_\sigma}\bar{F}\indices{_1^{\rho\sigma}} + \bar{F}\indices{_2_\rho_\sigma}\bar{F}\indices{_2^{\rho\sigma}} -2\Lambda\Big)\left(\bar{g}^{\mu\alpha} \bar{g}^{\nu\beta} + \bar{g}^{\mu\beta} \bar{g}^{\nu\alpha}\right)\bigg) \hat{h}_{\alpha\beta}  \nonumber \\
		&\quad\qquad  + 2\hat{h}\Lambda\hat{h} - \tilde{\Phi} \Big({\kappa_1}^2 \bar{F}\indices{_1_\mu_\nu}\bar{F}\indices{_1^{\mu\nu}}  + {\kappa_2}^2 \bar{F}\indices{_2_\mu_\nu}\bar{F}\indices{_2^{\mu\nu}}\Big)\tilde{\Phi} - a_{1\mu} R^{\mu\nu}a_{1\nu} \nonumber \\
		&\quad\qquad - a_{2\mu} R^{\mu\nu}a_{2\nu} - 2i \Big(\bar{F}\indices{_1^{\mu\alpha}}\bar{F}\indices{_1^\nu_\alpha} + \bar{F}\indices{_2^{\mu\alpha}}\bar{F}\indices{_2^\nu_\alpha}\Big)\Big(\hat{h}_{\mu\nu}\hat{h} + \hat{h}\hat{h}_{\mu\nu}\Big)  \nonumber \\
		&\quad\qquad - 2\sqrt{2} \Big(\kappa_1\bar{F}\indices{_1^{\mu\alpha}}\bar{F}\indices{_1^\nu_\alpha} + \kappa_2\bar{F}\indices{_2^{\mu\alpha}}\bar{F}\indices{_2^\nu_\alpha}\Big)\Big(\hat{h}_{\mu\nu}\tilde{\Phi} + \tilde{\Phi}\hat{h}_{\mu\nu} \Big) \nonumber \\
		&\quad\qquad -\frac{\sqrt{2}}{2}\Big( D^\mu \bar{F}\indices{_1^{\nu\alpha}} + D^\nu \bar{F}\indices{_1^{\mu\alpha}} \Big)\Big(\hat{h}_{\mu\nu} a_{1\alpha} + a_{1\alpha} \hat{h}_{\mu\nu} \Big)\nonumber \\
		&\quad\qquad -\frac{\sqrt{2}}{2}\Big( D^\mu \bar{F}\indices{_2^{\nu\alpha}} + D^\nu \bar{F}\indices{_2^{\mu\alpha}} \Big)\Big(\hat{h}_{\mu\nu} a_{2\alpha} + a_{2\alpha} \hat{h}_{\mu\nu} \Big).
	\end{align}
}
From here, our objective involves utilizing all the matrix-valued data recorded in \cref{sdc11,sdc13,sdc14} to derive the components of the desired matrices $E$ and $\Omega_{\rho\sigma}$, which encode the complete details of all quadratic STU interactions in terms of the background metric and Maxwell fields. This process requires the formulation of additional terms that involve contraction and commutation operations between the connection $\omega_\rho$ and the covariant derivative $D_\rho$, as delineated by the formulas in \cref{comp6,comp7,commutations}.  Subsequently, we calculate the traces $\text{Tr}\left(E\right)$, $\text{Tr}\left(E^2\right)$, and $\text{Tr}\left(\Omega_{\rho\sigma}\Omega^{\rho\sigma}\right)$ over all relevant components. Although these calculations are highly intricate and laborious, they may provide valuable insights to interested readers. We have provided a systematic outline of the underlying steps in \cref{calcul}. The computed trace results are summarized as follows
{
	\allowdisplaybreaks
	\begin{align}
		\text{Tr}(E) &= -12\Lambda + 7 \left(\bar{F}\indices{_1_{\mu\nu}}\bar{F}\indices{_1^{\mu\nu}} + \bar{F}\indices{_2_{\mu\nu}}\bar{F}\indices{_2^{\mu\nu}}\right), \label{sdc15a}\\[8pt] 
		\text{Tr}(E^2) &= 3 R_{\mu\nu\rho\sigma}R^{\mu\nu\rho\sigma} - \Bigg(\frac{23}{4} + \frac{2\kappa_1^2}{\left(\kappa_1^4 + 1\right)}\Bigg)R_{\mu\nu}R^{\mu\nu} + \Bigg(31 + \frac{8\kappa_1^2}{\left(\kappa_1^4 + 1\right)}\Bigg)\Lambda^2 \nonumber \\
		&\enspace + 3R_{\mu\nu\rho\sigma} \left(\bar{F}\indices{_1^{\mu\nu}}\bar{F}\indices{_1^{\rho\sigma}} + \bar{F}\indices{_2^{\mu\nu}}\bar{F}\indices{_2^{\rho\sigma}}\right) + \frac{37}{4}\left(\bar{F}_{1\mu\nu}\bar{F}\indices{_1^{\mu\nu}} + \bar{F}_{2\mu\nu}\bar{F}\indices{_2^{\mu\nu}} \right)^2 \nonumber \\
		&\enspace  - \Lambda\Bigg(18 + \frac{4\kappa_1^2}{\left(\kappa_1^4 + 1\right)}\Bigg)\left(\bar{F}_{1\mu\nu}\bar{F}\indices{_1^{\mu\nu}} + \bar{F}_{2\mu\nu}\bar{F}\indices{_2^{\mu\nu}} \right), \label{sdc15b}\\[8pt] 
		\text{Tr}\left(\Omega_{\rho\sigma}\Omega^{\rho\sigma}\right) &= -8 R_{\mu\nu\rho\sigma}R^{\mu\nu\rho\sigma} + \Bigg(\frac{113}{2} + \frac{4\kappa_1^2}{\left(\kappa_1^4 + 1\right)}\Bigg)R_{\mu\nu}R^{\mu\nu} - \Bigg(226 + \frac{16\kappa_1^2}{\left(\kappa_1^4 + 1\right)}\Bigg)\Lambda^2 \nonumber \\
		&\enspace - 18 R_{\mu\nu\rho\sigma} \left(\bar{F}\indices{_1^{\mu\nu}}\bar{F}\indices{_1^{\rho\sigma}} + \bar{F}\indices{_2^{\mu\nu}}\bar{F}\indices{_2^{\rho\sigma}}\right) - \frac{111}{2}\left(\bar{F}_{1\mu\nu}\bar{F}\indices{_1^{\mu\nu}} + \bar{F}_{2\mu\nu}\bar{F}\indices{_2^{\mu\nu}} \right)^2 \nonumber \\
		&\enspace  + \Lambda\Bigg(60 + \frac{8\kappa_1^2}{\left(\kappa_1^4 + 1\right)}\Bigg)\left(\bar{F}_{1\mu\nu}\bar{F}\indices{_1^{\mu\nu}} + \bar{F}_{2\mu\nu}\bar{F}\indices{_2^{\mu\nu}} \right).\label{sdc15c}
\end{align}}
The aforementioned trace relations are simplified by incorporating appropriate on-shell identities (see \cref{iden}), expressing them solely in terms of distinct irreducible invariants of Einstein-Maxwell backgrounds for the two $U(1)^2$-charged STU models with $\kappa_1\kappa_2 = -1$. However, it has been observed that nearly all invariants involving background Maxwell field strengths in the specific trace data are precisely canceled out within the $a_4(x)$ formula \eqref{comp8}. The final Seeley-DeWitt result involves only $\Lambda\bar{F}_{1\mu\nu}\bar{F}\indices{_1^{\mu\nu}}$ and $\Lambda\bar{F}_{2\mu\nu}\bar{F}\indices{_2^{\mu\nu}}$, specifically for the case $(\kappa_1, \kappa_2) = (\sqrt{3}, -1/\sqrt{3})$ in the STU truncation. These can be collectively encoded inside the effective invariant $R \bar{F}_{\mu\nu}\bar{F}^{\mu\nu}$, where $R = 4\Lambda$, in accordance with the choice of the embedding condition \eqref{mod10}. All of this leads to the following result
\begin{align}\label{sdc16}
	(4\pi)^2 {a_4}^{\text{gauge-fixed}}(x) &= \frac{169}{180} R_{\mu\nu\rho\sigma}R^{\mu\nu\rho\sigma} + \Bigg(\frac{311}{180} - \frac{2\kappa_1^2}{3\left(\kappa_1^4 + 1\right)}\Bigg)R_{\mu\nu}R^{\mu\nu} \nonumber \\
	&\quad  - \Bigg(\frac{4}{9} - \frac{\kappa_1^2}{6\left(\kappa_1^4 + 1\right)}\Bigg)R^2 + \frac{\left(\kappa_1^2 - 1\right)^2}{6\left(\kappa_1^4 + 1\right)}R \bar{F}_{\mu\nu}\bar{F}^{\mu\nu},
\end{align}
where we have set $\chi = 1$ for the graviton and $U(1)$ or Maxwell fluctuations. It is worth noting some insightful remarks about the Seeley-DeWitt result \eqref{sdc16}. The coefficient of $R_{\mu\nu\rho\sigma}R^{\mu\nu\rho\sigma}$ remains robust for a theory with fixed off-shell degrees of freedom. In contrast, the specific coefficients for $R_{\mu\nu}R^{\mu\nu}$, $R^2$ and $R \bar{F}_{\mu\nu}\bar{F}^{\mu\nu}$ are not fundamental but rather sensitive to the type of non-minimal couplings and gauge interactions that are effective between fluctuations. Moreover, their respective contributions can be interchanged through the one-shell Einstein equation \eqref{mod8}. Therefore, at any stage, it is never advisable to disregard the background invariants proportional to $\bar{F}_{1\mu\nu}$ and $\bar{F}_{2\mu\nu}$, which were present in the trace data obtained in \cref{sdc15a,sdc15b,sdc15c} but canceled out inside the final result \eqref{sdc16}.

\subsubsection{Ghost contribution}
Now, we need to proceed with a similar heat kernel computation for the ghost term \eqref{sdc4}, which further refines the gauge-fixed contribution \eqref{sdc16} of the Seeley-DeWitt coefficient $a_4(x)$. To begin, we can express the ghost action as  
\begin{align}\label{sdc17}
	&\int \mathrm{d}^4x \sqrt{\det \bar{g}}\Big[ 2b_{I\mu}\left(\bar{g}^{\mu\nu}D_\rho D^\rho + R^{\mu\nu}\right) c_{J\nu}\delta_{IJ} + 2b_I D_\rho D^\rho c_J\delta_{IJ}  -4 b_I \bar{F}_J^{\rho\nu} D_\rho c_{K\nu}\delta_{IJK} \Big],
\end{align}
where $\delta_{IJ}$ and $\delta_{IJK}$ are the Kronecker deltas with indices $I, J, K = 1, 2$ enumerating all the ghost fields correspond to the two $U(1)$ gauge or Maxwell species. Clearly, the kinetic terms need to be diagonalized to make use of the Laplace-type quadratic operator for the heat kernel methodology. A convenient approach is to introduce the following redefinitions between the ghost species  
\begin{align}\label{sdc24}
	\begin{gathered}
		b_{I\mu} \to \frac{\sqrt{2}}{2}\left(c_{I\mu} - ib_{I\mu}\right), \quad c_{I\mu} \to \frac{\sqrt{2}}{2}\left(c_{I\mu} + ib_{I\mu}\right), \\
		b_I \to \frac{\sqrt{2}}{2}\left(c_I - ib_I\right), \quad c_I \to \frac{\sqrt{2}}{2}\left(c_I + ib_I\right). 
	\end{gathered}
\end{align}  
This allows us to extract the desired form of the kinetic operator (up to quadratic order) operating on the six ghost fluctuations $\tilde{\phi}_m = \big\lbrace b_{I\mu}, c_{I\mu}, b_I, c_I \big\rbrace$ as
{
	\allowdisplaybreaks
	\begin{subequations}\label{sdc25}
		\begin{align}
			\delta^2 \mathcal{S}_{\text{ghost}} = \int \mathrm{d}^4x \sqrt{\det \bar{g}}\, \tilde{\phi}_m \mathcal{H}^m_n\tilde{\phi}^n,
		\end{align}
		where
		\begin{align}
			\tilde{\phi}_m \mathcal{H}^m_n\tilde{\phi}^n &= b_{I\mu}\left(\bar{g}^{\mu\nu}D_\rho D^\rho + R^{\mu\nu}\right)b_{J\nu}\delta_{IJ} + c_{I\mu}\left(\bar{g}^{\mu\nu}D_\rho D^\rho + R^{\mu\nu}\right)c_{J\nu}\delta_{IJ} \nonumber \\
			&\quad + b_I D_\rho D^\rho b_J\delta_{IJ} + c_I D_\rho D^\rho c_J\delta_{IJ} + b_{I\mu} \bar{F}_J^{\rho\mu}\left( D_\rho b_K + i D_\rho c_K \right)\delta_{IJK} \nonumber\\
			&\quad  +  c_{I\mu} \bar{F}_J^{\rho\mu}\left(D_\rho c_K - i D_\rho b_K \right)\delta_{IJK} - b_I \bar{F}_J^{\rho\mu}\left(D_\rho b_{K\mu} -i D_\rho c_{K\mu}\right)\delta_{IJK} \nonumber\\
			& \quad  - c_I \bar{F}_J^{\rho\mu}\left(D_\rho c_{K\mu} + i D_\rho b_{K\mu}\right)\delta_{IJK}.
		\end{align}
\end{subequations}}
From here, the task is to write the forms of the following necessary heat kernel matrices (as defined in \cref{4dcomp})
{
	\allowdisplaybreaks
	\begin{align}\label{sdc26}
		\tilde{\phi}_m \mathcal{I}^{mn}\tilde{\phi}_n &=b_{I\mu}\left(\bar{g}^{\mu\nu}\delta_{IJ}\right)b_{J\nu} + c_{I\mu}\left(\bar{g}^{\mu\nu}\delta_{IJ}\right)c_{J\nu} + b_I\left(\delta_{IJ}\right) b_J + c_I\left(\delta_{IJ}\right) c_J,\\[4pt]
		\tilde{\phi}_m P^{mn}\tilde{\phi}_n &=b_{I\mu} \left(R^{\mu\nu} \delta_{IJ}\right) b_{J\nu} + c_{I\mu} \left(R^{\mu\nu} \delta_{IJ}\right) c_{J\nu},\\
		\tilde{\phi}_m (\omega^\rho)^{mn}\tilde{\phi}_n &=  \frac{1}{2} b_{I\mu} \left( \bar{F}_J^{\rho\mu}\delta_{IJK}\right) b_K - \frac{1}{2} b_I \left(\bar{F}_J^{\rho\mu} \delta_{IJK}\right) b_{K\mu}  + \frac{1}{2} b_{I\mu} \left(i \bar{F}_J^{\rho\mu}\delta_{IJK}\right) c_K  \nonumber\\
		&\enspace - \frac{1}{2} c_I \left(i\bar{F}_J^{\rho\mu}\delta_{IJK}\right) b_{K\mu} + \frac{1}{2} c_{I\mu} \left(\bar{F}_J^{\rho\mu}\delta_{IJK}\right) c_K - \frac{1}{2} c_I \left(\bar{F}_J^{\rho\mu}\delta_{IJK}\right) c_{K\mu}  \nonumber \\
		&\enspace + \frac{1}{2} c_{I\mu} \left(i \bar{F}_J^{\rho\mu}\delta_{IJK} \right) b_K - \frac{1}{2} b_I \left(i\bar{F}_J^{\rho\mu}\right) c_{K\mu}, 
\end{align}}    
which further provides,
{
	\allowdisplaybreaks
	\begin{align}\label{sdc27}
		\tilde{\phi}_m E^{mn}\tilde{\phi}_n &=b_{I\mu} \left(R^{\mu\nu} \delta_{IJ}\right) b_{J\nu} + c_{I\mu} \left(R^{\mu\nu} \delta_{IJ}\right) c_{J\nu},\\[4pt]
		\tilde{\phi}_m \left(\Omega_{\rho\sigma}\right)^{mn}\tilde{\phi}_n &=  b_{I\mu} \left(R\indices{^\mu^\nu_\rho_\sigma}\delta_{IJ}\right)b_{J\nu} + c_{I\mu} \left(R\indices{^\mu^\nu_\rho_\sigma}\delta_{IJ}\right)c_{J\nu} \nonumber\\
		&\quad - \frac{1}{2}b_{I\mu}\left(D^\mu \bar{F}_{J\rho\sigma}\delta_{IJK}\right)b_K + \frac{1}{2} b_I \left(D^\mu \bar{F}_{J\rho\sigma}\delta_{IJK}\right)b_{K\mu}   \nonumber\\
		&\quad - \frac{1}{2}b_{I\mu} \left(iD^\mu \bar{F}_{J\rho\sigma}\delta_{IJK}\right)c_K + \frac{1}{2} c_I \left(i D^\mu \bar{F}_{J\rho\sigma}\delta_{IJK}\right)b_{K\mu}   \nonumber\\
		&\quad + \frac{1}{2}c_{I\mu} \left(iD^\mu \bar{F}_{J\rho\sigma}\delta_{IJK}\right)b_K - \frac{1}{2} b_I \left(iD^\mu \bar{F}_{J\rho\sigma}\delta_{IJK}\right)c_{K\mu}  \nonumber \\
		&\quad - \frac{1}{2}c_{I\mu} \left(D^\mu \bar{F}_{J\rho\sigma}\delta_{IJK}\right)c_K + \frac{1}{2} c_I \left( D^\mu \bar{F}_{J\rho\sigma}\delta_{IJK}\right)c_{K\mu}.
\end{align}}
It is essential to recognize that all other matrix components that are technically valid but nullified by the application of the equations of motion \eqref{mod3} to \eqref{mod5} for our specific choice of EM background embedded STU supergravity model. Taking into account the derived components for the matrices $\mathcal{I}$, $E$ and $\Omega_{\rho\sigma}$, we can determine the following traces for the ghosts
\begin{align}\label{sdc28}
	\begin{split}
		\text{Tr}(\mathcal{I}) &= 4\delta_{II} + 4\delta_{II} + \delta_{II} + \delta_{II} = 20, \\
		\text{Tr}(E) &= R\delta_{II} + R\delta_{II}   = 16\Lambda, \\
		\text{Tr}(E^2) &= 2R_{\mu\nu}R^{\mu\nu}\delta_{IJ}\delta_{JI} = 4R_{\mu\nu}R^{\mu\nu}, \\
		\text{Tr}\left(\Omega_{\rho\sigma}\Omega^{\rho\sigma}\right) &= -2 R_{\mu\nu\rho\sigma}R^{\mu\nu\rho\sigma}\delta_{IJ}\delta_{JI} = -4 R_{\mu\nu\rho\sigma}R^{\mu\nu\rho\sigma}.
	\end{split}
\end{align}
This trace data is sufficient to determine the ghost contribution to the third-order Seeley-DeWitt coefficient $a_4(x)$,
\begin{align}\label{sdc29}
	(4\pi)^2 {a_4}^{\text{ghost}}(x) &= \frac{2}{9} R_{\mu\nu\rho\sigma}R^{\mu\nu\rho\sigma} - \frac{17}{9}R_{\mu\nu}R^{\mu\nu} - \frac{17}{18} R^2.
\end{align}
One must note that this result incorporates $\chi = -1$ in the formula \eqref{comp8} due to the inclusion of bosonic ghosts where the negative signature arises as the ghosts follow a reverse spin-statistics compared to the physical fields.  

\subsection{Results}\label{res1}

In gravitational theories with quantum fluctuations, the logarithmic correction to black hole entropy induced by the one-loop effective action is well-known to be associated with trace anomalies (e.g., see \cite{Keeler:2014nn, Xiao:2021zly, Aros:2013taa}). Within the framework of effective action formulated by the heat kernel expansion and the background field formalism (as detailed in \cref{setup}), the trace anomalies explicitly appear as terms within the heat expansion coefficients and depend on the background curvature invariants. Specifically, the trace anomalies are encoded within the third-order Seeley-DeWitt coefficient $a_4(x)$ for four-dimensional theories.

Our current objective is to unveil trace anomalies within $a_4(x)$ for the ongoing analysis of the quantized STU supergravity model around Einstein-Maxwell-AdS backgrounds. Specifically, we will demonstrate that the square of the Weyl tensor $W_{\mu\nu\rho\sigma}W^{\mu\nu\rho\sigma}$ and the four-dimensional Euler-Gauss-Bonnet density $E_4$ respectively appear as the type-B and type-A trace anomalies, defined as 
\begin{align}\label{sdc31}
	\begin{split}
		W_{\mu\nu\rho\sigma}W^{\mu\nu\rho\sigma} &= R_{\mu\nu\rho\sigma}R^{\mu\nu\rho\sigma}-2R_{\mu\nu}R^{\mu\nu} + \frac{1}{3}R^2,\\
		E_4 &= R_{\mu\nu\rho\sigma}R^{\mu\nu\rho\sigma}-4R_{\mu\nu}R^{\mu\nu} + R^2.
	\end{split}
\end{align} 
We need to combine the computed gauge-fixed contribution (\ref{sdc16}) and ghost contribution (\ref{sdc29}). These contributions are computed separately since the ghosts do not interact with the gauge-fixed sector of the theory containing physical modes. The combination of gauge-fixed and ghost parts is straightforward, resulting in
\begin{align}\label{sdc30}
	(4\pi)^2 {a_4}^{\text{STU}}(x) &= \frac{209}{180} R_{\mu\nu\rho\sigma}R^{\mu\nu\rho\sigma} - \Bigg(\frac{29}{180} + \frac{2\kappa_1^2}{3\left(\kappa_1^4 + 1\right)}\Bigg)R_{\mu\nu}R^{\mu\nu} \nonumber \\
	&\quad  - \Bigg(\frac{25}{18} - \frac{\kappa_1^2}{6\left(\kappa_1^4 + 1\right)}\Bigg)R^2 + \frac{\left(\kappa_1^2 - 1\right)^2}{6\left(\kappa_1^4 + 1\right)}R \bar{F}_{\mu\nu}\bar{F}^{\mu\nu}.
\end{align}  
For the solution of the STU supergravity model embedded with Einstein-Maxwell backgrounds in AdS space (or with a negative cosmological constant $\Lambda$), the Seeley-DeWitt coefficient $a_4(x)$ can be decomposed in terms of a sum of four-derivative background terms, including the trace anomalies:
{
	\allowdisplaybreaks
	\begin{subequations}\label{sdc32}
		\begin{align}\label{sdc32a}
			(4\pi)^2 {a_4}^{\text{STU}}(x) &=  \mathfrak{c} W_{\mu\nu\rho\sigma}W^{\mu\nu\rho\sigma} - \mathfrak{a} E_4 + \mathfrak{b}_1 R^2 + \mathfrak{b}_2R \bar{F}_{\mu\nu}\bar{F}^{\mu\nu},
		\end{align}
		where the coefficients $\mathfrak{c}$, $\mathfrak{a}$, $\mathfrak{b}_1$, and $\mathfrak{b}_2$ can be extracted from the relation \eqref{sdc30}, derived using the two-derivative action of the present theory. The coefficients $\mathfrak{c}$ and $\mathfrak{a}$, respectively multiplying $W_{\mu\nu\rho\sigma}W^{\mu\nu\rho\sigma}$ and $E_4$, are recognized as the two central charges of the conformal anomaly in 4D. The results of the coefficients in \eqref{sdc32a} are summarized as follows
		\begin{align}\label{sdc32b}
			\mathfrak{c} &=  \frac{1}{3}\left(\frac{269}{40} - \frac{\kappa_1^2}{\kappa_1^4 + 1} \right),\\
			\mathfrak{a} &=  \frac{1}{3}\left(\frac{389}{120} - \frac{\kappa_1^2}{\kappa_1^4 + 1} \right),\\
			\mathfrak{b}_1 &= -\frac{1}{18}\left(19 + \frac{\kappa_1^2}{\kappa_1^4 + 1} \right),\\
			\mathfrak{b}_2 &= \frac{\left(\kappa_1^2 - 1\right)^2}{6\left(\kappa_1^4 + 1\right)}  .
		\end{align}
	\end{subequations}
}
It is important to note that the absence of the dilaton coupling constant $\kappa_2$ from the results presented in \cref{sdc30,sdc32,sdc16} should not raise concerns. The $\kappa_2$ contributions are consistently taken into account in all relevant places. We substituted them using the specific choice $\kappa_2 = -1/\kappa_1$ in order to simplify the forms of the heat kernel results computed for the two truncation cases of the STU model, i.e., when $(\kappa_1, \kappa_2) = (1, -1)$ and $(\sqrt{3}, -1/\sqrt{3})$ (refer to \cref{embeddingEM}). While it is feasible to rewrite the results in terms of relatively complex forms involving both $\kappa_1$ and $\kappa_2$ at any point, such a task is not straightforward and requires the use of trace data and identities provided in \cref{tracing,iden}.

The derived trace anomaly data \eqref{sdc32} is specific to the chosen theory, relying entirely on the one-loop fluctuations and embedded asymptotically-AdS EM background in the STU supergravity model. Equivalent results for the $U(1)^2$-charged STU model in asymptotically-flat EM backgrounds can be easily obtained by setting the cosmological constant to zero, $\Lambda=0$. In this flat space limit, terms such as $R^2$ and $R \bar{F}_{\mu\nu}\bar{F}^{\mu\nu}$ vanish, and the relation \eqref{sdc30} for $a_4(x)$ is solely governed by the Weyl anomaly $W_{\mu\nu\rho\sigma}W^{\mu\nu\rho\sigma}$ and the Euler density $E_4$, with non-vanishing coefficients $\mathfrak{c}$ and $\mathfrak{a}$. Moreover, the $a_4(x)$ coefficient in the flat-space limit becomes invariant under electromagnetic duality, as it is independent of terms proportional to the effective field strength $\bar{F}_{\mu\nu}$ (as defined in \cref{mod10}). Interestingly, this property persists even when transitioning beyond the flat limit and considering AdS backgrounds embedded in the $(\kappa_1, \kappa_2) = (1, -1)$ case of the STU theory, resulting in a vanishing $\mathfrak{b}_2$. However, the STU model with $(\kappa_1, \kappa_2) = (\sqrt{3}, -1/\sqrt{3})$ breaks the electromagnetic duality invariant nature of $a_4(x)$ due to the induction of a non-zero $\mathfrak{b}_2$.

Before delving into the implications of the heat kernel and trace anomaly data for computing the logarithmic correction to the entropy of embedded black holes (refer to \cref{logresult}), it is essential to assess their consistency with existing literature. While the current STU supergravity model stands as the most natural and non-trivial generalization of previously explored bosonic supergravity models (for example, refer to the review \cite{Bhattacharyya:2012ss,Karan:2021teq,Sen:2013ns,Charles:2015nn,Karan:2019sk,Karan:2020sk,David:2021eoq,Sen:2012qq,Keeler:2014nn,Castro:2018tg,Karan:2022dfy}), there is no direct path leading to these available results by setting appropriate limits in the relations presented in \cref{sdc30,sdc32}. Technically, this is because the dilaton-coupling-independent terms in the heat kernel or trace anomaly data are sensitive to the specific off-shell degrees of freedom and their non-minimal interactions within the theory, making them less controllable. Nevertheless, we have conducted meticulous checks, starting with the original form of trace data recorded in \cref{tracing} and proceeding with top-down manipulations. This approach has yielded exact results for the fluctuated Einstein-Maxwell theory \cite{Bhattacharyya:2012ss,Karan:2021teq,Sen:2013ns} or the boson sector of $\mathcal{N}=2$ supergravity (for $\kappa_1=\kappa_2=0$) \cite{Charles:2015nn,Karan:2019sk,Karan:2020sk,David:2021eoq,Sen:2012qq,Keeler:2014nn}, the Kaluza-Klein system (for $\kappa_1=\sqrt{3}, \kappa_2=0, \Lambda =0$) \cite{Castro:2018tg}, and Einstein-Maxwell-dilaton models (for $\kappa_1 = 1, \sqrt{3}, 1/\sqrt{3}$ and $\kappa_2=0$) \cite{Karan:2022dfy}. Each of these emerges as a distinct and consistent limiting case of the current STU model \eqref{mod2} heat kernel results obtained in \cref{sdc30,sdc32}.


\section{Logarithmic corrections to AdS$_4$ and flat$_4$ black hole entropy}\label{logresult}
In this section, we delve into the derivation of the logarithmic correction to the entropy of black holes embedded in the truncated four-dimensional STU supergravity models discussed in \cref{setup}. These backgrounds encompass all the Kerr-Newman-AdS, Reissner-Nordström-AdS, Kerr-AdS, and Schwarzschild-AdS black holes, along with the Kerr-Newman, Reissner-Nordström, Kerr, and Schwarzschild black holes, which are generic solutions of the Einstein-Maxwell theory evolving with and without a negative cosmological constant, respectively. Here, we interpret or uplift them as the background solutions of the $U(1)^2$-charged EMD-AdS and EMD theories intersecting with the gauged and ungauged STU supergravities in four spacetime dimensions.

At this stage, we apply the framework established in \cref{setup}, which necessitates the utilization of the Seeley-DeWitt coefficient $a_4(x)$ calculated in \cref{SDCcomputation,res1}. Specifically, our current objective is to provide the trace anomaly form \eqref{sdc32} of $a_4(x)$ as a precursor for determining the logarithmic corrections for both non-extremal and extremal black holes embedded in the STU models. In the local contribution $\mathcal{C}_{\text{local}}$, all the four-derivative background invariants, including the $W_{\mu\nu\rho\sigma}W^{\mu\nu\rho\sigma}$ and $E_4$ anomalies, need to be integrated over the appropriate part of the geometry of the concerned black hole backgrounds. This leads us to the following general formula for obtaining the local part of logarithmic entropy corrections
\begin{align}\label{lr1}
	\mathcal{C}_{\text{local}} &= \frac{\mathfrak{c}}{16\pi^2}\int \mathrm{d}^4x \sqrt{\bar{g}} W_{\mu\nu\rho\sigma}W^{\mu\nu\rho\sigma} - \frac{\mathfrak{a}}{16\pi^2}\int \mathrm{d}^4x \sqrt{\bar{g}} E_4 \nonumber \\
	&\enspace + \frac{\mathfrak{b}_1}{16\pi^2}\int \mathrm{d}^4x \sqrt{\bar{g}} R^2 + \frac{\mathfrak{b}_2}{16\pi^2}\int \mathrm{d}^4x \sqrt{\bar{g}} R \bar{F}_{\mu\nu}\bar{F}^{\mu\nu},
\end{align}
where the coefficients $\mathfrak{a}$, $\mathfrak{c}$, $\mathfrak{b}_1$ and $\mathfrak{b}_2$ do not affect the integrations since their values, as obtained in \cref{sdc32b}, are constants fixed by the choice of theory. For non-extremal or finite-temperature AdS$_4$ and flat$_4$ black holes, the integrations need to be performed over the entire background geometry $\bar{g}_{\mu\nu}$. Meanwhile, for extremal black holes, we observe that the contributions to $\mathcal{C}_{\text{local}}$ remain the same whether the integration \eqref{lr1} is executed over the full geometry or only the finite piece of near-horizon geometry.

In the following subsections, we systematically derive specific relations for computing contributions to $\mathcal{C}_{\text{local}}$ in STU models. These relations are expressed in terms of various dimensionless ratios involving different black hole parameters and the anomaly coefficients $\mathfrak{a}$, $\mathfrak{c}$, $\mathfrak{b}_1$ and $\mathfrak{b}_2$. Throughout this process, our primary focus is on evaluating the curvature invariant integrations, as detailed in relation \eqref{lr1}, for the most generic non-extremal background of a four-dimensional Kerr-Newman-AdS black hole. Subsequently, we emphasize on the extremal and near-horizon cases, along with their asymptotically-flat limits. Furthermore, the same analysis is extended to other AdS$_4$ and flat$_4$ counterparts of Reissner-Nordström, Kerr, and Schwarzschild black holes. As a final output, we present the explicit results for the logarithmic correction to the entropy of all black holes embedded in both the gauged and ungauged versions of the STU supergravity theory.  

\subsection{Black hole backgrounds in STU supergravity and $\mathcal{C}_{\text{local}}$ contributions}\label{local}
\subsubsection{Kerr-Newman-AdS$_4$ black hole}\label{kna}  
We start with a four-dimensional non-extremal Kerr-Newman-AdS (KNAdS) black hole, representing the most interesting and generic Einstein-Maxwell (EM) background in our chosen STU models. In terms of the standard Boyer-Lindquist type coordinates, the metric for this charged and rotating background is given by \cite{Caldarelli:1999x}  
\begin{align}\label{kna1}
	\bar{g}_{\mu\nu}\mathrm{d}x^\mu \mathrm{d}x^\nu &= - \frac{\Delta _r}{\rho^2} \left( \mathrm{d}t - \frac{a \sin^2 \theta }{\Xi} \mathrm{d}\phi \right)^2 + \frac{\rho^2}{\Delta_r}\mathrm{d}r^2  + \frac{\rho^2}{\Delta_\theta} \mathrm{d}\theta^2 \nonumber\\
	&\qquad + \frac{\Delta_\theta \sin^2 \theta}{\rho^2} \left(a\,\mathrm{d}t - \frac{r^2+a^2}{\Xi} \mathrm{d}\phi \right)^2,
\end{align}
where we set $G_N = 1$. The parameters $\Delta_r$, $\Delta_\theta$, $\rho$, and $\Xi$ are related to the physical mass $M$, angular momentum $J$, and electric charge $Q$ of the black hole, as well as the boundary radius $\ell$ of AdS$4$ space, via:
\begin{align}\label{kna2}
	\begin{gathered}
		M = \frac{m}{\Xi^2}, \enspace J= \frac{ma}{\Xi^2}, \enspace Q = \frac{q}{\Xi},\enspace \Xi = 1 -\frac{a^2}{\ell^2},\\[5pt]
		\Delta_r = (r^2+a^2) \left(1+\frac{r^2}{\ell^2}\right) - 2m r + q^2,\\
		\Delta_\theta = 1 -\frac{a^2}{\ell^2} \cos^2 \theta, \enspace \rho^2 = r^2 + a^2 \cos^2 \theta.
	\end{gathered}
\end{align}
It is important to note that this background setup is valid only when the rotational parameter $a$ satisfies $a^2 < \ell^2$ and becomes singular in the limit $a^2 = \ell^2$. Throughout our analysis, we consider $a \geq 0$, and obtain all interested results by replacing $a\to \left|a\right|$ everywhere without any loss of generality. Furthermore, we consistently turn off terms related to the magnetic charge parameters in order to embed the charged EM backgrounds into the current STU supergravity models. The metric \eqref{kna1} satisfies the background EM field equations \eqref{mod8} for $R=4\Lambda = -\dfrac{12}{\ell^2}$ with an electrically-charged vector potential $\bar{A}_\mu$ and an associated Maxwell field strength tensor $\bar{F}_{\mu\nu}$, given respectively by
\begin{align}\label{kna3}
	\bar{A} = -\frac{qr}{\rho^2}\left(\mathrm{d}t - \frac{a \sin^2 \theta }{\Xi} \mathrm{d}\phi\right),
\end{align}
and
\begin{align}\label{kna4}
	\bar{F}_{\mu\nu}\bar{F}^{\mu\nu} = -\frac{2q^2}{\left(r^2 + a^2 \cos ^2\theta\right)^4}\left(r^4- 6a^2r^2\cos^2\theta + a^4\cos^4\theta\right).
\end{align}
As per the specific embedding choices \eqref{mod9} and \eqref{mod10}, the KNAdS background is interpreted as a solution of the truncated STU models \eqref{mod2} for an effective electric charge parameter,
\begin{align}\label{kna5}
	q = \sqrt{q_1^2 + q_2^2} \quad \text{or} \enspace Q = \sqrt{Q_1^2 + Q_2^2},
\end{align}   
where $q_1$ and $q_2$ represent the respective charge parameters of the two $U(1)$ gauge fields, satisfying \eqref{kna3} and \eqref{kna4}. If the outer event horizon of the KNAdS black hole \eqref{kna1} is located at $r=r_+$, then one obtains $r_+$ by finding the largest real root of $g_{00}=\Delta_r =0$ and fixing the mass parameter as
\begin{align}\label{kna6}
	m= \frac{r_+}{2}\left(1+ \frac{a^2}{\ell^2} + \frac{a^2 + q^2}{r_+^2} + \frac{r_+^2}{\ell^2}\right).
\end{align}
At this level, it is standard practice to consider the analytical continuation of the Lorentzian metric \eqref{kna1} by $t \to i\tau$ and $a \to ia$ for obtaining the Euclidean structure of the KNAdS black hole. The associated Bekenstein-Hawking entropy is given by
\begin{align}
	S_{\text{BH}} = 4\pi \frac{\left(r_+^2 + a^2\right)}{\left(1 -\frac{a^2}{\ell^2}\right)}.
\end{align}
Also, the regularity at $r=r_+$ requires identifying $\tau \sim \tau + \beta$ where inverse of $\beta$ is a measure of the Hawking temperature of the black hole,
\begin{align}\label{kna7}
	T_{\text{bh}} = \beta^{-1} = \frac{r_+}{4\pi \left(r_+^2 + a^2\right)} \left(1 + \frac{a^2}{\ell^2} + \frac{3 r_+^2}{\ell^2}-\frac{a^2 + q^2}{r_+^2}\right).
\end{align}
With the above setup, the four curvature invariants necessary for \eqref{lr1} in the electrically-charged and rotating KNAdS black hole case are expressed as follows   
{\allowdisplaybreaks 
	\begin{subequations}\label{kna8}
		\begin{align}
			W_{\mu\nu\rho\sigma}W^{\mu\nu\rho\sigma} &= \frac{48}{\left(r^2 + a^2 \cos ^2\theta\right)^6} \bigg[ 8 r^4 \left(q^2 - 2mr\right)^2 - m^2 \left(r^2 + a^2 \cos ^2\theta\right)^3 \nonumber\\
			&\qquad - 8r^2 \left(q^2 - 3mr\right)\left(q^2 - 2mr\right)\left(r^2 + a^2 \cos ^2\theta\right) \nonumber\\
			& \qquad + \Big(q^4 - 10 m r q^2+ 18 m^2r^2\Big)\left(r^2 + a^2 \cos ^2\theta\right)^2 \bigg],\\[5pt]
			E_4 &= \frac{24}{\ell^4} + \frac{8}{\left(r^2 + a^2 \cos ^2\theta\right)^6} \bigg[ 48 r^4 \left(q^2 - 2mr\right)^2 - 6m^2 \left(r^2 + a^2 \cos ^2\theta\right)^3 \nonumber\\
			&\qquad - 48r^2 \left(q^2 - 3mr\right)\left(q^2 - 2mr\right)\left(r^2 + a^2 \cos ^2\theta\right) \nonumber\\
			& \qquad + \Big(5q^4 - 60 m r q^2+ 108 m^2r^2\Big)\left(r^2 + a^2 \cos ^2\theta\right)^2 \bigg],\\[5pt]
			R^2 &= \frac{144}{\ell^4},\\[5pt]
			R\bar{F}_{\mu\nu}\bar{F}^{\mu\nu} &= \frac{24q^2}{\ell^2\left(r^2 + a^2 \cos ^2\theta\right)^4}\left(r^4- 6a^2r^2\cos^2\theta + a^4\cos^4\theta\right).
		\end{align}
\end{subequations}	 }
Next, we proceed to integrate all the aforementioned invariants over the Euclideanized KNAdS$_4$ black hole geometry. However, these integrations pose challenges due to divergence caused by the infinite volume of AdS$_4$. In this paper, we address this issue by employing holographic renormalization principles \cite{Skenderis:2002wp}. Specifically, we introduce a cutoff at $r=r_c$ on the boundary of the KNAdS$_4$ geometry \eqref{kna1}. Following this, we introduce a holographic counterterm,
\begin{align}\label{kna9}
	\mathcal{C}_{\text{HCT}} = \int_{\partial(\text{AdS$_4$})} \mathrm{d}^3y \sqrt{\det\gamma} \left(c_1 + c_2 \mathscr{R}\right). 
\end{align}
Here, $\mathscr{R}$ represents the Ricci scalar associated with the metric $\gamma_{\mu\nu}$ characterizing the KNAdS$_4$ boundary geometry,
{\allowdisplaybreaks 
	\begin{subequations}
		\begin{align}\label{holo1}
			\gamma_{\mu\nu}\mathrm{d}y^\mu \mathrm{d}y^\nu &= - \frac{\Delta_{r_c}}{\rho_c^2} \left( i\mathrm{d}\tau + \frac{a \sin^2 \theta }{\Xi} \mathrm{d}\phi \right)^2  + \frac{\rho_c^2}{\Delta_\theta} \mathrm{d}\theta^2  + \frac{\Delta_\theta \sin^2 \theta}{\rho_c^2} \left(ia\,\mathrm{d}\tau + \frac{r_c^2+a^2}{\Xi} \mathrm{d}\phi \right)^2,
		\end{align}
		where
		\begin{align}\label{holo2}
			\begin{gathered}
				\Delta_{r_c} = (r_c^2+a^2) \left(1+\frac{r_c^2}{\ell^2}\right) - 2m {r_c} + q^2, \quad {\rho_c}^2 = r_c^2 + a^2 \cos^2 \theta.
			\end{gathered}
		\end{align}
\end{subequations}}  
For this boundary geometry, we determine
{\allowdisplaybreaks 
	\begin{subequations} 
		\begin{align}
			\det \gamma &= \frac{1}{\Xi^2}\left[q^2 -2mr_c + \left(r_c^2 + a^2\right)\left(1 + \frac{r_c^2}{\ell^2}\right) \right]\rho_c^2 \sin^2\theta,\\[5pt]
			\mathscr{R} &= \frac{2}{\rho_c^6}\bigg[r_c^4 + \left(r_c^2-2m{r_c} + q^2\right)a^2\cos^2 \theta +\frac{a^2}{\ell^2}\Big(r_c^2\left(1-5\cos^2\theta\right)-3a^2\cos^4\theta\Big)\rho_c^2\bigg],
		\end{align}
\end{subequations}}
and derive the following form of the holographic counterterm \eqref{kna9},
\begin{align}\label{kna10}
	\mathcal{C}_{\text{HCT}} = \frac{4\pi\beta}{\Xi\ell}\left[c_1 r_c^3 + \frac{1}{6\ell^2}\Big(c_1\left(4a^2 + 3\ell^2\right)\ell^2 + 4c_2\left(3\ell^2 - 2a^2\right)\Big)r_c - c_1m\ell^2\right] + \mathcal{O}\left(r_c^{-1}\right).
\end{align}
The integration range is defined as $0 \leq \tau \leq \beta$, $0 \leq \theta \leq \pi$, and $0 \leq \phi \leq 2\pi$. Subsequently, the boundary term \eqref{kna10} needs to be combined with the bulk $\mathcal{C}_{\text{local}}$ term \eqref{lr1}, involving integrations over the KNAdS$_4$ background invariants. The resulting bulk-boundary combinations contain $\mathcal{O}\left(r_c^{-1}\right)$ order terms that vanish as $r_c \to \infty$. The next step involves appropriately setting the coefficients $c_1$ and $c_2$ to cancel the $r_c^3$ and $r_c$ divergences. This procedure yields a finite contribution of all the integrated background invariants, leading to a regularized $\mathcal{C}_{\text{local}}$ given by,
\begin{align}\label{kna11}
	\mathcal{C}_{\text{local}} = \lim_{r_c \to \infty}\left[\frac{1}{16\pi^2}\int_0^\beta \mathrm{d}\tau \int_{r_+}^{r_c} \mathrm{d}r \int_0^{\pi} \mathrm{d}\theta \int_0^{2\pi} \mathrm{d}\phi\, \sqrt{\det\bar{g}}\,{a_4}(\tau, r, \theta, \phi) + \mathcal{C}_{\text{HCT}}\right].
\end{align} 
Here, the Seeley-DeWitt coefficient ${a_4}(\tau, r, \theta, \phi)$ appears as a function of the invariants $W_{\mu\nu\rho\sigma}W^{\mu\nu\rho\sigma}$, $E_4$, $R^2$, and $ R\bar{F}_{\mu\nu}\bar{F}^{\mu\nu}$ around the Euclideanized KNAdS$_4$ geometry \eqref{kna1}. The integrated invariants, as required in \eqref{lr1}, take the following particular forms,
{\allowdisplaybreaks
	\begin{subequations}\label{kna12} 
		\begin{align}\label{nbh9a}	
			\frac{1}{16\pi^2}\int \mathrm{d}^4x \sqrt{\det\bar{g}}\, W_{\mu\nu\rho\sigma}W^{\mu\nu\rho\sigma} &=  W_1 + \beta W_2 + \frac{W_3}{\beta},\\
			\frac{1}{16\pi^2}\int \mathrm{d}^4x \sqrt{\det\bar{g}}\, E_4 &= 4, \\
			\frac{1}{16\pi^2}\int \mathrm{d}^4x \sqrt{\det\bar{g}}\, R^2 &= R_1 + \beta R_2,\\
			\frac{1}{16\pi^2}\int \mathrm{d}^4x \sqrt{\det\bar{g}}\, R\bar{F}_{\mu\nu}\bar{F}^{\mu\nu} &= F_1 + \beta F_2,
		\end{align}
		where the pieces $W_i$, $R_i$, and $F_i$ are isolated based on their dependency on the inverse temperature parameter $\beta$. In particular, they are expressed as
		\begin{align}\label{kna13} 
			W_1 &=  \frac{\left(a^2+r_+^2\right)}{2\Xi a^5 \ell^2r_+^3 }\bigg[\Big(3 a^4 \left(\ell ^2-r_+^2\right) + 4 a^2\ell ^2 r_+^2 -3 \left(\ell ^2 + 3 r_+^2\right)r_+^4 \Big)a r_+ \nonumber\\
			&\quad - 3 \left(r_+^4- a^4\right) \Big(a^2 \left(\ell ^2 - r_+^2\right) - \left(\ell ^2 + 3 r_+^2\right)r_+^2 \Big) \arctan\left(\frac{a}{r_+}\right)\bigg], \\[7pt]
			W_2 &=  \frac{1}{16\pi \Xi a^5 \ell^4\left(r_+^2 + a^2\right)r_+^4  } \bigg[3 a^9 \left(\ell ^2-r_+^2\right)^2r_+ -4 a^7 \left(3 r_+^4 + 12 \ell ^2r_+^2 +\ell ^4\right)r_+^3 \nonumber\\
			&\quad +2 a^5 \left(5 r_+^4 -14\ell ^2 r_+^2 +\ell ^4\right)r_+^5  -4 a^3 \left(\ell ^4-9 r_+^4\right)r_+^7  +3 a \left(\ell ^2 + 3 r_+^2\right)^2r_+^9 \nonumber\\
			& \quad - 3 \left(r_+^4 - a^4\right) \left(a^2+r_+^2\right) \Big( a^2 \left(\ell ^2 - r_+^2\right) - \left(\ell ^2 + 3 r_+^2\right)r_+^2  \Big)^2 \arctan\left(\frac{a}{r_+}\right)\bigg], \\[7pt]
			W_3 &= \frac{\pi \left(a^2+r_+^2\right)}{\Xi a^5 r_+^2 } \bigg[ar_{+}\left(3a^4 + 2a^2r_+^2 + 3 r_+^4\right)  - 3 \left(r_+^4 - a^4\right) \left(a^2+r_+^2\right)\arctan\left(\frac{a}{r_+}\right)\bigg], \\
			R_1 &= -\frac{24}{\Xi\ell^2}\left(r_+^2 + a^2\right), \quad R_2 = \frac{12r_+}{\pi\Xi\ell^4}\left(\ell^2 + r_+^2\right), \\[5pt]
			F_1 &= -\frac{24 r_+^2}{\Xi\ell^2}, \quad F_2 = \frac{6 r_+ }{\pi  \Xi  \ell ^4\left(a^2+r_+^2\right)}\left[3 r_+^4 + \left(\ell^2 + a^2\right)r_+^2 - a^2\ell ^2 \right].
		\end{align}
\end{subequations} }
The regularization procedure outlined above is applicable to all background geometries considered in this paper. The holographic renormalization prescription employed is natural, producing consistent and unambiguous results by isolating the finite piece of the bulk effective action from divergent terms. In contrast, any naive regularization treatment that merely removes the divergent term would fail to yield physically sensible results, as demonstrated in \cref{kna11,kna12}. As an illustrative example, consider verifying the Euler characteristic for AdS$_4$ spacetime obtained through this holographic renormalization procedure:
\begin{align}\label{nbh14}
	\chi = \lim_{r_c \to \infty}\left[\frac{1}{32\pi^2}\int_{\text{KNAdS$_4$}} \mathrm{d}^4x \sqrt{\det\bar{g}}\, E_4 + \frac{1}{32\pi^2}\int_{\partial(\text{KNAdS$_4$})} \mathrm{d}^3y \sqrt{\det\gamma} \left(c_1 + c_2 \mathscr{R}\right)\right] = 2,
\end{align}
where $c_1 = -\frac{8}{\ell^3}$ and $c_2 = \frac{2}{\ell}$. Remarkably, this result aligns precisely with the prediction from the integrated four-dimensional Euler density $E_4$ using the Gauss-Bonnet-Chern theorem \cite{Chern:1945wp}. This alignment validates the current renormalization procedure, and for more discussions, we refer the reader to \cite{David:2021eoq,Karan:2022dfy}.

Finally, the integration of the invariants \eqref{kna12} culminates in the ultimate expression for $\mathcal{C}_{\text{local}}$ for the non-extremal Kerr-Newman-AdS$_4$ black hole in STU models,
\begin{align}\label{nbh15}
	\mathcal{C}_{\text{local}}^{\text{(KN-AdS)}} = \frac{\mathfrak{c}W_3}{\beta}  - \left(4\mathfrak{a} -\mathfrak{c}W_1 - \mathfrak{b}_1R_1 - \mathfrak{b}_2F_1\right) + \beta \left(\mathfrak{c}W_2 + \mathfrak{b}_1R_2 + \mathfrak{b}_2 F_2\right),
\end{align}
where the non-topological terms $W_i$, $R_i$ and $F_i$ are expressed in terms of different black hole parameters, as detailed in \cref{kna13}. Notably, the relation \eqref{nbh15} precisely aligns with the results obtained in \cite{David:2021eoq} for vanishing magnetic charges on the KNAdS$_4$ background. Furthermore, we verify the consistency of the $\mathcal{C}_{\text{local}}$ contributions derived for the other black hole backgrounds considered in this paper as distinct special cases.


\subsubsection{Reissner-Nordström-AdS$_4$ black hole}\label{rn}
The background metric for a 4D non-extremal Reissner-Nordström-AdS (RNAdS$_4$) black hole solution of the STU field equations \eqref{mod3} is expressed as follows
{\allowdisplaybreaks
	\begin{subequations}\label{rn1}
		\begin{align}
			\bar{g}_{\mu\nu}\mathrm{d}x^\mu \mathrm{d}x^\nu &= - f(r) \mathrm{d}t^2 + \frac{\mathrm{d}r^2}{f(r)} + r^2 \left(\mathrm{d}\theta^2 + \sin^2\theta\mathrm{d}\phi^2\right), \enspace \bar{A} = \bar{A}_t(r)\mathrm{d}t = -\frac{q}{r}\mathrm{d}t,
		\end{align}
		where the function $f(r)$ is expressed in terms of black hole mass parameter $m$, effective electric charge $q$ (refer to \eqref{kna5}) and AdS$_4$ radius $\ell$,
		\begin{align}
			f(r) = 1 - \frac{2 m}{r} + \frac{q^2}{r^2} + \frac{r^2}{\ell^2}.
		\end{align}
\end{subequations}}
Upon Euclidean continuation ($t \to i\tau$), the corresponding Hawking temperature is given by 
\begin{align}\label{rn2}
	T_{\text{bh}} = \beta^{-1} = \frac{1}{4\pi r_+}\left(1 - \frac{q^2}{r_+^2} + \frac{3r_+^2}{\ell^2}\right),
\end{align}
where $r_+$ represents the outer horizon radius at which $g_{00}(r_+) = f(r_+) = 0$. This yields the mass parameter as
\begin{align}\label{rn3}
	m= \frac{r_+}{2}\left(1+ \frac{q^2}{r_+^2} + \frac{r_+^2}{\ell^2}\right).
\end{align}
For the Euclideanized background geometry \eqref{rn1}, the required invariants are given by
{\allowdisplaybreaks 
	\begin{subequations}\label{rn4}
		\begin{align}
			W_{\mu\nu\rho\sigma}W^{\mu\nu\rho\sigma} &= \frac{48}{r^{8}}\left(m r-q^{2}\right)^{2},\\[3pt]
			E_4 &= \frac{24}{\ell^{4}}+\frac{8}{r^{8}}\left(6 m^{2} r^{2}-12 mq^{2} r+5q^{4}\right),\\[3pt]
			R^2 &= \frac{144}{\ell^4},\\[3pt]
			R\bar{F}_{\mu\nu}\bar{F}^{\mu\nu} &=\frac{24q^{2}}{\ell^2r^{4}}.
		\end{align}
\end{subequations}	 }
We then employ the holographic renormalization procedure as outlined in \cref{kna} to derive regulated values for the integration of the above invariants on the RNAdS$_4$. The results are as follows
{\allowdisplaybreaks
	\begin{subequations}\label{rn5} 
		\begin{align}
			\frac{1}{16\pi^2}\int \mathrm{d}^4x \sqrt{\bar{g}}\, W_{\mu\nu\rho\sigma}W^{\mu\nu\rho\sigma} &= \frac{ 2\beta r_+^3}{5 \pi \ell^{4}} \left(4 +\frac{\ell^{2}}{r_+^2}  + \frac{\ell^{4}}{r_+^4}\right) + \frac{4}{5} \left(1-\frac{7 r_+^{2}}{\ell^{2}}\right) + \frac{32 \pi r_{+}}{5 \beta},\\
			\frac{1}{16\pi^2}\int \mathrm{d}^4x \sqrt{\det\bar{g}}\, E_4 &= 4, \\
			\frac{1}{16\pi^2}\int \mathrm{d}^4x \sqrt{\det\bar{g}}\, R^2 &= \frac{12\beta r_+^3}{\pi \ell^{4}}\left(1 +\frac{\ell^{2}}{r_+^2}\right) -\frac{24 r_{+}^{2}}{\ell^{2}} ,\\[2pt]
			\frac{1}{16\pi^2}\int \mathrm{d}^4x \sqrt{\det\bar{g}}\, R\bar{F}_{\mu\nu}\bar{F}^{\mu\nu} &= \frac{6\beta r_+^3}{\pi \ell^{4}}\left(3 +\frac{\ell^{2}}{r_+^2}\right) -\frac{24 r_{+}^{2}}{\ell^{2}}.
		\end{align}
\end{subequations}}	
These results lead to the final form of $\mathcal{C}_{\text{local}}$ for the non-extremal Reissner-Nordström-AdS$_4$ black hole embedded in the STU models, given by
\begin{align}\label{rn6}
	\mathcal{C}_{\text{local}}^{\text{(RN-AdS)}} &= \frac{4}{5} \left(\mathfrak{c}-5\mathfrak{a}\right) - \frac{4}{5} \bigg[\left(7\mathfrak{c} + 30\mathfrak{b}_1 + 30\mathfrak{b}_2\right)\frac{r_{+}^{2}}{\ell^{2}} - 8\pi\mathfrak{c}\frac{r_+}{\beta} \nonumber\\
	&\enspace  -\frac{\beta r_+^3}{2\pi\ell ^4} \left( \left(4\mathfrak{c} + 30 \mathfrak{b}_1 + 45 \mathfrak{b}_2\right)  + \left(\mathfrak{c} + 30 \mathfrak{b}_1 + 15 \mathfrak{b}_2\right)\frac{\ell^2}{r_+^2}  + \frac{\mathfrak{c}\ell^4}{r_+^4}\right) \bigg].
\end{align}
We have verified that the above relation is in perfect agreement with \eqref{nbh15} for a vanishing rotation parameter, i.e., $a=0$. Notably, there exists an additional topological term $\frac{4}{5}\mathfrak{c}$ when transitioning into the non-rotating but charged background.


\subsubsection{Kerr-AdS$_4$ black hole}\label{ka}

In the Boyer-Lindquist coordinate system, the background metric describing the Kerr-AdS black hole is presented as a truncation of the geometry \eqref{kna1} by setting $q=0$. Specifically, the parameter $\Delta_r$ takes a modified form,
\begin{align}\label{ka1}
	\Delta_r = (r^2+a^2) \left(1+\frac{r^2}{\ell^2}\right) - 2m r.
\end{align}
For the outer horizon at $r=r_+$ with $\Delta_r (r_+)= 0$, and with the same Euclideanized setup, the mass parameter $m$ and the inverse Hawking temperature $\beta$ are characterized as
\begin{align}\label{ka2}
	\begin{split}
		m &= \frac{r_+}{2}\left(1+ \frac{a^2}{\ell^2} + \frac{a^2}{r_+^2} + \frac{r_+^2}{\ell^2}\right),\\[5pt]
		\beta^{-1} &= \frac{r_+}{4\pi \left(r_+^2 + a^2\right)} \left(1 + \frac{a^2}{\ell^2} + \frac{3 r_+^2}{\ell^2}-\frac{a^2}{r_+^2}\right). 
	\end{split}
\end{align}
The necessary background invariants on the KAdS$_4$ geometry are derived as
{\allowdisplaybreaks 
	\begin{subequations}\label{ka3}
		\begin{align}
			W_{\mu\nu\rho\sigma}W^{\mu\nu\rho\sigma} &= \frac{48m^2}{\left(r^2 + a^2 \cos ^2\theta\right)^6} \bigg[r^6 - 15r^2a^2\left(r^2 - a^2\cos ^2\theta\right)\cos ^2\theta  - a^6 \cos ^6\theta \bigg],\\[5pt]
			E_4 &= \frac{24}{\ell^4} + \frac{48m^2}{\left(r^2 + a^2 \cos ^2\theta\right)^6} \bigg[r^6 - 15r^2a^2\left(r^2 - a^2\cos ^2\theta\right)\cos ^2\theta  - a^6 \cos ^6\theta \bigg],\\[3pt]
			R^2 &= \frac{144}{\ell^4},\qquad
			R\bar{F}_{\mu\nu}\bar{F}^{\mu\nu} = 0,
		\end{align}
\end{subequations}	 } 
followed by their regulated integration (via holographic renormalization) relations,
{\allowdisplaybreaks
	\begin{subequations}\label{ka4} 
		\begin{align}
			\frac{1}{16\pi^2}\int \mathrm{d}^4x \sqrt{\det\bar{g}}\, W_{\mu\nu\rho\sigma}W^{\mu\nu\rho\sigma} &= 4- \frac{\beta r_+ }{\pi \left(\ell ^2-a^2\right)} \left(1+ \frac{a^2}{r_+^2} -\frac{a^2 + r_+^2}{\ell ^2}\right),\\
			\frac{1}{16\pi^2}\int \mathrm{d}^4x \sqrt{\det\bar{g}}\, E_4 &= 4, \\
			\frac{1}{16\pi^2}\int \mathrm{d}^4x \sqrt{\det\bar{g}}\, R^2 &= \frac{6 \beta r_+ }{\pi \left(\ell ^2-a^2\right)} \left(1+ \frac{a^2}{r_+^2} -\frac{a^2 + r_+^2}{\ell ^2}\right),\\
			\frac{1}{16\pi^2}\int \mathrm{d}^4x \sqrt{\det\bar{g}}\, R\bar{F}_{\mu\nu}\bar{F}^{\mu\nu} &= 0.
		\end{align}
\end{subequations}}	
All this provides the following simplified $\mathcal{C}_{\text{local}}$ relation for the non-extremal Kerr-AdS$_4$ black hole in the STU models
\begin{align}\label{ka5}
	\mathcal{C}_{\text{local}}^{\text{(Kerr-AdS)}} &= 4\left(\mathfrak{c}- \mathfrak{a}\right) -  \frac{\left(\mathfrak{c} - 6\mathfrak{b}_1\right)\beta r_+ }{\pi \left(\ell ^2-a^2\right)} \left(1+ \frac{a^2}{r_+^2} -\frac{a^2 + r_+^2}{\ell ^2}\right).
\end{align}
It can be verified that the above result matches the expression \eqref{nbh15} by managing all the Kerr-Newman-AdS$_4$ terms $W_i$, $R_i$, and $F_i$ for a vanishing $q=0$.   

\subsubsection{Schwarzschild-AdS$_4$ black hole}\label{sch}
We derive the background metric for a Schwarzschild-AdS$_4$ (SchAdS$_4$) black hole as a charged-truncation (i.e., $q\to0$) of the RNAdS$_4$ geometry \eqref{rn1}. In this context, the corresponding background metric function $f(r)$ captures a specific form,
\begin{align}\label{sch1}
	f(r) = 1 - \frac{2 m}{r} + \frac{r^2}{\ell^2}.
\end{align}
The inverse Hawking temperature parameter for the SchAdS$_4$ black hole is given by
\begin{align}\label{sch2}
	\beta^{-1} = \frac{1}{4\pi r_+}\left(1 + \frac{3r_+^2}{\ell^2}\right),
\end{align}
where $r_+$ is the horizon radius such that $f(r_+)=0$, and it relates the mass parameter $m$ via,
\begin{align}\label{sch3}
	m= \frac{r_+}{2}\left(1 + \frac{r_+^2}{\ell^2}\right).
\end{align} 
The background invariants are computed as
{\allowdisplaybreaks 
	\begin{subequations}\label{sch4}
		\begin{align}
			W_{\mu\nu\rho\sigma}W^{\mu\nu\rho\sigma} &= \frac{48m^2}{r^{6}},\\[3pt]
			E_4 &= \frac{24}{\ell^{4}} + \frac{48m^2}{r^{6}},\\[3pt]
			R^2 &= \frac{144}{\ell^4},\quad
			R\bar{F}_{\mu\nu}\bar{F}^{\mu\nu} =0.
		\end{align}
\end{subequations}	 } 
To derive the regulated values of the above integrated invariants, we proceed with the same holographic renormalization procedure detailed in \cref{kna}. The results are
{\allowdisplaybreaks
	\begin{subequations}\label{sch5} 
		\begin{align}
			\frac{1}{16\pi^2}\int \mathrm{d}^4x \sqrt{\det\bar{g}}\, W_{\mu\nu\rho\sigma}W^{\mu\nu\rho\sigma} &= 4- \frac{4 r_+^2}{\left(\ell^2+3 r_+^{2}\right)}\left(1-\frac{r_+^2}{\ell^{2}}\right),\\
			\frac{1}{16\pi^2}\int \mathrm{d}^4x \sqrt{\det\bar{g}}\, E_4 &= 4, \\
			\frac{1}{16\pi^2}\int \mathrm{d}^4x \sqrt{\det\bar{g}}\, R^2 &= \frac{24 r_+^2}{\left(\ell^2+3 r_+^{2}\right)}\left(1-\frac{r_+^2}{\ell^{2}}\right),\\[2pt]
			\frac{1}{16\pi^2}\int \mathrm{d}^4x \sqrt{\det\bar{g}}\, R\bar{F}_{\mu\nu}\bar{F}^{\mu\nu} &= 0.
		\end{align}
\end{subequations}}
Finally, we find the relation to obtain the $\mathcal{C}_{\text{local}}$ for the non-extremal SchAdS$_4$ black hole in the STU models,
\begin{align}\label{sch6}
	\mathcal{C}_{\text{local}}^{\text{(Sch-AdS)}} & =4\left(\mathfrak{c}- \mathfrak{a}\right) -  \frac{4\left(\mathfrak{c} - 6\mathfrak{b}_1\right) r_+^2}{\left(\ell^2+3 r_+^{2}\right)}\left(1-\frac{r_+^2}{\ell^{2}}\right).
\end{align}
The expression is found to match the results \eqref{kna13}, \eqref{rn6}, and \eqref{ka5} by appropriately setting $q=0$ and $a=0$.  


\subsubsection{Extremal limits and near-horizon geometry}\label{elimit}

Now, let us delve into the extremal limit of black holes, denoted by $\beta\to\infty$, wherein the Hawking temperatures vanish, implying $T_{\text{bh}} = \beta^{-1} = 0$. Applying the extremal limit to the aforementioned non-extremal backgrounds and their setups might, naively, lead to divergences. However, this challenge is effectively addressed by employing the quantum entropy function (QEF) formalism \cite{Sen:2008wa,Sen:2009wb,Sen:2009wc}, succinctly outlined in \cref{ext}. The analysis is specifically confined to the ``finite'' part of the extremal near-horizon (ENH) geometry, which includes an AdS$_2$ component. 

At this stage of our work, our objective is to derive the invariants $W_{\mu\nu\rho\sigma}W^{\mu\nu\rho\sigma}$, $E_4$, $R^2$, and $ R\bar{F}_{\mu\nu}\bar{F}^{\mu\nu}$ (as outlined in the $\mathcal{C}_{\text{local}}$ formula \eqref{lr1}), along with their regulated integration results around the Euclideanized extremal near-horizon (ENH) geometry of the Kerr-Newman-AdS$_4$, Reissner-Nordström-AdS$_4$, and Kerr-AdS$_4$ black holes. Notably, the extremal Schwarzschild background is not a valid geometry.\footnote{As an illustration, setting the Schwarzschild limit, i.e., $a=0$ and $q=0$ in the relation \eqref{foot1} yields a vanishing extremal horizon $r_0$.} The specific process is detailed as follows.     

To structure the extremal near-horizon geometry of the KNAdS$_4$ metric \eqref{kna1}, we introduce a new parameter $\lambda$ and new coordinates $\tilde{t}$, $\tilde{r}$, and $\tilde{\phi}$ as
\begin{align}\label{el1}
	\begin{gathered}
		r = r_0 + \lambda \tilde{r}, \quad  t =\frac{\ell_2^2}{\lambda} \tilde{t}, \quad \phi = \tilde{\phi } + \frac{a \left(\ell ^2-a^2\right)}{\ell^2 \left(a^2+r_0^2\right)} t,
	\end{gathered}
\end{align}
Here, $r_0$ represents the location of the extremal horizon, satisfying $\beta^{-1}(r_0)=0$, imposing the following extremality bounds on the KNAdS$_4$ black hole parameters,
\begin{align}\label{el2}
	\begin{gathered}
		a^2+ q^2  = r_0^2\left(1 + \frac{a^2}{\ell^2} + \frac{3r_0^2}{\ell^2}\right),\\
		m= r_0\left(1 + \frac{a^2}{\ell^2} + \frac{2r_0^2}{\ell^2}\right).
	\end{gathered}
\end{align}
The parameter $\ell_2$ will be later identified as the radius of AdS$_2$ space characterizing the ENH AdS$_4$ geometry of the KNAdS$_4$ black hole and can be expressed as
\begin{align}\label{el3}
	\ell_2 = \ell \sqrt{\frac{a^2 + r_0^2}{a^2 + \ell^2 + 6r_0^2}} = \ell \sqrt{\frac{2r_0^4 + \ell^2 \left(2r_0^2- q^2\right)}{\ell^4 + \ell^2\left(6r_0^2-q^2\right)-3r_0^4}}.
\end{align}
Notably, the condition $\lambda \to 0$ in transformations \eqref{el1} represents a combined limit of the near-horizon and extremality, while keeping $\tilde{r}$, $\tilde{t}$, and $\tilde{\phi}$ fixed. In this limit, along with the Euclidean continuation $\tilde{t} \to -i\tilde{\tau}$, the near-horizon coordinate transformations \eqref{el1} restructure the KNAdS$_4$ metric \eqref{kna1} into the forms,
{\allowdisplaybreaks
	\begin{subequations}\label{el4}
		\begin{align}\label{el4a}
			\tilde{g}_{\mu\nu}\mathrm{d}\tilde{x}^\mu \mathrm{d}\tilde{x}^\nu
			&= \frac{\ell_2^2\rho_0^2}{a^2+r_0^2}\left(\tilde{r}^2 \mathrm{d}\tilde{\tau}^2+\frac{\mathrm{d} \tilde{r}^2}{\tilde{r}^2}\right)+\frac{\rho_0^2}{\Delta_\theta} \mathrm{d}\theta^2 \nonumber\\
			& +\frac{\left(a^2+r_0^2\right)^2\Delta_\theta \sin ^2 \theta}{\Xi^2\rho_0^2}\left(\mathrm{d} \tilde{\phi} + \frac{2\Xi a r_0\ell_2^2 }{\left(a^2+r_0^2\right)^2} i \tilde{r} \mathrm{d}\tilde{\tau}\right)^2,
		\end{align}
		where 
		\begin{align}\label{el4b}
			\Delta_\theta = 1 -\frac{a^2}{\ell^2} \cos^2 \theta, \quad \rho_0^2 = r_0^2 + a^2 \cos^2 \theta, \quad \Xi = 1 -\frac{a^2}{\ell^2}.
		\end{align}
\end{subequations}}
Similarly, one obtains the background Maxwell field \eqref{kna3} supplying the charges to the above extremal near-horizon geometry as
\begin{align}\label{el5}
	\tilde{A}= -\frac{q}{\rho_0^2}\left[\frac{i\ell_2^2}{a^2 + r_0^2}\left(r_0^2-a^2\cos^2\theta\right)\tilde{r}\mathrm{d}\tilde{\tau} - \frac{ar_0\sin^2\theta}{\Xi}\mathrm{d}\tilde{\phi}\right].
\end{align}
From here, the Euclidean ENH metric \eqref{el4} can be structured into a wrapped version of AdS$_2$ fibered over a compact space $K$, as expressed by (for instance, refer to \cite{Sen:2008wa,Sen:2009wb,Sen:2009wc}): 
\begin{align}\label{el6}
	\tilde{g}_{\mu\nu}\mathrm{d}\tilde{x}^\mu \mathrm{d}\tilde{x}^\nu = \ell_2^2 f(y)\left[\left(\tilde{r}^2-1\right)\mathrm{d}\tilde{\tau}^2 + \frac{\mathrm{d}\tilde{r}^2}{\left(\tilde{r}^2-1\right)}\right] + \mathrm{d}s^2_K.
\end{align}
Here, $\left(\tilde{r},\thinspace \tilde{\tau}\right)$ represent coordinates describing AdS$_2$ within the range $1 \leq \tilde{r}<\infty$ and $\tilde{\tau} \equiv 0 + 2\pi$. Additionally, $f(y)$ is a function of coordinates $(\theta,\thinspace \tilde{\phi})$ along the $K$ space, collectively denoted by $y$. Importantly, the structure of $\mathrm{d}s^2_K$ involves the remaining differentials that are invariant under the $SL(2,R)$ isometry of AdS$_2$. It follows that,
\begin{align}\label{el7}
	\sqrt{\det\tilde{g}}= G(y) = \frac{\ell ^2 \ell _2^2 }{(\ell ^2-a^2)}\left(r_0^2 + a^2\cos^2\theta\right) \sin \theta,
\end{align}  
where $G(y)$ is a function that has no dependence on the AdS$_2$ coordinates.

Now, our task is to evaluate the contribution of $\mathcal{C}_{\text{local}}$ for extremal black holes by integrating the $a_4(x)$ coefficient or the related background invariants over the ENH geometry, as structured in \cref{el4,el6}. However, these integrations suffer from divergences due to the infinite volume of the AdS$_2$ piece. As per the QEF prescription discussed in \cref{ext}, we must ignore these divergences and extract only the divergent-insensitive finite part of the ENH geometry. This finite part encodes the quantum degrees of freedom for the extremal black holes and serves as the exact result for the $\mathcal{C}_{\text{local}}$ contribution. Importantly, we do not need to employ any specific regularization scheme, such as the holographic renormalization considered for the non-extremal case in \cref{kna}. As a way forward, we introduce an infrared cut-off at $\tilde{r} =\tilde{r}_c$ and expand the $\mathcal{C}_{\text{local}}$ formula in terms of the $a_4(x)$ coefficient as
\begin{align}\label{el8}
	\mathcal{C}_{\text{local}} &= \int_{1}^{\tilde{r}_c} \mathrm{d}\tilde{r} \int_{0}^{2\pi}\mathrm{d}\tilde{\tau}\int \mathrm{d}\theta \mathrm{d}\tilde{\phi} \sqrt{\det\tilde{g}}\,a_4(x) \nonumber \\ 
	&= 2\pi\left(\tilde{r}_c - 1\right)\int_{\text{ENH}} \mathrm{d}^2y \,G(y)a_4(y).
\end{align}  
The term proportional to $\tilde{r}_c$ is interpreted as an infinite shift in the ground state energy in the extremal limit \cite{Sen:2008wa,Sen:2009wb,Sen:2009wc}. On the other hand, the cut-off independent finite piece is the true and unambiguous contribution to the one-loop correction for the quantum horizon degeneracy of extremal black holes. Thus, we need to discard the divergent part and identify the finite term, involving a $-2\pi$ prefactor, as the desired $\mathcal{C}_{\text{local}}$ contribution for extremal black holes. This yields the following revised formula:
\begin{align}\label{el9}
	\mathcal{C}_{\text{local}} = \frac{\left(-2\pi\right)}{\left(4\pi\right)^2}\int_{\text{ENH}} \mathrm{d}^2y \,G(y)\left[\mathfrak{c} W_{\mu\nu\rho\sigma}W^{\mu\nu\rho\sigma} - \mathfrak{a} E_4 + \mathfrak{b}_1 R^2 + \mathfrak{b}_2R \bar{F}_{\mu\nu}\bar{F}^{\mu\nu}\right],
\end{align}
where we have utilized the trace anomaly relation \eqref{sdc32} for $a_4$ and the integration range is $0 \leq \theta \leq \pi$ and $0 \leq \tilde{\phi} \leq 2\pi$ for the $y$ coordinates on the ENH geometry \eqref{el4}. The final form of the background invariants, as well as their integrations over specific parts of the ENH geometry, are derived and separately listed in \cref{enhii}. This results in the following expression for $\mathcal{C}_{\text{local}}$ for the extremal Kerr-Newman-AdS$_4$ black hole in STU models:
{\allowdisplaybreaks
	\begin{align}\label{el10}
		\mathcal{C}_{\text{local}}^{\text{(ext,KN-AdS)}} &= -4\mathfrak{a} - \frac{\ell_2^2}{2 a\Xi \ell^4 r_0^5\left(a^2+r_0^2\right)^2}\bigg[3 a^7 r_0^5\left((\mathfrak{c} + 16 \mathfrak{b}_1) - \frac{2\mathfrak{c}\ell^2}{r_0^2} + \frac{\mathfrak{c}\ell^4}{r_0^4}\right) \nonumber \\
		& + a^5r_0^7 \left(\left(39\mathfrak{c} + 240 \mathfrak{b}_1 - 24 \mathfrak{b}_2\right) - \left(22\mathfrak{c}-24\mathfrak{b}_2\right)\frac{\ell^2}{r_0^2} - \frac{\mathfrak{c}\ell^4}{r_0^4}\right) \nonumber \\
		& + a^3r_0^9 \left(\left(49\mathfrak{c} + 336 \mathfrak{b}_1 - 48 \mathfrak{b}_2\right) - \left(50\mathfrak{c}+ 48\mathfrak{b}_2\right)\frac{\ell^2}{r_0^2} - \frac{15\mathfrak{c}\ell^4}{r_0^4}\right) \nonumber \\
		& + 3ar_0^{11} \left(\left(7\mathfrak{c} + 48 \mathfrak{b}_1 + 24 \mathfrak{b}_2\right) - \left(6\mathfrak{c} - 8\mathfrak{b}_2\right)\frac{\ell^2}{r_0^2} - \frac{\mathfrak{c}\ell^4}{r_0^4}\right) \nonumber \\
		& + 3 \mathfrak{c}r_0^{12} \left(1 + \frac{a^2}{r_0^2}\right)^2 \left( 3 + \frac{\ell^2 + a^2}{r_0^2} - \frac{a^2\ell^2}{r_0^4}\right)^2 \arctan\left(\frac{a}{r_0}\right) \bigg].
\end{align}}
Next, we compute the same integrated invariants on the near-horizon background of extremal Kerr-AdS$_4$ and Reissner-Nordström-AdS$_4$ black holes in STU models (refer to \cref{enhii}). The corresponding regulated and finite contributions to $\mathcal{C}_{\text{local}}$ are presented as follows
{\allowdisplaybreaks
	\begin{align}
		\mathcal{C}_{\text{local}}^{\text{(ext,Kerr-AdS)}} &= 4\left(\mathfrak{c}- \mathfrak{a}\right) -  \frac{2\left(\mathfrak{c} - 6\mathfrak{b}_1\right)\ell_2^2 }{\left(\ell ^2-a^2\right)}\left(1 - \frac{a^2 + 3 r_0^2}{\ell^2} \right), \label{el11}\\[8pt]
		\mathcal{C}_{\text{local}}^{\text{(ext,RN-AdS)}} &= \left(\frac{4}{3}\mathfrak{c}- 4\mathfrak{a} + 4\mathfrak{b}_1\right)- \left(\frac{2}{3}\mathfrak{c} + 2\mathfrak{b}_1 - \mathfrak{b}_2\right)\frac{\ell_2^2}{r_0^2} - \left(\frac{2}{3}\mathfrak{c} + 2\mathfrak{b}_1 + \mathfrak{b}_2\right)\frac{r_0^2}{\ell_2^2}.\label{el12}
\end{align}}
Note that with appropriate limits, particularly $q = 0$ and $a=0$, one can successively reproduce the above formulas as consistent truncations of the Kerr-Newman-AdS$_4$ relation \eqref{el10}.

To ensure the consistency of our setup, we have also taken an alternative approach to obtain the $\mathcal{C}_{\text{local}}$ contributions by directly considering the extremal limit, $\beta \to \infty$, on the finite temperature backgrounds discussed in \cref{kna,rn,ka}. This method enables us to circumvent naive divergences associated with taking the $\beta \to \infty$ limit. In the initial steps, we treat the outer horizon $r_+$ as an explicit function of the inverse temperature $\beta$, while keeping all other parameters and charges fixed. By employing a low-temperature expansion, we obtain
\begin{align}\label{el13}
	r_+ = r_0 + \frac{2 \pi \ell_2^2}{\beta} + \mathcal{O}\left(\beta^{-2}\right).
\end{align} 
Here, $\ell_2$ represents the radius of AdS$_2$ space that naturally emerges in the structure of extremal black holes. On the other hand, the extremal horizon $r_0$ is interpreted as a finite component within the non-extremal horizon radius $r_+$ as $\beta \to \infty$. For a generic extremal Kerr-Newman-AdS$_4$ black hole, $\ell_2$ is given in \cref{el3}, while the typical form of $r_0$ is expressed as
\begin{align}\label{foot1}
	r_0^2 = \frac{\ell^2}{6}\left(\sqrt{1 + \frac{2}{\ell^2}\left(6q^2 + 7a^2\right)}- \frac{a^2}{\ell^2}-1\right).
\end{align}
Continuing with the expansion \eqref{el13}, we can expand the $\mathcal{C}_{\text{local}}$ contribution as
\begin{align}\label{el14}
	\lim_{\beta \to \infty} \mathcal{C}_{\text{local}} = \mathcal{C}_1\beta + \mathcal{C}_0 + \mathcal{O}\left(\beta^{-1}\right).
\end{align}
In the extremal limit as $\beta \to \infty$, the terms with inverse powers of $\beta$ vanish, while the first term, linear in $\beta$, diverges. This divergence can be interpreted as an infinite shift in the ground state energy due to one-loop fluctuations. Therefore, we can safely disregard this linear and divergent term and consider the finite constant term $\mathcal{C}_0$ as an unambiguous and effective value for the $\mathcal{C}_{\text{local}}$ contribution of extremal black holes. The results obtained by following this procedure align precisely with those derived in \cref{el10,el11,el12} via the QEF formalism.

\subsubsection{Asymptotically-flat limits}\label{flimit}

We will now explore the $\mathcal{C}_{\text{local}}$ contributions for the asymptotically flat counterpart of the previously discussed non-extremal and extremal AdS$_4$ black holes embedded in STU supergravity models with a vanishing cosmological constant, i.e., $\Lambda = 0$. The flat-space limit is achieved by setting $\ell \to \infty$ in the relations given in \cref{nbh15,rn6,ka5,sch6,el10,el11,el12}. In this scenario, the outer-horizon and inverse Hawking-temperature for the non-extremal Kerr-Newman black hole are given by,
\begin{align}\label{fl1}
	r_+ = m + \sqrt{m^2 - q^2 - a^2}, \quad \beta = \frac{4\pi r_+ \left(r_+^2 + a^2\right)}{\left(r_+^2 - q^2 - a^2\right)},
\end{align}
while $r_0$ and $\ell_2$, as well as the revised Kerr parameter $b=\frac{a}{q}$ controlling the related extremality, are constrained as
\begin{align}\label{fl2}
	r_0 = \sqrt{q^2 + a^2}, \quad \ell_2 =  \sqrt{q^2 + 2a^2}, \quad b = \frac{r_0}{q}\sqrt{\frac{\ell_2^2}{r_0^2} - 1}.
\end{align}
Following that, the $\mathcal{C}_{\text{local}}$ formulas for the non-extremal and extremal Kerr-Newman black holes embedded in STU models are expressed as
{\allowdisplaybreaks
	\begin{align}\label{fl3}
		\mathcal{C}_{\text{local}}^{\text{(KN)}} &= 4\left(\mathfrak{c}- \mathfrak{a}\right) + \frac{\mathfrak{c} \beta q^4}{16\pi}\bigg[ \frac{\left(3 a^4 + 2a^2 r_+^2 + 3 r_+^4\right)}{a^4 r_+^3(a^2+r_+^2)}  + \frac{3(a^2-r_+^2)(a^2 + r_+^2)}{a^5 r_+^4} \arctan \left(\frac{a}{r_+}\right)\bigg], \nonumber\\[7pt]	
		\mathcal{C}_{\text{local}}^{\text{(ext,KN)}} &= -4\mathfrak{a} + \frac{\mathfrak{c}}{2}\bigg[\frac{\left(3 + 24b^2 + 40b^4 + 16b^6 \right)}{(b^2+1)^2(2b^2+1)} -\frac{3(2b^2+1)}{b{\left(b^2 + 1\right)^{5/2}}}\arctan\left(\frac{b}{\sqrt{b^2 + 1}}\right)\bigg].
\end{align}	}
Similarly, we successively derive the local contributions of non-extremal and extremal Schwarzschild ($q=0$, $a=0$), Reissner-Nordström ($a=0$, $\ell_2 = r_0$, and $b = 0$) and Kerr ($q=0$, $\ell_2 = \sqrt{2}r_0$, and $b \to \infty$) black holes as follows
{\allowdisplaybreaks
	\begin{align}
		\mathcal{C}_{\text{local}}^{\text{(Sch)}} &= 4\left(\mathfrak{c}- \mathfrak{a}\right), \label{fl4}\\[3pt]
		\mathcal{C}_{\text{local}}^{\text{(Kerr)}} &= 4\left(\mathfrak{c}- \mathfrak{a}\right),\quad \mathcal{C}_{\text{local}}^{\text{(ext,Kerr)}} = 4\left(\mathfrak{c}- \mathfrak{a}\right),\label{fl5}\\[3pt]
		\mathcal{C}_{\text{local}}^{\text{(RN)}} &= 4\left(\mathfrak{c}- \mathfrak{a}\right) + \frac{2\mathfrak{c} \beta q^4}{5\pi r_+^5},\quad \mathcal{C}_{\text{local}}^{\text{(ext,RN-AdS)}} = -4\mathfrak{a}.\label{fl6}
\end{align}}
We have verified that the above $\mathcal{C}_{\text{local}}$ expressions are consistent with the results obtained in \cite{Bhattacharyya:2012ss,Sen:2013ns, Charles:2015nn,Karan:2019sk,Banerjee:2020wbr,Karan:2020sk,Karan:2021teq} for asymptotically-flat$_4$ black holes embedded in Einstein-Maxwell (super)gravity theories.

\subsection{Results}\label{res2}
This section provides a comprehensive overview of the final outcomes of our research, specifically focusing on the logarithmic corrections to black hole entropy within two distinct scenarios of STU supergravity intersecting with $U(1)^2$-charged EMD theories in four-dimensional spacetime (as detailed in \cref{model}). Our analysis is grounded in the central formula outlined in \cref{comp1}. The local component of the logarithmic corrections is derived from the corresponding $\mathcal{C}_{\text{local}}$ formulas, as detailed in \cref{local}, while considering trace anomaly data from \eqref{sdc32} for the two STU truncations: $(\kappa_1, \kappa_2) = (1, -1)$ and $(\sqrt{3}, -1/\sqrt{3})$. In contrast, the zero-mode contributions for various black hole backgrounds are computed using the global relation \eqref{comp1c}. For the specific black holes considered in this study, we refer to the $\mathcal{C}_{\text{zm}}$ data listed in \cref{czero}. We will further elaborate on the typical nature and significance of these logarithmic correction results in \cref{discuss}.

\subsubsection{Quantum corrected \boldmath AdS$_4$ black holes}\label{ra}

\subsubsection*{Case I: \boldmath $(\kappa_1, \kappa_2) = (1, -1)$}
For the STU supergravity model truncated into a $U(1)^2$-charged EMD-AdS theory with dilaton couplings $\kappa_1 = 1, \kappa_2 = -1$, the logarithmic entropy correction results for Kerr-Newman-AdS$_4$, Reissner-Nordström-AdS$_4$, Kerr-AdS$_4$, and Schwarzschild-AdS$_4$ black holes are calculated as
{\allowdisplaybreaks
	\begin{align}
		\Delta S_{\text{BH}}^{\text{(Sch-AdS)}} & = \bigg[\frac{37}{45} -  \frac{343 r_+^2}{20\left(\ell^2+3 r_+^{2}\right)}\left(1-\frac{r_+^2}{\ell^{2}}\right)\bigg]\ln \mathcal{A}_{H},\label{raI1}\\[6pt]
		\Delta S_{\text{BH}}^{\text{(Kerr-AdS)}} &= \bigg[\frac{82}{45} -  \frac{343\beta r_+ }{80\pi \left(\ell ^2-a^2\right)} \left(1+ \frac{a^2}{r_+^2} -\frac{a^2 + r_+^2}{\ell ^2}\right)\bigg]\ln \mathcal{A}_{H},\label{raI2}\\[6pt]
		\Delta S_{\text{BH}}^{\text{(RN-AdS)}} &= \bigg[-\frac{562}{225} +\frac{719}{100}\frac{r_{+}^{2}}{\ell^{2}} + \frac{166\pi r_+}{25\beta}  - \frac{\beta r_+^3}{25\pi\ell ^4} \left( 121  + \frac{1217\ell^2}{8r_+^2}  - \frac{83\ell^4}{8 r_+^4}\right) \bigg]\ln \mathcal{A}_{H},\label{raI3}\\
		\Delta S_{\text{BH}}^{\text{(KN-AdS)}} &= \bigg[-\frac{419}{180} + \frac{ a^4}{32 \Xi  r_+^2 \left(r_+^2 + a^2\right)} \bigg\lbrace  \frac{249}{5} + \frac{1831 r_+^2}{5\ell ^2} + \frac{2r_+^2}{a^2} \left({83} + \frac{1831r_+^2 }{5\ell ^2} \right) \nonumber\\
		&  + \frac{8 r_+^4}{5a^4} \left(83 + \frac{271r_+^2}{2\ell ^2} \right) - \frac{166r_+^6}{5a^6} \left(1 + \frac{9  r_+^2}{\ell ^2}\right) + \frac{\beta}{40\pi r_+} \bigg({249}  \left(1- \frac{r_+^2}{\ell^2}\right)^2 \nonumber\\
		& - \frac{16 r_+^2}{a^2} \left(\frac{83}{4} + \frac{769r_+^2}{\ell ^2} + \frac{2329r_+^4}{4\ell ^4}  \right) + \frac{20 r_+^4}{a^4} \left(\frac{83}{10} - \frac{2661r_+^2}{5\ell ^2} - \frac{749r_+^4}{2\ell ^4}  \right) \nonumber\\
		& - \frac{332 r_+^6}{a^6} \left(1 - \frac{9r_+^4}{\ell ^4} \right)\bigg) + \frac{166\pi r_+}{5\beta}\left(3 + \frac{8r_+^2}{a^2} + \frac{10r_+^4}{a^4} + \frac{8 r_+^6}{a^6}\right) \nonumber \\
		& + \frac{249\beta r_+^7}{40\pi a^8}\bigg( 1 - \frac{4\pi r_+}{\beta } + \frac{3r_+^2}{\ell^2}\bigg)^2 + \frac{249\beta r_+^2}{40\pi a^3 } \left(1- \frac{r_+^2}{a^2}\right)\left(1 + \frac{r_+^2}{a^2}\right)^2  \nonumber\\
		&  \bigg(1 - \frac{a^2}{r_+^2} + \frac{\left(a^2 + 3r_+^2\right)}{\ell^2} - \frac{4\pi r_+}{\beta}\left(1 + \frac{a^2}{r_+^2}\right) \bigg)^2 \arctan\left(\frac{a}{r_+}\right) \bigg\rbrace \bigg]\ln \mathcal{A}_{H}. \label{raI4}
\end{align}}
Similarly, in the extremal limit and near-horizon analysis (see \cref{elimit}) for the above backgrounds, we obtain
{\allowdisplaybreaks
	\begin{align}
		\Delta S_{\text{BH}}^{\text{(ext,Kerr-AdS)}} &=  \bigg[\frac{29}{90} -  \frac{343\ell_2^2 }{40\left(\ell ^2-a^2\right)}\left(1 - \frac{a^2 + 3 r_0^2}{\ell^2} \right)\bigg]\ln \mathcal{A}_{H}, \label{raI5}\\[6pt]
		\Delta S_{\text{BH}}^{\text{(ext,RN-AdS)}} &= \bigg[-\frac{101}{18}+ \frac{47}{120}\left(\frac{\ell_2^2}{r_0^2} + \frac{r_0^2}{\ell_2^2}\right)\bigg]\ln \mathcal{A}_{H}, \label{raI6}\\[6pt]
		\Delta S_{\text{BH}}^{\text{(ext,KN-AdS)}} &= \bigg[-\frac{689}{180} + \frac{\ell_2^2}{80 \Xi a\ell^4 r_0^5\left(r_0^2 + a^2\right)^2}\bigg\lbrace {3 a^7 r_0^5}\left(\frac{1831}{6} + \frac{83\ell^2}{r_0^2} - \frac{83\ell^4}{2r_0^4}\right) \nonumber \\
		& + {a^5r_0^7} \left(\frac{7163}{2} + \frac{913\ell^2}{r_0^2} + \frac{83\ell^4}{2r_0^4}\right) + {5a^3r_0^9} \left(\frac{10493}{10} + \frac{415\ell^2}{r_0^2} + \frac{249\ell^4}{2r_0^4}\right) \nonumber \\
		& + {3ar_0^{11}} \left(\frac{1499}{2} + \frac{249\ell^2}{r_0^2} + \frac{83\ell^4}{2r_0^4}\right) - \frac{249r_0^{12}}{2} \left(1 + \frac{a^2}{r_0^2}\right)^2 \nonumber \\
		&  \left( 3 + \frac{\ell^2 + a^2}{r_0^2} - \frac{a^2\ell^2}{r_0^4}\right)^2 \arctan\left(\frac{a}{r_0}\right) \bigg\rbrace\bigg]\ln \mathcal{A}_{H}. \label{raI7}
\end{align}}

\subsubsection*{Case II: \boldmath $(\kappa_1, \kappa_2) = (\sqrt{3}, -1/\sqrt{3})$}
For the STU supergravity model intersecting with a $U(1)^2$-charged EMD-AdS theory, with dilaton couplings $\kappa_1 = \sqrt{3}$ and $\kappa_2 = -1/\sqrt{3}$, the logarithmic correction to the entropy of Kerr-Newman-AdS$_4$, Reissner-Nordström-AdS$_4$, Kerr-AdS$_4$, and Schwarzschild-AdS$_4$ black holes are computed as
{\allowdisplaybreaks
	\begin{align}
		\Delta S_{\text{BH}}^{\text{(Sch-AdS)}} & = \bigg[\frac{37}{45} -  \frac{343 r_+^2}{20\left(\ell^2+3 r_+^{2}\right)}\left(1-\frac{r_+^2}{\ell^{2}}\right)\bigg]\ln \mathcal{A}_{H},\label{raII1}\\[6pt]
		\Delta S_{\text{BH}}^{\text{(Kerr-AdS)}} &= \bigg[\frac{82}{45} -  \frac{343\beta r_+ }{80\pi \left(\ell ^2-a^2\right)} \left(1+ \frac{a^2}{r_+^2} -\frac{a^2 + r_+^2}{\ell ^2}\right)\bigg]\ln \mathcal{A}_{H},\label{raII2}\\[6pt]
		\Delta S_{\text{BH}}^{\text{(RN-AdS)}} &= \bigg[-\frac{586}{225} + \frac{607}{100}\frac{r_{+}^{2}}{\ell^{2}} + \frac{514\pi r_+}{75\beta}  - \frac{\beta r_+^3}{25\pi\ell ^4} \left( 103  + \frac{1161\ell^2}{8r_+^2}  - \frac{257\ell^4}{24 r_+^4}\right) \bigg]\ln \mathcal{A}_{H},\label{raII3}\\
		\Delta S_{\text{BH}}^{\text{(KN-AdS)}} &= \bigg[-\frac{443}{180} + \frac{ a^4}{32 \Xi  r_+^2 \left(r_+^2 + a^2\right)} \bigg\lbrace  \frac{257}{5} + \frac{1081 r_+^2}{3\ell ^2} + \frac{2r_+^2}{3a^2} \left({257} + \frac{5213r_+^2 }{5\ell ^2} \right) \nonumber\\
		&  + \frac{8 r_+^4}{15a^4} \left({257} +  \frac{677r_+^2}{2\ell ^2} \right) - \frac{514 r_+^6}{15a^6} \left(1 + \frac{9  r_+^2}{\ell ^2}\right) + \frac{\beta}{40\pi r_+} \bigg({257}  \left(1- \frac{r_+^2}{\ell^2}\right)^2 \nonumber\\
		& - \frac{80 r_+^2}{3a^2} \left(\frac{257}{20} + \frac{2363r_+^2}{5\ell ^2} + \frac{1351r_+^4}{4\ell ^4}  \right) + \frac{20r_+^4}{a^4} \left(\frac{257}{30} -  \frac{7783r_+^2}{15\ell ^2} - \frac{661r_+^4}{2\ell ^4}  \right) \nonumber\\
		& - \frac{1028 r_+^6}{3a^6} \left(1 - \frac{9r_+^4}{\ell ^4} \right)\bigg) + \frac{514\pi r_+}{15\beta}\left(3 + \frac{8r_+^2}{a^2} + \frac{10r_+^4}{a^4} + \frac{8 r_+^6}{a^6}\right) \nonumber \\
		& + \frac{257\beta r_+^7}{40\pi a^8}\bigg( 1 - \frac{4\pi r_+}{\beta } + \frac{3r_+^2}{\ell^2}\bigg)^2 + \frac{257\beta r_+^2}{40\pi a^3 } \left(1- \frac{r_+^2}{a^2}\right)\left(1 + \frac{r_+^2}{a^2}\right)^2 \nonumber\\
		&  \bigg(1 - \frac{a^2}{r_+^2}  + \frac{\left(a^2 + 3r_+^2\right)}{\ell^2} - \frac{4\pi r_+}{\beta}\left(1 + \frac{a^2}{r_+^2}\right) \bigg)^2 \arctan\left(\frac{a}{r_+}\right) \bigg\rbrace \bigg]\ln \mathcal{A}_{H}.\label{raII4}
\end{align}}
On the other hand, considering the near-horizon limit in the extremal case for the above backgrounds yields
{\allowdisplaybreaks
	\begin{align}
		\Delta S_{\text{BH}}^{\text{(ext,Kerr-AdS)}} &=  \bigg[\frac{29}{90} -  \frac{343\ell_2^2 }{40\left(\ell ^2-a^2\right)}\left(1 - \frac{a^2 + 3 r_0^2}{\ell^2} \right)\bigg]\ln \mathcal{A}_{H}, \label{raII5}\\[6pt]
		\Delta S_{\text{BH}}^{\text{(ext,RN-AdS)}} &= \bigg[-\frac{511}{90} + \frac{47}{120}\frac{\ell_2^2}{r_0^2} + \frac{13}{40}\frac{r_0^2}{\ell_2^2}\bigg]\ln \mathcal{A}_{H}, \label{raII6}\\[6pt]
		\Delta S_{\text{BH}}^{\text{(ext,KN-AdS)}} &= \bigg[-\frac{713}{180} + \frac{\ell_2^2}{16 \Xi a\ell^4 r_0^5\left(r_0^2 + a^2\right)^2}\bigg\lbrace {a^7 r_0^5}\left(\frac{1081}{6} + \frac{257\ell^2}{5r_0^2} - \frac{257\ell^4}{10r_0^4}\right) \nonumber \\
		& + \frac{a^5r_0^7}{15} \left(\frac{21049}{2} + \frac{2731\ell^2}{r_0^2} + \frac{257\ell^4}{2r_0^4}\right) + {a^3r_0^9} \left(\frac{10341}{10} + \frac{6617\ell^2}{15r_0^2} + \frac{257\ell^4}{2r_0^4}\right) \nonumber \\
		& + {3ar_0^{11}} \left(\frac{279}{2} + \frac{739\ell^2}{15r_0^2} + \frac{257\ell^4}{30r_0^4}\right)  - \frac{257r_0^{12}}{10} \left(1 + \frac{a^2}{r_0^2}\right)^2\nonumber \\
		& \left( 3 + \frac{\ell^2 + a^2}{r_0^2} - \frac{a^2\ell^2}{r_0^4}\right)^2 \arctan\left(\frac{a}{r_0}\right) \bigg\rbrace\bigg]\ln \mathcal{A}_{H}.\label{raII7}
\end{align}} 
All the results presented above are entirely non-topological as they are intricate functions of the respective black hole parameters and dimensionless ratios, denoted as $\left\lbrace a, r_+, r_0, \ell, \ell_2, \beta \right\rbrace$. These parameters are distinct for each background, representing direct functions of the black hole mass, charge, and angular momentum (as detailed in \cref{local}). Notably, the formulas derived for uncharged Kerr-AdS$_4$ and Schwarzschild-AdS$_4$ black holes in both extremal and non-extremal limits are dilaton coupling independent and identical. These quantum entropy corrections are novel and represent a significant achievement in this work.

\subsubsection{Quantum corrected \boldmath flat$_4$ black holes}\label{rf}

\subsubsection*{Case I: \boldmath $(\kappa_1, \kappa_2) = (1, -1)$}
For the STU supergravity model truncated into a $U(1)^2$-charged EMD theory with dilaton couplings $\kappa_1 = 1, \kappa_2 = -1$, the logarithmic entropy correction results for asymptotically-flat Kerr-Newman, Reissner-Nordström, Kerr and Schwarzschild black holes are calculated as  
{\allowdisplaybreaks
	\begin{align}
		\Delta S_{\text{BH}}^{\text{(Sch)}} &=  \frac{37}{45}\ln \mathcal{A}_{H}, \label{rfI1}\\[4pt]
		\Delta S_{\text{BH}}^{\text{(Kerr)}} &= \frac{82}{45}\ln \mathcal{A}_{H},\quad
		\Delta S_{\text{BH}}^{\text{(ext,Kerr)}} = \frac{29}{90}\ln \mathcal{A}_{H},\label{rfI2}\\[6pt]
		\Delta S_{\text{BH}}^{\text{(RN)}} &= \left[ \frac{37}{45} + \frac{83 \beta q^4}{200\pi r_+^5}\right]\ln \mathcal{A}_{H},\quad
		\Delta S_{\text{BH}}^{\text{(ext,RN)}} = -\frac{869}{180}\ln \mathcal{A}_{H},\label{rfI3}\\[6pt]
		\Delta S_{\text{BH}}^{\text{(KN)}} &= \bigg[\frac{82}{45} + \frac{83 \beta q^4}{1280\pi}\bigg\lbrace \frac{\left(3 a^4 + 2a^2 r_+^2 + 3 r_+^4\right)}{a^4 r_+^3(a^2+r_+^2)} \nonumber \\
		&\qquad + \frac{3(a^2-r_+^2)(a^2 + r_+^2)}{a^5 r_+^4} \arctan \left(\frac{a}{r_+}\right)\bigg\rbrace\bigg]\ln \mathcal{A}_{H}, \label{rfI4}\\[6pt]
		\Delta S_{\text{BH}}^{\text{(ext,KN)}} &= \bigg[-\frac{689}{180} + \frac{83}{160}\bigg\lbrace\frac{\left(3 + 24b^2 + 40b^4 + 16b^6 \right)}{(b^2+1)^2(2b^2+1)} \nonumber \\
		&\qquad -\frac{3(2b^2+1)}{b{\left(b^2 + 1\right)^{5/2}}}\arctan\left(\frac{b}{\sqrt{b^2 + 1}}\right)\bigg\rbrace\bigg]\ln \mathcal{A}_{H}.\label{rfI5}
\end{align}}

\subsubsection*{Case II: \boldmath $(\kappa_1, \kappa_2) = (\sqrt{3}, -1/\sqrt{3})$}
For the STU supergravity model intersecting with a $U(1)^2$-charged EMD theory, with dilaton couplings $\kappa_1 = \sqrt{3}$ and $\kappa_2 = -1/\sqrt{3}$, the logarithmic correction to the entropy of Kerr-Newman, Reissner-Nordström, Kerr, and Schwarzschild black holes are computed as 
{\allowdisplaybreaks
	\begin{align}
		\Delta S_{\text{BH}}^{\text{(Sch)}} &=  \frac{37}{45}\ln \mathcal{A}_{H}, \label{rfII1}\\[4pt]
		\Delta S_{\text{BH}}^{\text{(Kerr)}} &= \frac{82}{45}\ln \mathcal{A}_{H},\quad
		\Delta S_{\text{BH}}^{\text{(ext,Kerr)}} = \frac{29}{90}\ln \mathcal{A}_{H},\label{rfII2}\\[6pt]
		\Delta S_{\text{BH}}^{\text{(RN)}} &= \left[ \frac{37}{45} + \frac{257 \beta q^4}{600\pi r_+^5}\right]\ln \mathcal{A}_{H},\quad 
		\Delta S_{\text{BH}}^{\text{(ext,RN)}} = -\frac{893}{180}\ln \mathcal{A}_{H},\label{rfII3}\\[6pt]
		\Delta S_{\text{BH}}^{\text{(KN)}} &= \bigg[\frac{82}{45} + \frac{257 \beta q^4}{3840\pi}\bigg\lbrace \frac{\left(3 a^4 + 2a^2 r_+^2 + 3 r_+^4\right)}{a^4 r_+^3(a^2+r_+^2)}  \nonumber \\
		&\qquad + \frac{3(a^2-r_+^2)(a^2 + r_+^2)}{a^5 r_+^4} \arctan \left(\frac{a}{r_+}\right)\bigg\rbrace\bigg]\ln \mathcal{A}_{H}, \label{rfII4}\\[6pt]
		\Delta S_{\text{BH}}^{\text{(ext,KN)}} &= \bigg[-\frac{713}{180} + \frac{257}{480}\bigg\lbrace\frac{\left(3 + 24b^2 + 40b^4 + 16b^6 \right)}{(b^2+1)^2(2b^2+1)} \nonumber \\
		&\qquad -\frac{3(2b^2+1)}{b{\left(b^2 + 1\right)^{5/2}}}\arctan\left(\frac{b}{\sqrt{b^2 + 1}}\right)\bigg\rbrace\bigg]\ln \mathcal{A}_{H}.\label{rfII5}
\end{align}}
In contrast to the AdS$_4$ results, the logarithmic correction formulas obtained for the flat$_4$ backgrounds are less complicated. Interestingly, their non-topological nature, which was unavoidable for the AdS$_4$ cases, exhibits a specific pattern for the asymptotically-flat backgrounds. For charged backgrounds, the results remain non-topological but are considerably more streamlined in terms of background parameters and dimensionless ratios, as illustrated in \cref{flimit}. However, in the extremal limit, the Reissner-Nordström formulas become expressions in terms of numerical ratios, completely independent of geometric parameters. Similarly, in the limit of vanishing charge ($q = 0, b \to \infty$), the logarithmic correction relations for Kerr and Schwarzschild black holes become entirely topological, both in the extremal and non-extremal limits. Readers familiar with the literature may recognize this characteristic pattern, as observed in previous works such as \cite{Karan:2021teq, Bhattacharyya:2012ss, Sen:2013ns}. Additionally, the Kerr-Newman results are always found to be transcendental instead of rational. Further discussion of the non-topological versus topological logarithmic correction results as well as their implications are presented in the next section.


\section{Summary and discussions}\label{discuss}

In this paper, we computed the logarithmic corrections to the Bekenstein-Hawking entropy of black holes in 4D STU supergravity \cite{Cremmer:1984hj,Duff:1995sm}. Specifically, this study considers the scenario where STU models are viewed as a $U(1)^4$-charged EMD theory \cite{Cvetic:1999xp,Clement:2013fc,Chow:2014cca,Cvetic:2014vsa,Cvetic:2021lss,Anabalon:2022aig} and further truncated into two specific $U(1)^2$-charged EMD systems \cite{Clement:2013fc,Chow:2014cca,Cvetic:2014vsa,Lu:2013eoa}. These systems are characterized by one dilaton coupled non-minimally to two Maxwell fields, with the specific coupling constant values $(\kappa_1, \kappa_2) = (1, -1)$ and $(\sqrt{3}, -1/\sqrt{3})$ in the action \eqref{mod2}. Subsequently, we demonstrated how setting appropriate constraints on the two $U(1)$ Maxwell backgrounds/charges, i.e., ${Q_2} = {\kappa_1}{Q_1} = -\frac{{Q_1}}{{\kappa_2}}$, uplifts the entire Kerr-Newman-AdS family of black holes and their flat counterparts, embedding them within the two cases of truncated STU systems. This technically defines our choice of EM embedding, where the Einstein-Maxwell backgrounds (both cosmological and flat) non-trivially satisfy the $U(1)^2$-charged STU equations of motion for a null dilaton background. In the future, we aspire to surmount the obstacles associated with going beyond the charge-constrained or EM-embedding limit and explore the quantum black hole entropy within STU models characterized by a non-vanishing dilaton background.

Notably, one should not be concerned about the absence of any fermionic contribution in the main results presented in \cref{res1,res2}. The current STU models are essentially a bosonic truncation of $\mathcal{N}=8$ supergravity, which includes multiple gauge and scalar moduli fields \cite{Cremmer:1978ds,Cremmer:1979up}. As discussed, these models have undergone further truncations into $U(1)^4$ and $U(1)^2$-charged EMD systems \eqref{mod1} and \eqref{mod2}, forming the central or exact bosonic sector of all $\mathcal{N} \geq 2$ supergravity theories and providing them with general black hole solutions \cite{Cvetic:1996zq}. In essence, STU models are of particular interest as they serve as universal bosonic building blocks within supergravity/low-energy string vacua. However, one can utilize the current STU results at any point and explore the full supergravity theories (minimal as well as matter-coupled), where the $a_4(x)$ and logarithmic correction relations receive contributions from additional one-loop fermionic and bosonic fluctuations in supergravity, vector, and other matter multiplets. When considering the embedding of the same black holes, it is essential to remember that the additional fermionic and bosonic content must have a vanishing background but fluctuate around the common EM backgrounds (for details, see \cite{Banerjee:2011pp,Sen:2012qq,Charles:2015nn,Karan:2019sk,Banerjee:2020wbr,Karan:2020sk,David:2021eoq}).

The logarithmic correction relations for the embedded AdS$_4$ and flat$_4$ black holes, as detailed in \cref{raI1,raI2,raI3,raI4,raI5,raI6,raI7,rfI1,rfI2,rfI3,rfI4,rfI5,raII1,raII2,raII3,raII4,raII5,raII6,raII7,rfII1,rfII2,rfII3,rfII4,rfII5}, are computed using Euclidean quantum gravity setups \cite{Sen:2008wa,Sen:2009wb,Sen:2009wc,Sen:2013ns} as elaborated in \cref{setup}, followed by the heat kernel method-based Seeley-DeWitt expansion \cite{Vassilevich:2003ll}. In the local part, the four dimensional $\mathcal{C}_{\text{local}}$ contributions depend only on the $a_4(x)$ coefficient. The relevant data \eqref{sdc30} is obtained by fluctuating the STU content around the embedded EM backgrounds and is re-expressed in terms of appropriate geometric invariants and conformal anomalies in \cref{sdc32}. The related central charges and anomaly coefficients are determined in terms of generic formulas \eqref{sdc32b} involving dilaton couplings $\kappa_1$ or $\kappa_2$. In special limiting cases, the obtained heat kernel data aligns perfectly with previously available results for the Einstein-Maxwell theory \cite{Bhattacharyya:2012ss,Karan:2021teq,Sen:2013ns}, or the bosonic sector of $\mathcal{N}=2$ supergravity ($\kappa_1=\kappa_2=0$) \cite{Charles:2015nn,Karan:2019sk,Karan:2020sk,David:2021eoq,Sen:2012qq,Keeler:2014nn}, the Kaluza-Klein system ($\kappa_1=\sqrt{3}, \kappa_2=0, \Lambda =0$) \cite{Castro:2018tg}, and Einstein-Maxwell-dilaton models ($\kappa_1 = 1, \sqrt{3}, 1/\sqrt{3}$ and $\kappa_2=0$) \cite{Karan:2022dfy}. Furthermore, we independently verified the computed $a_4(x)$ relation through manual calculation and by developing \textit{Mathematica} algorithms using xAct \cite{Garcia:2002wp} and xPert \cite{Brizuela:2008ra}. These validations enhance our confidence in the consistency of $a_4(x)$ derivations and logarithmic entropy correction results achieved in this work. 

All the results presented in \cref{ra,rf} are obtained by integrating the Weyl anomaly $W_{\mu\nu\rho\sigma}W^{\mu\nu\rho\sigma}$, the Euler density $E_4$, and other background invariants such as $R^2$ and $R \bar{F}_{\mu\nu}\bar{F}^{\mu\nu}$, which constitute the central $a_4(x)$ relation \eqref{sdc32}. Specifically, these integrated invariant relations, along with their central charge coefficients $(\mathfrak{c}, \mathfrak{a}, \mathfrak{b}_1, \mathfrak{b}_2)$, collectively evaluate the $\mathcal{C}_{\text{local}}$ formulas outlined in \cref{local}. For all AdS$_4$ black holes, this integration process necessitates the inclusion of a holographic boundary term \eqref{kna9} to regulate the divergences arising from their infinite volume. This choice of regularization naturally aligns with the Gauss-Bonnet-Chern theorem \cite{Chern:1945wp}, resulting in a consistent and unambiguous result by isolating the finite portion from the diverging bulk one-loop effective action. The holographic renormalization choice yields the consistent 4D Euler characteristic value of $\chi = 2$ when integrating $E_4$ around all AdS$_4$ backgrounds of interest. Furthermore, we have verified that the integrated AdS$_4$ invariants precisely match the established relations in \cite{Sen:2013ns,Charles:2015nn,Karan:2020sk,Karan:2021teq,Bhattacharyya:2012ss} for asymptotically-flat$_4$ black holes in the limit $\ell \to \infty$. Around these flat backgrounds, invariants such as $R^2$ and $R \bar{F}_{\mu\nu}\bar{F}^{\mu\nu}$ vanish, and $a_4(x)$ is solely governed by $W_{\mu\nu\rho\sigma}W^{\mu\nu\rho\sigma}$ and $E_4$ anomalies, along with non-vanishing central charge values $\mathfrak{c}$ and $\mathfrak{a}$. In this flat-space scenario, the integration of $W_{\mu\nu\rho\sigma}W^{\mu\nu\rho\sigma}$ and $E_4$ over the non-extremal backgrounds always traces out a finite volume and does not require the utilization of any regularization treatment or incorporation of boundary terms.

The regularization procedure of the integrated background invariants over all AdS$_4$ and flat$_4$ black holes in their extremal limit rely on the prescription of quantum entropy function (QEF) formalism \cite{Sen:2008wa,Sen:2009wb,Sen:2009wc}, as detailed in \cref{ext}. Here, the underlying analysis is confined to the cut-off independent ``finite'' part of the extremal near-horizon (ENH) geometry, which includes an AdS$_2$ component and suffers divergences due to its infinite volume. Following the QEF prescription, we ignored these divergences and extracted only the divergent-insensitive finite part of the ENH geometry. This treatment serves as the exact result for the $\mathcal{C}_{\text{local}}$ contributions provided in \cref{elimit}. Moreover, these extremal results are further verified by directly taking the $\beta \to \infty$ limit on all the non-extremal backgrounds considered in \cref{kna,rn,ka}. This ensures that the extremal logarithmic correction results presented in this paper are unambiguous and robust.  

However, there are instances reported in the literature \cite{Liu:2017vll,Jeon:2017ij} where the AdS$_4$ logarithmic corrections derived from extremal near-horizon analyses do not align with the results obtained from field theory computations. It is important to resolve this discrepancy and highlight that the $\mathcal{C}_{\text{local}}$ contributions remain consistent across all treatments---whether involving the full geometry or focusing on the near-horizon of extremal black holes. Actually, the differences arise in the zero-mode or $\mathcal{C}_{\text{zm}}$ contributions, leading to distinct total logarithmic correction outcomes when employing different approaches to analyze the two facets of extremal black hole geometry. In the future, it would be intriguing to investigate whether the degrees of freedom responsible for the zero-mode quantum entropy in extremal AdS black holes reside in the near-horizon region, encompass the full geometry, or exist elsewhere. Progress in this direction may require determining the appropriate ensemble and scalings for extremal AdS black hole backgrounds.   

In the current choice of STU supergravity models, we observe that all the AdS$_4$ logarithmic correction formulas in \cref{raI1,raI2,raI3,raI4,raI5,raI6,raI7,raII1,raII2,raII3,raII4,raII5,raII6,raII7} are guaranteed to be non-topological and represented by non-trivial functions of different black hole parameters. However, their flat-space counterparts in \cref{rfI1,rfI2,rfI3,rfI4,rfI5,rfII1,rfII2,rfII3,rfII4,rfII5} exhibit a relatively simplified yet contrasting nature. Specifically, in the absence of the AdS$_4$ boundary, a vanishing charge (as in the Kerr and Schwarzschild cases) ensures confirmed topological or black hole parameter-independent numerical results. Moreover, the extremal limit guarantees the same nature for all uncharged (e.g., Kerr) and non-rotating (e.g., Reissner-Nordström) backgrounds, while the charged-rotating Kerr-Newman result remains non-topological. Readers familiar with the literature may recognize a similar characteristic pattern of logarithmic corrections as reported in previous works \cite{Karan:2021teq,Bhattacharyya:2012ss,Sen:2013ns,Karan:2022dfy,David:2021eoq}.

The aforementioned observation might set a stringent criterion for four-dimensional supergravity models to admit a UV completion, which is highly sensitive to the microscopic analysis of the relevant black holes. In fact, we have already seen that the available microstate counting data \cite{Strominger:1996sh,Maldacena:1996gb,Horowitz:1996fn,Emparan:2006it,Sen:2014aja,Belin:2016knb,Benini:2019dyp,Gang:2019uay,PandoZayas:2020iqr,Liu:2017vbl,Liu:2017vll,Benini:2015eyy} is indeed topological. However, in all even dimensions, quantum black holes exhibiting non-topological logarithmic corrections are a generic and natural feature (e.g., see \cite{David:2021eoq,Karan:2021teq,Karan:2022dfy,Bhattacharyya:2012ss,Sen:2013ns}). Notably, for a parent theory in odd-$D$ dimensions, the Seeley-DeWitt coefficients $a_D(x)$ characterizing $\mathcal{C}_{\text{local}}$ vanish due to the lack of any diffeomorphism invariant scalar functions connected to the concerned background \cite{Vassilevich:2003ll}, and the $\Delta S_{\text{BH}}$ formula is entirely defined by the topological $\mathcal{C}_{\text{zm}}$ contributions. Such an expectation has also been confirmed in various 11D supergravity computations \cite{Benini:2019dyp,Gang:2019uay,PandoZayas:2020iqr,Liu:2017vbl}.  


In order to anticipate a topological logarithmic correction in even dimensions following the current setup of this study, one needs to either set $\mathfrak{c}= \mathfrak{b}_1 = \mathfrak{b}_2 = 0$ or somehow constrain the integrated invariants, such as $W_{\mu\nu\rho\sigma}W^{\mu\nu\rho\sigma}$, $R^2$ and $R \bar{F}_{\mu\nu}\bar{F}^{\mu\nu}$, to have a vanishing or topological value.\footnote{Note that the integrated $E_4$ always defines a topological Euler characteristic value.} Typically, these two classes of criteria, as well as their possible overlaps, are found to be fulfilled while considering the supergravity backgrounds with anomaly and other non-trivial cancellations \cite{Charles:2015nn,Karan:2019sk,Karan:2020sk,Castro:2018tg,David:2021eoq,Bobev:2023dwx}, adhering to extremality conditions, approaching the flat-space limit, having vanishing charge or rotation parameters, turning off the Maxwell backgrounds, etc.\footnote{For example, readers are encouraged to review \cite{Bhattacharyya:2012ss,Sen:2013ns,Charles:2015nn,Castro:2018tg,Karan:2019sk,Karan:2020sk,Karan:2021teq,Karan:2022dfy,Banerjee:2020wbr,Banerjee:2021pdy}, as well as \cref{flimit,enhii} of this paper.} However, as soon as one considers even-dimensional AdS backgrounds, the logarithmic correction formulas are guaranteed to have non-topological characteristics due to the appearance of a natural and robust boundary $\ell$ (with $\ell^2 =-12/R =-3/\Lambda $), which eventually declines or neutralize the simultaneous occurrence of available topological constraints \cite{David:2021eoq,Karan:2022dfy,Bobev:2023dwx}. Arguably, this reasoning provides a consistent explanation for the specific nature of logarithmic correction results obtained for all Kerr-Newman-AdS and Kerr-Newman families of black holes embedded within the four-dimensional STU supergravity models. 

However, discrepancies have been reported in certain cases where logarithmic corrections derived through different computational methods deviate from the expected non-topological nature. For instance, in the context of ten-dimensional massive IIA supergravity, the microscopic computations for the logarithmic entropy corrections of AdS$_4$ black holes appear to be topological \cite{Liu:2018bac}. To address this puzzle, we suspect that incorporating matter multiplets induced from the full Kaluza-Klein (KK) tower modes could be crucial in higher-dimensional supergravity theories to preserve the true character of logarithmic correction relations. Similarly, in the computation of logarithmic corrections for BPS black holes in four-dimensional $\mathcal{N}= 2$ gauged supergravity presented in \cite{Hristov:2021zai}, the results are reported to be topological, which contradicts the outcomes in \cite{David:2021eoq} and our expectation. As a potential resolution, we argue that the topological characteristic in \cite{Hristov:2021zai} explicitly emerges from the considered Euler term. However, it might be possible to recover the true non-topological nature of logarithmic corrections by incorporating the $\eta$-invariant term as a correction due to the presence of a boundary \cite{Atiyah:1975jf}.   

In conclusion, the logarithmic correction to the entropy of extremal and non-extremal AdS black holes embedded in any even-dimensional parent theory confirms a non-topological nature compared to their flat-space counterparts. This insight provides a natural and more comprehensive ``infrared window into the microstates'' of the black holes. We argue that the presence of AdS boundary must be interpreted as a robust and general criterion for this `non-topological' characteristic, which is also sensitive to the microscopic details of the black holes. Therefore, the current study sheds light on novel aspects of black hole properties in this specific class of STU supergravity theories and significantly contributes to understanding low-energy effective string theory models in 4D. It would be intriguing to extend this topological vs. non-topological analysis to the recently explored near-extremal black holes \cite{Iliesiu:2022onk,Banerjee:2023quv,Kapec:2023ruw,Rakic:2023vhv,Banerjee:2023gll}, even in higher-dimensional supergravity. A more challenging task would involve realizing and interpreting the non-topological nature of logarithmic corrections through progress in microstate counting within the framework of UV-completed string theory counterparts. We aim to explore some of these advancements in future research.

\begin{acknowledgments}

	This project has been a long time in the making, and we express our gratitude for the hospitality extended by IBS Daejeon, APCTP, Tokyo Tech, YITP, IMSc, and CMI over the past two years. Especially, SK is thankful to IISER Bhopal, where the early-stage of this project was developed and collaboration happened. 
	
	We are also grateful to Nabamita Banerjee, Suvankar Dutta, and Bibhas Ranjan Majhi for their unwavering support throughout the course of this project. Special thanks to Binata Panda for insightful discussions and comments. SK is supported by IIT Guwahati through the Institute Post-Doctoral Fellowship project IITG/R\&D/IPDF/2023-24/20230922P120.
	
	
\end{acknowledgments}

\appendix
\section{Heat kernel trace computations for $a_4(x)$ in STU models}\label{calcul}

\subsection{Notation and background identities}\label{iden}
In this paper, we considered the two Maxwell field strengths, denoted as $\bar{F}\indices{_1_{\mu\nu}}$ and $\bar{F}\indices{_2_{\mu\nu}}$, which play a crucial role in embedding EM backgrounds into the $U(1)^2$-charged STU supergravity models for two sets of coupling constants: $(\kappa_1, \kappa_2) = (1, -1)$ and $(\sqrt{3}, -1/\sqrt{3})$. They lead to solely electrically charged configurations that satisfy either $Q_2 = \kappa_1 Q_1$ or $Q_1 = \kappa_2 Q_2$ (for detailed discussions, see \cref{setupmodel,embeddingEM}). Throughout our analysis, we maintained strict control over the Maxwell field strengths by imposing a set of constraint relations:
{
	\allowdisplaybreaks
	\begin{align}
		&\kappa_1\kappa_2 = -1, \quad \kappa_1\bar{F}_{1\mu\nu}\bar{F}\indices{_1^\mu^\nu} + \kappa_2\bar{F}_{2\mu\nu}\bar{F}\indices{_2^\mu^\nu} = 0, \\[5pt]
		& \kappa_1^2\bar{F}_{1\mu\nu}\bar{F}\indices{_1^\mu^\nu} + \kappa_2^2\bar{F}_{2\mu\nu}\bar{F}\indices{_2^\mu^\nu} = \bar{F}_{1\mu\nu}\bar{F}\indices{_1^\mu^\nu} + \bar{F}_{2\mu\nu}\bar{F}\indices{_2^\mu^\nu},  \\[5pt]
		&\kappa_1^2	\left(\bar{F}_{1\mu\nu}\bar{F}\indices{_1^{\mu\nu}}\right)^2 + \kappa_2^2\left(\bar{F}_{2\mu\nu}\bar{F}\indices{_2^{\mu\nu}} \right)^2 = - 2\kappa_1\kappa_2\bar{F}_{1\mu\nu}\bar{F}\indices{_1^{\mu\nu}}\bar{F}_{2\mu\nu}\bar{F}\indices{_2^{\mu\nu}}, \\[5pt]
		& \left(\bar{F}_{1\mu\nu}\bar{F}\indices{_2^{\mu\nu}}\right)^2 = \bar{F}_{1\mu\nu}\bar{F}\indices{_1^{\mu\nu}}\bar{F}_{2\rho\sigma}\bar{F}\indices{_2^{\rho\sigma}} = \bar{F}_{1\mu\nu}\bar{F}\indices{_2^{\mu\nu}}\bar{F}_{1\rho\sigma}\bar{F}\indices{_2^{\rho\sigma}}.
\end{align}}	
The above constraints, along with the following Einstein and Maxwell evolution equations
{
	\allowdisplaybreaks
	\begin{align} 
		&\bar{F}_{1\mu\rho}\bar{F}\indices{_1_\nu^\rho} + \bar{F}_{2\mu\rho}\bar{F}\indices{_2_\nu^\rho} = \frac{1}{2}\left(R_{\mu\nu} - \Lambda\bar{g}_{\mu\nu}\right) +\frac{1}{4}\bar{g}_{\mu\nu}\left(\bar{F}_{1\mu\nu}\bar{F}\indices{_1^\mu^\nu} + \bar{F}_{2\mu\nu}\bar{F}\indices{_2^\mu^\nu}\right), \\[4pt]
		& R= 4\Lambda, \quad D_\mu \bar{F}\indices{_1^\mu^\nu} = 0, \quad D_\mu \bar{F}\indices{_2^\mu^\nu} = 0, \quad D_{[\mu}\bar{F}_{1\rho\sigma]}=0,\quad D_{[\mu}\bar{F}_{2\rho\sigma]}=0.
\end{align}}
lead to several induced on-shell identities that are presented as
{
	\allowdisplaybreaks
	\begin{align}
		{R}_{\mu\nu}\bar{F}\indices{_1^{\mu\rho}}\bar{F}\indices{_1^\nu_\rho}&= \frac{1}{\kappa_1^4}{R}_{\mu\nu}\bar{F}\indices{_2^{\mu\rho}}\bar{F}\indices{_2^\nu_\rho} \nonumber \\
		&= \frac{1}{2\left(\kappa_1^4 + 1\right)}\Big\lbrace R_{\mu\nu}R^{\mu\nu}- 4\Lambda^2  + 2\Lambda \left(\bar{F}_{1\mu\nu}\bar{F}\indices{_1^{\mu\nu}} + \bar{F}_{2\mu\nu}\bar{F}\indices{_2^{\mu\nu}}\right) \Big\rbrace, \\[6pt]
		\bar{F}\indices{_1^{\mu\rho}}\bar{F}\indices{_1^\nu_\rho} \bar{F}\indices{_1_{\mu\sigma}}\bar{F}\indices{_1_\nu^\sigma} &= \frac{1}{\kappa_1^4}\bar{F}\indices{_2^{\mu\rho}}\bar{F}\indices{_2^\nu_\rho} \bar{F}\indices{_2_{\mu\sigma}}\bar{F}\indices{_2_\nu^\sigma} \nonumber \\
		& = \frac{1}{\kappa_1^2}\bar{F}\indices{_1^{\mu\rho}}\bar{F}\indices{_2^\nu_\rho} \bar{F}\indices{_1_{\mu\sigma}}\bar{F}\indices{_2_\nu^\sigma} \nonumber \\
		& = \frac{1}{\kappa_1^2}\bar{F}\indices{_1^{\mu\rho}}\bar{F}\indices{_2^\nu^\sigma} \bar{F}\indices{_1_{\mu\nu}}\bar{F}\indices{_2_\rho_\sigma} \nonumber \\
		&= \frac{1}{4\left(\kappa_1^2 + 1\right)^2}\left\lbrace R_{\mu\nu}R^{\mu\nu}- 4\Lambda^2  +  \left(\bar{F}_{1\mu\nu}\bar{F}\indices{_1^{\mu\nu}} + \bar{F}_{2\mu\nu}\bar{F}\indices{_2^{\mu\nu}}\right)^2 \right\rbrace, \\[6pt]
		R_{\mu\rho\nu\sigma}\bar{F}\indices{_1^{\mu\nu}}\bar{F}\indices{_1^{\rho\sigma}} &= \frac{1}{2}R_{\mu\nu\rho\sigma}\bar{F}\indices{_1^{\mu\nu}}\bar{F}\indices{_1^{\rho\sigma}}, \quad R_{\mu\rho\nu\sigma}\bar{F}\indices{_2^{\mu\nu}}\bar{F}\indices{_2^{\rho\sigma}} = \frac{1}{2}R_{\mu\nu\rho\sigma}\bar{F}\indices{_2^{\mu\nu}}\bar{F}\indices{_2^{\rho\sigma}}, \\[6pt]
		\left(D_\rho \bar{F}_{1\mu\nu}\right)\left(D^\rho \bar{F}\indices{_1^{\mu\nu}}\right) &= 2\left(D_\mu \bar{F}\indices{_1_\rho^\nu}\right)\left(D_\nu \bar{F}\indices{_1^{\rho\mu}}\right) = R_{\mu\nu\rho\sigma}\bar{F}\indices{_1^{\mu\nu}}\bar{F}\indices{_1^{\rho\sigma}} - 2 {R}_{\mu\nu}\bar{F}\indices{_1^{\mu\rho}}\bar{F}\indices{_1^\nu_\rho}, \\[8pt]
		\left(D_\rho \bar{F}_{2\mu\nu}\right)\left(D^\rho \bar{F}\indices{_2^{\mu\nu}}\right) &= 2\left(D_\mu \bar{F}\indices{_2_\rho^\nu}\right)\left(D_\nu \bar{F}\indices{_2^{\rho\mu}}\right) = R_{\mu\nu\rho\sigma}\bar{F}\indices{_2^{\mu\nu}}\bar{F}\indices{_2^{\rho\sigma}} - 2 {R}_{\mu\nu}\bar{F}\indices{_2^{\mu\rho}}\bar{F}\indices{_2^\nu_\rho}.
	\end{align}
}
In deriving the above identities, we have considered the gravitational Bianchi identities and related relations whenever needed:
\begin{align}
	R_{\mu[\nu\rho\sigma]}=0, \qquad R_{\mu\rho\nu\sigma}R^{\mu\nu\rho\sigma} =\frac{1}{2}R_{\mu\nu\rho\sigma}R^{\mu\nu\rho\sigma}.
\end{align} 
Notably, we have adjusted terms involving covariant derivatives on the Maxwell field strengths up to total derivatives, accompanied by the commutation relation of covariant derivatives for a rank-2 tensor, as follows (for both $I = 1, 2$):
\begin{align}
	\left(D_\rho \bar{F}_{I\mu\nu}\right)\left(D^\rho \bar{F}\indices{_I^{\mu\nu}}\right) &= 2\bar{F}\indices{_I^\mu_\nu} D_\rho D_\mu \bar{F}\indices{_I^{\nu\rho}}\nonumber \\
	&= 2\bar{F}\indices{_I^\mu_\nu} [D_\rho,D_\mu] \bar{F}\indices{_I^{\nu\rho}} \nonumber \\
	&= 2 \bar{F}\indices{_I^\mu_\nu} \left(R\indices{^\nu_\sigma_\mu_\rho}\bar{F}\indices{_I^{\rho\sigma}} + R_{\mu\rho}\bar{F}\indices{_I^{\nu\rho}}\right).
\end{align}

\subsection{Components of heat kernel matrices $E$ and $\Omega_{\rho\sigma}$}\label{comp}
In this section, we demonstrate the derivation of the components of matrices $E$ and $\Omega_{\rho\sigma}$, which are essential for proceeding with the heat kernel treatment discussed in \cref{SDCcomputation}. To achieve this, we employ the formulas presented in \cref{comp7,comp6}. For the current configuration of STU fluctuations denoted as $\tilde{\phi}_m = \big\lbrace \hat{h}_{\mu\nu}, \hat{h}, a_{1\mu}, a_{2\mu}, \tilde{\Phi} \big\rbrace$, we utilize all the matrix-valued data of $\mathcal{I}\indices{^{\tilde{\phi}_m}^{\tilde{\phi}_n}}$, $\left(\omega^\rho\right)\indices{^{\tilde{\phi}_m}^{\tilde{\phi}_n}}$ and $P\indices{^{\tilde{\phi}_m}^{\tilde{\phi}_n}}$ recorded in \cref{sdc11,sdc13,sdc14}. Technically, we have identified 5 diagonal and 14 off-diagonal valid components of the effective potential matrix $E$, which are listed as follows:
{\allowdisplaybreaks
	\begin{align}
		E\indices{^{\hat{h}_{\mu\nu}}^{\hat{h}_{\alpha\beta}}} &= P\indices{^{\hat{h}_{\mu\nu}}^{\hat{h}_{\alpha\beta}}} - \left(\omega^\rho\right)\indices{^{\hat{h}_{\mu\nu}}^{a_{1\gamma}}}\left(\omega_\rho\right)\indices{^{a_{1\delta}}^{\hat{h}_{\alpha\beta}}}\mathcal{I}\indices{_{a_{1\gamma}}_{a_{1\delta}}}- \left(\omega^\rho\right)\indices{^{\hat{h}_{\mu\nu}}^{a_{2\gamma}}}\left(\omega_\rho\right)\indices{^{a_{2\delta}}^{\hat{h}_{\alpha\beta}}}\mathcal{I}\indices{_{a_{2\gamma}}_{a_{2\delta}}}, \\[3pt]
		E\indices{^{\hat{h}}^{\hat{h}}} &=  P\indices{^{\hat{h}}^{\hat{h}}}, \\[3pt]
		E\indices{^{a_{1\alpha}}^{a_{1\beta}}} &= P\indices{^{a_{1\alpha}}^{a_{1\beta}}} - \left(\omega^\rho\right)\indices{^{a_{1\alpha}}^{\hat{h}_{\mu\nu}}}\left(\omega_\rho\right)\indices{^{\hat{h}_{\gamma\delta}}^{a_{1\beta}}}\mathcal{I}_{\hat{h}_{\mu\nu}\hat{h}_{\gamma\delta}}- \left(\omega^\rho\right)\indices{^{a_{1\alpha}}^{\tilde{\Phi}}}\left(\omega_\rho\right)\indices{^{\tilde{\Phi}}^{a_{1\beta}}}\mathcal{I}\indices{_{\tilde{\Phi}}_{\tilde{\Phi}}},\\[3pt]
		E\indices{^{a_{2\alpha}}^{a_{2\beta}}} &= P\indices{^{a_{2\alpha}}^{a_{2\beta}}} - \left(\omega^\rho\right)\indices{^{a_{2\alpha}}^{\hat{h}_{\mu\nu}}}\left(\omega_\rho\right)\indices{^{\hat{h}_{\gamma\delta}}^{a_{2\beta}}}\mathcal{I}_{\hat{h}_{\mu\nu}\hat{h}_{\gamma\delta}}- \left(\omega^\rho\right)\indices{^{a_{2\alpha}}^{\tilde{\Phi}}}\left(\omega_\rho\right)\indices{^{\tilde{\Phi}}^{a_{2\beta}}}\mathcal{I}\indices{_{\tilde{\Phi}}_{\tilde{\Phi}}},\\[3pt]
		E\indices{^{\tilde{\Phi}}^{\tilde{\Phi}}} &= P\indices{^{\tilde{\Phi}}^{\tilde{\Phi}}} - \left(\omega^\rho\right)\indices{^{\tilde{\Phi}}^{a_{1\alpha}}}\left(\omega_\rho\right)\indices{^{a_{1\beta}}^{\tilde{\Phi}}} \mathcal{I}\indices{_{a_{1\alpha}}_{a_{1\beta}}} - \left(\omega^\rho\right)\indices{^{\tilde{\Phi}}^{a_{2\alpha}}}\left(\omega_\rho\right)\indices{^{a_{2\beta}}^{\tilde{\Phi}}}\mathcal{I}\indices{_{a_{2\alpha}}_{a_{2\beta}}}, \\[3pt]
		E\indices{^{\hat{h}_{\mu\nu}}^{\hat{h}}} &= E\indices{^{\hat{h}}^{\hat{h}_{\mu\nu}}} = P\indices{^{\hat{h}_{\mu\nu}}^{\hat{h}}} = P\indices{^{\hat{h}}^{\hat{h}_{\mu\nu}}}, \\[3pt]
		E\indices{^{a_{1\alpha}}^{a_{2\beta}}} &=  - \left(\omega^\rho\right)\indices{^{a_{1\alpha}}^{\hat{h}_{\mu\nu}}}\left(\omega_\rho\right)\indices{^{\hat{h}_{\gamma\delta}}^{a_{2\beta}}} \mathcal{I}_{\hat{h}_{\mu\nu}\hat{h}_{\gamma\delta}}- \left(\omega^\rho\right)\indices{^{a_{1\alpha}}^{\tilde{\Phi}}}\left(\omega_\rho\right)\indices{^{\tilde{\Phi}}^{a_{2\beta}}}\mathcal{I}\indices{_{\tilde{\Phi}}_{\tilde{\Phi}}}, \\[3pt]
		E\indices{^{a_{2\alpha}}^{a_{1\beta}}} &=  - \left(\omega^\rho\right)\indices{^{a_{2\alpha}}^{\hat{h}_{\mu\nu}}}\left(\omega_\rho\right)\indices{^{\hat{h}_{\gamma\delta}}^{a_{1\beta}}}\mathcal{I}_{\hat{h}_{\mu\nu}\hat{h}_{\gamma\delta}}- \left(\omega^\rho\right)\indices{^{a_{2\alpha}}^{\tilde{\Phi}}}\left(\omega_\rho\right)\indices{^{\tilde{\Phi}}^{a_{1\beta}}}\mathcal{I}\indices{_{\tilde{\Phi}}_{\tilde{\Phi}}}, \\[3pt]
		E\indices{^{\hat{h}_{\mu\nu}}^{\tilde{\Phi}}} &= P\indices{^{\hat{h}_{\mu\nu}}^{\tilde{\Phi}}}  - \left(\omega^\rho\right)\indices{^{\hat{h}_{\mu\nu}}^{a_{1\gamma}}}\left(\omega_\rho\right)\indices{^{a_{1\delta}}^{\tilde{\Phi}}} \mathcal{I}\indices{_{a_{1\gamma}}_{a_{1\delta}}}- \left(\omega^\rho\right)\indices{^{\hat{h}_{\mu\nu}}^{a_{2\gamma}}}\left(\omega_\rho\right)\indices{^{a_{2\delta}}^{\tilde{\Phi}}}\mathcal{I}\indices{_{a_{2\gamma}}_{a_{2\delta}}}, \\[3pt]
		E\indices{^{\tilde{\Phi}}^{\hat{h}_{\mu\nu}}} &= P\indices{^{\tilde{\Phi}}^{\hat{h}_{\mu\nu}}}  - \left(\omega^\rho\right)\indices{^{\tilde{\Phi}}^{a_{1\gamma}}}\left(\omega_\rho\right)\indices{^{a_{1\delta}}^{\hat{h}_{\mu\nu}}}\mathcal{I}\indices{_{a_{1\gamma}}_{a_{1\delta}}}- \left(\omega^\rho\right)\indices{^{\tilde{\Phi}}^{a_{2\gamma}}}\left(\omega_\rho\right)\indices{^{a_{2\delta}}^{\hat{h}_{\mu\nu}}}\mathcal{I}\indices{_{a_{2\gamma}}_{a_{2\delta}}}, \\[3pt]
		E\indices{^{\hat{h}_{\mu\nu}}^{a_{1\alpha}}} &= P\indices{^{\hat{h}_{\mu\nu}}^{a_{1\alpha}}} - \left(D_\rho\omega^\rho\right)\indices{^{\hat{h}_{\mu\nu}}^{a_{1\alpha}}}, \\[3pt]
		E\indices{^{\hat{h}_{\mu\nu}}^{a_{2\alpha}}} &= P\indices{^{\hat{h}_{\mu\nu}}^{a_{2\alpha}}} - \left(D_\rho\omega^\rho\right)\indices{^{\hat{h}_{\mu\nu}}^{a_{2\alpha}}}, \\[3pt]
		E\indices{^{\hat{h}_{\mu\nu}}^{a_{2\alpha}}} &= P\indices{^{\hat{h}_{\mu\nu}}^{a_{2\alpha}}} - \left(D_\rho\omega^\rho\right)\indices{^{\hat{h}_{\mu\nu}}^{a_{2\alpha}}}, \\[3pt]
		E\indices{^{a_{2\alpha}}^{\hat{h}_{\mu\nu}}} &= P\indices{^{a_{2\alpha}}^{\hat{h}_{\mu\nu}}} - \left(D_\rho\omega^\rho\right)\indices{^{a_{2\alpha}}^{\hat{h}_{\mu\nu}}}, \\[3pt]
		E\indices{^{a_{1\alpha}}^{\tilde{\Phi}}} &=  - \left(D_\rho\omega^\rho\right)\indices{^{a_{1\alpha}}^{\tilde{\Phi}}}, \qquad E\indices{^{\tilde{\Phi}}^{a_{1\alpha}}} =  - \left(D_\rho\omega^\rho\right)\indices{^{\tilde{\Phi}}^{a_{1\alpha}}}, \\[3pt]
		E\indices{^{a_{2\alpha}}^{\tilde{\Phi}}} &=  - \left(D_\rho\omega^\rho\right)\indices{^{a_{2\alpha}}^{\tilde{\Phi}}}, \qquad E\indices{^{\tilde{\Phi}}^{a_{2\alpha}}} =  - \left(D_\rho\omega^\rho\right)\indices{^{\tilde{\Phi}}^{a_{2\alpha}}}.
	\end{align}
}
Notice that the aforementioned relations rely on the $\mathcal{I}\indices{_{\phi_m}_{\phi_n}}$ form of the projection matrices, which are expressed as follows
\begin{align}
	\begin{split}
		I_{\hat{h}\hat{h}} &= I\indices{_{\tilde{\Phi}}_{\tilde{\Phi}}} = 1,\quad I\indices{_{a_{1\mu}}_{a_{1\nu}}} = I\indices{_{a_{2\mu}}_{a_{2\nu}}} = \bar{g}_{\mu\nu},\\[5pt]	
		I_{\hat{h}_{\mu\nu}\hat{h}_{\alpha\beta}} &= \frac{1}{2}\Big(\bar{g}_{\mu\alpha} \bar{g}_{\nu\beta} + \bar{g}_{\mu\beta} \bar{g}_{\nu\alpha}-\frac{1}{2}\bar{g}_{\mu\nu}\bar{g}_{\alpha\beta}\Big).
	\end{split}
\end{align}
Similarly, we derive the effective curvature commutator matrix $\Omega_{\rho\sigma}$, which consists of 4 diagonal and 12 off-diagonal components. These components are summarized as follows:
{\allowdisplaybreaks
	\begin{align}
		\left(\Omega_{\rho\sigma}\right)\indices{^{\hat{h}_{\mu\nu}}^{\hat{h}_{\alpha\beta}}} &= \frac{1}{2}\left(\bar{g}^{\mu\alpha}R\indices{^\nu^\beta_\rho_\sigma}+ \bar{g}^{\mu\beta}R\indices{^\nu^\alpha_\rho_\sigma} +\bar{g}^{\nu\alpha}R\indices{^\mu^\beta_\rho_\sigma}+\bar{g}^{\nu\beta}R\indices{^\mu^\alpha_\rho_\sigma} \right)\nonumber \\
		&\quad\enspace + \left(\left(\omega_\rho\right)\indices{^{\hat{h}_{\mu\nu}}^{a_{1\gamma}}}\left(\omega_\sigma\right)\indices{^{a_{1\delta}}^{\hat{h}_{\alpha\beta}}}- \left(\omega_\sigma\right)\indices{^{\hat{h}_{\mu\nu}}^{a_{1\gamma}}}\left(\omega_\rho\right)\indices{^{a_{1\delta}}^{\hat{h}_{\alpha\beta}}}\right)\mathcal{I}\indices{_{a_{1\gamma}}_{a_{1\delta}}}\nonumber \\
		&\quad\enspace + \left(\left(\omega_\rho\right)\indices{^{\hat{h}_{\mu\nu}}^{a_{2\gamma}}}\left(\omega_\sigma\right)\indices{^{a_{2\delta}}^{\hat{h}_{\alpha\beta}}}- \left(\omega_\sigma\right)\indices{^{\hat{h}_{\mu\nu}}^{a_{2\gamma}}}\left(\omega_\rho\right)\indices{^{a_{2\delta}}^{\hat{h}_{\alpha\beta}}}\right)\mathcal{I}\indices{_{a_{2\gamma}}_{a_{2\delta}}}, \\[6pt]	
		\left(\Omega_{\rho\sigma}\right)\indices{^{a_{1\alpha}}^{a_{1\beta}}} &=  R\indices{^\alpha^\beta_\rho_\sigma} + \left( \left(\omega_\rho\right)\indices{^{a_{1\alpha}}^{\tilde{\Phi}}}\left(\omega_\sigma\right)\indices{^{\tilde{\Phi}}^{a_{1\beta}}} - \left(\omega_\sigma\right)\indices{^{a_{1\alpha}}^{\tilde{\Phi}}}\left(\omega_\rho\right)\indices{^{\tilde{\Phi}}^{a_{1\beta}}}\right)\mathcal{I}\indices{_{\tilde{\Phi}}_{\tilde{\Phi}}} \nonumber \\
		&\quad\enspace + \left(\left(\omega_\rho\right)\indices{^{a_{1\alpha}}^{\hat{h}_{\mu\nu}}}\left(\omega_\sigma\right)\indices{^{\hat{h}_{\gamma\delta}}^{a_{1\beta}}} - \left(\omega_\sigma\right)\indices{^{a_{1\alpha}}^{\hat{h}_{\mu\nu}}}\left(\omega_\rho\right)\indices{^{\hat{h}_{\gamma\delta}}^{a_{1\beta}}} \right)\mathcal{I}_{\hat{h}_{\mu\nu}\hat{h}_{\gamma\delta}}, \\[6pt]	
		\left(\Omega_{\rho\sigma}\right)\indices{^{a_{2\alpha}}^{a_{2\beta}}} &=  R\indices{^\alpha^\beta_\rho_\sigma} + \left( \left(\omega_\rho\right)\indices{^{a_{2\alpha}}^{\tilde{\Phi}}}\left(\omega_\sigma\right)\indices{^{\tilde{\Phi}}^{a_{2\beta}}} - \left(\omega_\sigma\right)\indices{^{a_{2\alpha}}^{\tilde{\Phi}}}\left(\omega_\rho\right)\indices{^{\tilde{\Phi}}^{a_{2\beta}}}\right)\mathcal{I}\indices{_{\tilde{\Phi}}_{\tilde{\Phi}}} \nonumber \\
		&\quad\enspace + \left(\left(\omega_\rho\right)\indices{^{a_{2\alpha}}^{\hat{h}_{\mu\nu}}}\left(\omega_\sigma\right)\indices{^{\hat{h}_{\gamma\delta}}^{a_{2\beta}}} - \left(\omega_\sigma\right)\indices{^{a_{2\alpha}}^{\hat{h}_{\mu\nu}}}\left(\omega_\rho\right)\indices{^{\hat{h}_{\gamma\delta}}^{a_{2\beta}}} \right)\mathcal{I}_{\hat{h}_{\mu\nu}\hat{h}_{\gamma\delta}}, \\[6pt]
		\left(\Omega_{\rho\sigma}\right)\indices{^{\tilde{\Phi}}^{\tilde{\Phi}}} &= \left(\left(\omega_\rho\right)\indices{^{\tilde{\Phi}}^{a_{1\alpha}}}\left(\omega_\sigma\right)\indices{^{a_{1\beta}}^{\tilde{\Phi}}} - \left(\omega_\sigma\right)\indices{^{\tilde{\Phi}}^{a_{1\alpha}}}\left(\omega_\rho\right)\indices{^{a_{1\beta}}^{\tilde{\Phi}}}\right) \mathcal{I}\indices{_{a_{1\alpha}}_{a_{1\beta}}} \nonumber \\
		&\quad\enspace + \left(\left(\omega_\rho\right)\indices{^{\tilde{\Phi}}^{a_{2\alpha}}}\left(\omega_\sigma\right)\indices{^{a_{2\beta}}^{\tilde{\Phi}}} - \left(\omega_\sigma\right)\indices{^{\tilde{\Phi}}^{a_{2\alpha}}}\left(\omega_\rho\right)\indices{^{a_{2\beta}}^{\tilde{\Phi}}}\right) \mathcal{I}\indices{_{a_{2\alpha}}_{a_{2\beta}}}, \\[6pt]
		\left(\Omega_{\rho\sigma}\right)\indices{^{\hat{h}_{\mu\nu}}^{\tilde{\Phi}}} &= \left(\left(\omega_\rho\right)\indices{^{\hat{h}_{\mu\nu}}^{a_{1\alpha}}}\left(\omega_\sigma\right)\indices{^{a_{1\beta}}^{\tilde{\Phi}}} - \left(\omega_\sigma\right)\indices{^{\hat{h}_{\mu\nu}}^{a_{1\alpha}}}\left(\omega_\rho\right)\indices{^{a_{1\beta}}^{\tilde{\Phi}}}\right) \mathcal{I}\indices{_{a_{1\alpha}}_{a_{1\beta}}} \nonumber \\
		&\quad\enspace + \left(\left(\omega_\rho\right)\indices{^{\hat{h}_{\mu\nu}}^{a_{2\alpha}}}\left(\omega_\sigma\right)\indices{^{a_{2\beta}}^{\tilde{\Phi}}} - \left(\omega_\sigma\right)\indices{^{\hat{h}_{\mu\nu}}^{a_{2\alpha}}}\left(\omega_\rho\right)\indices{^{a_{2\beta}}^{\tilde{\Phi}}}\right) \mathcal{I}\indices{_{a_{2\alpha}}_{a_{2\beta}}}, \\[6pt]
		\left(\Omega_{\rho\sigma}\right)\indices{^{\tilde{\Phi}}^{\hat{h}_{\mu\nu}}} &= \left(\left(\omega_\rho\right)\indices{^{\tilde{\Phi}}^{a_{1\alpha}}}\left(\omega_\sigma\right)\indices{^{a_{1\beta}}^{\hat{h}_{\mu\nu}}} - \left(\omega_\sigma\right)\indices{^{\tilde{\Phi}}^{a_{1\alpha}}}\left(\omega_\rho\right)\indices{^{a_{1\beta}}^{\hat{h}_{\mu\nu}}}\right)\mathcal{I}\indices{_{a_{1\alpha}}_{a_{1\beta}}} \nonumber \\
		&\quad\enspace + \left(\left(\omega_\rho\right)\indices{^{\tilde{\Phi}}^{a_{2\alpha}}}\left(\omega_\sigma\right)\indices{^{a_{2\beta}}^{\hat{h}_{\mu\nu}}} - \left(\omega_\sigma\right)\indices{^{\tilde{\Phi}}^{a_{2\alpha}}}\left(\omega_\rho\right)\indices{^{a_{2\beta}}^{\hat{h}_{\mu\nu}}}\right)\mathcal{I}\indices{_{a_{2\alpha}}_{a_{2\beta}}}, \\[6pt]
		\left(\Omega_{\rho\sigma}\right)\indices{^{a_{1\alpha}}^{a_{2\beta}}} &= \left(\left(\omega_\rho\right)\indices{^{a_{1\alpha}}^{\hat{h}_{\mu\nu}}}\left(\omega_\sigma\right)\indices{^{\hat{h}_{\gamma\delta}}^{a_{2\beta}}} - \left(\omega_\sigma\right)\indices{^{a_{1\alpha}}^{\hat{h}_{\mu\nu}}}\left(\omega_\rho\right)\indices{^{\hat{h}_{\gamma\delta}}^{a_{2\beta}}}\right)\mathcal{I}_{\hat{h}_{\mu\nu}\hat{h}_{\gamma\delta}} \nonumber \\
		&\quad \enspace + \left(\left(\omega_\rho\right)\indices{^{a_{1\alpha}}^{\tilde{\Phi}}}\left(\omega_\sigma\right)\indices{^{\tilde{\Phi}}^{a_{2\beta}}} - \left(\omega_\sigma\right)\indices{^{a_{1\alpha}}^{\tilde{\Phi}}}\left(\omega_\rho\right)\indices{^{\tilde{\Phi}}^{a_{2\beta}}} \right)\mathcal{I}\indices{_{\tilde{\Phi}}_{\tilde{\Phi}}}, \\[6pt]
		\left(\Omega_{\rho\sigma}\right)\indices{^{a_{2\alpha}}^{a_{1\beta}}} &= \left(\left(\omega_\rho\right)\indices{^{a_{2\alpha}}^{\hat{h}_{\mu\nu}}}\left(\omega_\sigma\right)\indices{^{\hat{h}_{\gamma\delta}}^{a_{1\beta}}} - \left(\omega_\sigma\right)\indices{^{a_{2\alpha}}^{\hat{h}_{\mu\nu}}}\left(\omega_\rho\right)\indices{^{\hat{h}_{\gamma\delta}}^{a_{1\beta}}}\right)\mathcal{I}_{\hat{h}_{\mu\nu}\hat{h}_{\gamma\delta}} \nonumber \\
		&\quad \enspace + \left(\left(\omega_\rho\right)\indices{^{a_{2\alpha}}^{\tilde{\Phi}}}\left(\omega_\sigma\right)\indices{^{\tilde{\Phi}}^{a_{1\beta}}} - \left(\omega_\sigma\right)\indices{^{a_{2\alpha}}^{\tilde{\Phi}}}\left(\omega_\rho\right)\indices{^{\tilde{\Phi}}^{a_{1\beta}}} \right)\mathcal{I}\indices{_{\tilde{\Phi}}_{\tilde{\Phi}}}, \\[5pt]
		\left(\Omega_{\rho\sigma}\right)\indices{^{\hat{h}_{\mu\nu}}^{a_{1\alpha}}} &=  \left(D_\rho\omega_\sigma\right)\indices{^{\hat{h}_{\mu\nu}}^{a_{1\alpha}}}- \left(D_\sigma\omega_\rho\right)\indices{^{\hat{h}_{\mu\nu}}^{a_{1\alpha}}}, \\[3pt]
		\left(\Omega_{\rho\sigma}\right)\indices{^{a_{1\alpha}}^{\hat{h}_{\mu\nu}}} &=  \left(D_\rho\omega_\sigma\right)\indices{^{a_{1\alpha}}^{\hat{h}_{\mu\nu}}} - \left(D_\sigma\omega_\rho\right)\indices{^{a_{1\alpha}}^{\hat{h}_{\mu\nu}}},\\[3pt]
		\left(\Omega_{\rho\sigma}\right)\indices{^{\hat{h}_{\mu\nu}}^{a_{2\alpha}}} &=  \left(D_\rho\omega_\sigma\right)\indices{^{\hat{h}_{\mu\nu}}^{a_{2\alpha}}}- \left(D_\sigma\omega_\rho\right)\indices{^{\hat{h}_{\mu\nu}}^{a_{2\alpha}}}, \\[3pt]
		\left(\Omega_{\rho\sigma}\right)\indices{^{a_{2\alpha}}^{\hat{h}_{\mu\nu}}} &=  \left(D_\rho\omega_\sigma\right)\indices{^{a_{2\alpha}}^{\hat{h}_{\mu\nu}}} - \left(D_\sigma\omega_\rho\right)\indices{^{a_{2\alpha}}^{\hat{h}_{\mu\nu}}},\\[3pt]
		\left(\Omega_{\rho\sigma}\right)\indices{^{a_{1\alpha}}^{\tilde{\Phi}}} &=  \left(D_\rho\omega_\sigma\right)\indices{^{a_{1\alpha}}^{\tilde{\Phi}}}- \left(D_\sigma\omega_\rho\right)\indices{^{a_{1\alpha}}^{\tilde{\Phi}}},\\[3pt]
		\left(\Omega_{\rho\sigma}\right)\indices{^{\tilde{\Phi}}^{a_{1\alpha}}} &=  \left(D_\rho\omega_\sigma\right)\indices{^{\tilde{\Phi}}^{a_{1\alpha}}} - \left(D_\sigma\omega_\rho\right)\indices{^{\tilde{\Phi}}^{a_{1\alpha}}},\\[3pt] 
		\left(\Omega_{\rho\sigma}\right)\indices{^{a_{2\alpha}}^{\tilde{\Phi}}} &=  \left(D_\rho\omega_\sigma\right)\indices{^{a_{2\alpha}}^{\tilde{\Phi}}}- \left(D_\sigma\omega_\rho\right)\indices{^{a_{2\alpha}}^{\tilde{\Phi}}},\\[3pt]
		\left(\Omega_{\rho\sigma}\right)\indices{^{\tilde{\Phi}}^{a_{2\alpha}}} &=  \left(D_\rho\omega_\sigma\right)\indices{^{\tilde{\Phi}}^{a_{2\alpha}}} - \left(D_\sigma\omega_\rho\right)\indices{^{\tilde{\Phi}}^{a_{2\alpha}}}.
	\end{align}
}
It is important to emphasize that during the evaluation of $\left[D_\rho,D_\sigma\right]$ contributions for the aforementioned components, we applied the covariant derivative commutation operations, as depicted in \cref{ct,cv,cs}. In this process, we treated the traceless graviton $\hat{h}_{\mu\nu}$ as a two-rank tensor, the Maxwell fluctuations $\left(a_{1\alpha}, a_{2\alpha}\right)$ as vectors, and the trace graviton $\hat{h}$ and dilaton $\tilde{\Phi}$ as scalars.

\subsection{Traces and background invariants}\label{tracing}
In this section, we finally turn to our goal of computing the traces $\text{Tr}(E)$, $\text{Tr}(E^2)$ and $\text{Tr}\left(\Omega_{\rho\sigma}\Omega^{\rho\sigma}\right)$ for the fluctuated STU models encountered in this paper. In terms of the components of matrices $E$ and $\Omega_{\rho\sigma}$, these traces can be defined as
{\allowdisplaybreaks
	\begin{align}
		\text{Tr}(E) &= E\indices{^{\hat{h}_{\mu\nu}}_{\hat{h}_{\mu\nu}}}+ E\indices{^{\hat{h}}_{\hat{h}}} + E\indices{^{a_{1\alpha}}_{a_{1\alpha}}} + E\indices{^{a_{2\alpha}}_{a_{2\alpha}}} + E\indices{^{\tilde{\Phi}}_{\tilde{\Phi}}},\label{calcul6a}\\[5pt]
		\text{Tr}\left(E^2\right) &= E\indices{^{\hat{h}_{\mu\nu}}_{\hat{h}_{\alpha\beta}}}E\indices{^{\hat{h}_{\alpha\beta}}_{\hat{h}_{\mu\nu}}} + E\indices{^{a_{1\alpha}}_{a_{1\beta}}}E\indices{^{a_{1\beta}}_{a_{1\alpha}}} + E\indices{^{a_{2\alpha}}_{a_{2\beta}}}E\indices{^{a_{2\beta}}_{a_{2\alpha}}} + E\indices{^{\hat{h}}_{\hat{h}}}E\indices{^{\hat{h}}_{\hat{h}}} \nonumber\\
		&\enspace +E\indices{^{\tilde{\Phi}}_{\tilde{\Phi}}}E\indices{^{\tilde{\Phi}}_{\tilde{\Phi}}} + E\indices{^{a_{1\alpha}}_{a_{2\beta}}}E\indices{^{a_{2\beta}}_{a_{1\alpha}}}+ E\indices{^{a_{2\alpha}}_{a_{1\beta}}}E\indices{^{a_{1\beta}}_{a_{2\alpha}}} +  E\indices{^{\hat{h}_{\mu\nu}}_{\hat{h}}}E\indices{^{\hat{h}}_{\hat{h}_{\mu\nu}}}  \nonumber\\
		&\enspace  + E\indices{^{\hat{h}}_{\hat{h}_{\mu\nu}}}E\indices{^{\hat{h}_{\mu\nu}}_{\hat{h}}} +  E\indices{^{\hat{h}_{\mu\nu}}_{a_{1\alpha}}}E\indices{^{a_{1\alpha}}_{\hat{h}_{\mu\nu}}} + E\indices{^{a_{1\alpha}}_{\hat{h}_{\mu\nu}}}E\indices{^{\hat{h}_{\mu\nu}}_{a_{1\alpha}}}  \nonumber \\
		&\enspace +  E\indices{^{\hat{h}_{\mu\nu}}_{a_{2\alpha}}}E\indices{^{a_{2\alpha}}_{\hat{h}_{\mu\nu}}} + E\indices{^{a_{2\alpha}}_{\hat{h}_{\mu\nu}}}E\indices{^{\hat{h}_{\mu\nu}}_{a_{2\alpha}}} +  E\indices{^{\hat{h}_{\mu\nu}}_{\tilde{\Phi}}}E\indices{^{\tilde{\Phi}}_{\hat{h}_{\mu\nu}}} + E\indices{^{\tilde{\Phi}}_{\hat{h}_{\mu\nu}}}E\indices{^{\hat{h}_{\mu\nu}}_{\tilde{\Phi}}} \nonumber \\
		&\enspace  +E\indices{^{a_{1\alpha}}_{\tilde{\Phi}}}E\indices{^{\tilde{\Phi}}_{a_{1\alpha}}} +  E\indices{^{\tilde{\Phi}}_{a_{1\alpha}}}E\indices{^{a_{1\alpha}}_{\tilde{\Phi}}} + +E\indices{^{a_{2\alpha}}_{\tilde{\Phi}}}E\indices{^{\tilde{\Phi}}_{a_{2\alpha}}} +  E\indices{^{\tilde{\Phi}}_{a_{2\alpha}}}E\indices{^{a_{2\alpha}}_{\tilde{\Phi}}},\label{calcul6b}\\[5pt]
		\text{Tr}\left(\Omega_{\rho\sigma}\Omega^{\rho\sigma}\right) &= {\left(\Omega_{\rho\sigma}\right)}\indices{^{\hat{h}_{\mu\nu}}_{\hat{h}_{\alpha\beta}}}{\left(\Omega^{\rho\sigma}\right)}\indices{^{\hat{h}_{\alpha\beta}}_{\hat{h}_{\mu\nu}}} + {\left(\Omega_{\rho\sigma}\right)}\indices{^{\tilde{\Phi}}_{\tilde{\Phi}}}{\left(\Omega^{\rho\sigma}\right)}\indices{^{\tilde{\Phi}}_{\tilde{\Phi}}} \nonumber \\
		&\quad + {\left(\Omega_{\rho\sigma}\right)}\indices{^{a_{1\alpha}}_{a_{1\beta}}}{\left(\Omega^{\rho\sigma}\right)}\indices{^{a_{1\beta}}_{a_{1\alpha}}} + {\left(\Omega_{\rho\sigma}\right)}\indices{^{a_{2\alpha}}_{a_{2\beta}}}{\left(\Omega^{\rho\sigma}\right)}\indices{^{a_{2\beta}}_{a_{2\alpha}}} \nonumber \\
		&\quad   + {\left(\Omega_{\rho\sigma}\right)}\indices{^{a_{1\alpha}}_{a_{2\beta}}}{\left(\Omega^{\rho\sigma}\right)}\indices{^{a_{2\beta}}_{a_{1\alpha}}}+ {\left(\Omega_{\rho\sigma}\right)}\indices{^{a_{2\alpha}}_{a_{1\beta}}}{\left(\Omega^{\rho\sigma}\right)}\indices{^{a_{1\beta}}_{a_{2\alpha}}} \nonumber \\
		&\quad +  {\left(\Omega_{\rho\sigma}\right)}\indices{^{\hat{h}_{\mu\nu}}_{a_{1\alpha}}}{\left(\Omega^{\rho\sigma}\right)}\indices{^{a_{1\alpha}}_{\hat{h}_{\mu\nu}}}+ {\left(\Omega_{\rho\sigma}\right)}\indices{^{a_{1\alpha}}_{\hat{h}_{\mu\nu}}}{\left(\Omega^{\rho\sigma}\right)}\indices{^{\hat{h}_{\mu\nu}}_{a_{1\alpha}}} \nonumber\\
		&\quad  +  {\left(\Omega_{\rho\sigma}\right)}\indices{^{\hat{h}_{\mu\nu}}_{a_{2\alpha}}}{\left(\Omega^{\rho\sigma}\right)}\indices{^{a_{2\alpha}}_{\hat{h}_{\mu\nu}}} +{\left(\Omega_{\rho\sigma}\right)}\indices{^{a_{2\alpha}}_{\hat{h}_{\mu\nu}}}{\left(\Omega^{\rho\sigma}\right)}\indices{^{\hat{h}_{\mu\nu}}_{a_{2\alpha}}} \nonumber \\
		&\quad +  {\left(\Omega_{\rho\sigma}\right)}\indices{^{\hat{h}_{\mu\nu}}_{\tilde{\Phi}}}{\left(\Omega^{\rho\sigma}\right)}\indices{^{\tilde{\Phi}}_{\hat{h}_{\mu\nu}}} +{\left(\Omega_{\rho\sigma}\right)}\indices{^{\tilde{\Phi}}_{\hat{h}_{\mu\nu}}}{\left(\Omega^{\rho\sigma}\right)}\indices{^{\hat{h}_{\mu\nu}}_{\tilde{\Phi}}}  \nonumber\\
		&\quad +{\left(\Omega_{\rho\sigma}\right)}\indices{^{a_{1\alpha}}_{\tilde{\Phi}}}{\left(\Omega^{\rho\sigma}\right)}\indices{^{\tilde{\Phi}}_{a_{1\alpha}}} +  {\left(\Omega_{\rho\sigma}\right)}\indices{^{\tilde{\Phi}}_{a_{1\alpha}}}{\left(\Omega^{\rho\sigma}\right)}\indices{^{a_{1\alpha}}_{\tilde{\Phi}}} \nonumber \\
		&\quad +{\left(\Omega_{\rho\sigma}\right)}\indices{^{a_{2\alpha}}_{\tilde{\Phi}}}{\left(\Omega^{\rho\sigma}\right)}\indices{^{\tilde{\Phi}}_{a_{2\alpha}}} +  {\left(\Omega_{\rho\sigma}\right)}\indices{^{\tilde{\Phi}}_{a_{2\alpha}}}{\left(\Omega^{\rho\sigma}\right)}\indices{^{a_{2\alpha}}_{\tilde{\Phi}}}.\label{calcul6c}
\end{align}}
Note that all the off-diagonal contributions in $\text{Tr}(E)$ vanish due to the absence of associated projection operators $\mathcal{I}$.  From here, we need to utilize all the relations derived in \cref{comp} for $E$ and $\Omega_{\rho\sigma}$ in order to compute the trace results required in the definitions mentioned above. This yields,
{\allowdisplaybreaks
	\begin{align}\label{calcul7}
		E\indices{^{\hat{h}_{\mu\nu}}_{\hat{h}_{\mu\nu}}}& =  E\indices{^{\hat{h}_{\mu\nu}}^{\hat{h}_{\alpha\beta}}}\mathcal{I}\indices{_{\hat{h}_{\mu\nu}}_{\hat{h}_{\alpha\beta}}} = -\frac{3}{2}R, \\
		E\indices{^{\hat{h}}_{\hat{h}}} &= E\indices{^{\hat{h}}^{\hat{h}}}\mathcal{I}\indices{_{\hat{h}}_{\hat{h}}}=2\Lambda, \\
		E\indices{^{\tilde{\Phi}}_{\tilde{\Phi}}} &= E\indices{^{\tilde{\Phi}}^{\tilde{\Phi}}}\mathcal{I}\indices{_{\tilde{\Phi}}_{\tilde{\Phi}}} = 0,\\
		E\indices{^{a_{1\alpha}}_{a_{1\alpha}}} &= E\indices{^{a_{1\alpha}}^{a_{1\beta}}}\mathcal{I}\indices{_{a_{1\alpha}}_{a_{1\beta}}} = -R + \left(\kappa_1^2 + 6\right)\bar{F}\indices{_1_{\mu\nu}}\bar{F}\indices{_1^{\mu\nu}}, \\
		E\indices{^{a_{2\alpha}}_{a_{2\alpha}}} &= E\indices{^{a_{1\alpha}}^{a_{2\beta}}}\mathcal{I}\indices{_{a_{2\alpha}}_{a_{1\beta}}} = -R + \left(\kappa_2^2 + 6\right)\bar{F}\indices{_2_{\mu\nu}}\bar{F}\indices{_2^{\mu\nu}}, 
\end{align}}
followed by,
{\allowdisplaybreaks
	\begin{align}\label{calcul8}
		E\indices{^{\hat{h}_{\mu\nu}}_{\hat{h}_{\alpha\beta}}}E\indices{^{\hat{h}_{\alpha\beta}}_{\hat{h}_{\mu\nu}}} &= E\indices{^{\hat{h}_{\mu\nu}}^{\hat{h}_{\gamma\delta}}}E\indices{^{\hat{h}_{\alpha\beta}}^{\hat{h}_{\theta\phi}}}\mathcal{I}\indices{_{\hat{h}_{\alpha\beta}}_{\hat{h}_{\gamma\delta}}}\mathcal{I}\indices{_{\hat{h}_{\mu\nu}}_{\hat{h}_{\theta\phi}}} \nonumber \\
		& =  2 R_{\mu\nu\rho\sigma}R^{\mu\nu\rho\sigma} + 2R_{\mu\rho\nu\sigma}R^{\mu\nu\rho\sigma} - 2 {R}_{\mu\nu}{R}^{\mu\nu} + \frac{1}{4}R^2, \\[3pt]
		E\indices{^{a_{1\alpha}}_{a_{1\beta}}}E\indices{^{a_{1\beta}}_{a_{1\alpha}}}& = E\indices{^{a_{1\alpha}}^{a_{1\gamma}}}E\indices{^{a_{1\beta}}^{a_{1\delta}}}\mathcal{I}\indices{_{a_{1\beta}}_{a_{1\gamma}}}\mathcal{I}\indices{_{a_{1\alpha}}_{a_{1\delta}}}\nonumber\\ 
		&= {R}_{\mu\nu}{R}^{\mu\nu} - 8\Lambda \bar{F}\indices{_1_{\mu\nu}}\bar{F}\indices{_1^{\mu\nu}} - \left(2\kappa_1^2 + 4\right){R}_{\mu\nu}\bar{F}\indices{_1^{\mu\rho}}\bar{F}\indices{_1^\nu_\rho} \nonumber \\
		&\quad + \left(2\kappa_1^2 + 8\right)\left(\bar{F}_{1\mu\nu}\bar{F}\indices{_1^{\mu\nu}} \right)^2 + \left(\kappa_1^2 + 2\right)^2\bar{F}\indices{_1^{\mu\rho}}\bar{F}\indices{_1^\nu_\rho} \bar{F}\indices{_1_{\mu\sigma}}\bar{F}\indices{_1_\nu^\sigma}, \\[5pt]
		E\indices{^{a_{2\alpha}}_{a_{2\beta}}}E\indices{^{a_{2\beta}}_{a_{2\alpha}}} & = E\indices{^{a_{2\alpha}}^{a_{2\gamma}}}E\indices{^{a_{2\beta}}^{a_{2\delta}}}\mathcal{I}\indices{_{a_{2\beta}}_{a_{2\gamma}}}\mathcal{I}\indices{_{a_{2\alpha}}_{a_{2\delta}}}\nonumber\\
		&= {R}_{\mu\nu}{R}^{\mu\nu} - 8\Lambda \bar{F}\indices{_2_{\mu\nu}}\bar{F}\indices{_2^{\mu\nu}} - \left(2\kappa_2^2 + 4\right){R}_{\mu\nu}\bar{F}\indices{_2^{\mu\rho}}\bar{F}\indices{_2^\nu_\rho} \nonumber \\
		&\quad + \left(2\kappa_2^2 + 8\right)\left(\bar{F}_{2\mu\nu}\bar{F}\indices{_2^{\mu\nu}} \right)^2 + \left(\kappa_2^2 + 2\right)^2\bar{F}\indices{_2^{\mu\rho}}\bar{F}\indices{_2^\nu_\rho} \bar{F}\indices{_2_{\mu\sigma}}\bar{F}\indices{_2_\nu^\sigma}, \\[3pt]
		E\indices{^{\hat{h}}_{\hat{h}}}E\indices{^{\hat{h}}_{\hat{h}}} & =E\indices{^{\hat{h}}^{\hat{h}}}E\indices{^{\hat{h}}^{\hat{h}}}\mathcal{I}\indices{_{\hat{h}}_{\hat{h}}}\mathcal{I}\indices{_{\hat{h}}_{\hat{h}}} =4\Lambda^2, \\[3pt] E\indices{^{\tilde{\Phi}}_{\tilde{\Phi}}}E\indices{^{\tilde{\Phi}}_{\tilde{\Phi}}} &= E\indices{^{\tilde{\Phi}}^{\tilde{\Phi}}}E\indices{^{\tilde{\Phi}}^{\tilde{\Phi}}}\mathcal{I}\indices{_{\tilde{\Phi}}_{\tilde{\Phi}}}\mathcal{I}\indices{_{\tilde{\Phi}}_{\tilde{\Phi}}}= 0, \\[3pt]
		E\indices{^{\hat{h}_{\mu\nu}}_{\tilde{\Phi}}}E\indices{^{\tilde{\Phi}}_{\hat{h}_{\mu\nu}}} &= E\indices{^{\tilde{\Phi}}_{\hat{h}_{\mu\nu}}}E\indices{^{\hat{h}_{\mu\nu}}_{\tilde{\Phi}}} = E\indices{^{\hat{h}_{\mu\nu}}^{\tilde{\Phi}}}E\indices{^{\tilde{\Phi}}^{\hat{h}_{\alpha\beta}}}\mathcal{I}\indices{_{\tilde{\Phi}}_{\tilde{\Phi}}} \mathcal{I}\indices{_{\hat{h}_{\mu\nu}}_{\hat{h}_{\alpha\beta}}} = 0,\\[3pt]
		E\indices{^{a_{1\alpha}}_{\tilde{\Phi}}}E\indices{^{\tilde{\Phi}}_{a_{1\alpha}}} &=  E\indices{^{\tilde{\Phi}}_{a_{1\alpha}}}E\indices{^{a_{1\alpha}}_{\tilde{\Phi}}} = E\indices{^{a_{1\alpha}}^{\tilde{\Phi}}}E\indices{^{\tilde{\Phi}}^{a_{1\beta}}}\mathcal{I}\indices{_{a_{1\alpha}}_{a_{1\beta}}}\mathcal{I}\indices{_{\tilde{\Phi}}_{\tilde{\Phi}}} =0,\\[3pt]
		E\indices{^{a_{2\alpha}}_{\tilde{\Phi}}}E\indices{^{\tilde{\Phi}}_{a_{2\alpha}}} &=  E\indices{^{\tilde{\Phi}}_{a_{2\alpha}}}E\indices{^{a_{2\alpha}}_{\tilde{\Phi}}} = E\indices{^{a_{2\alpha}}^{\tilde{\Phi}}}E\indices{^{\tilde{\Phi}}^{a_{2\beta}}}\mathcal{I}\indices{_{a_{2\alpha}}_{a_{2\beta}}}\mathcal{I}\indices{_{\tilde{\Phi}}_{\tilde{\Phi}}} =0,\\[3pt]
		E\indices{^{\hat{h}_{\mu\nu}}_{\hat{h}}}E\indices{^{\hat{h}}_{\hat{h}_{\mu\nu}}} &= E\indices{^{\hat{h}}_{\hat{h}_{\mu\nu}}}E\indices{^{\hat{h}_{\mu\nu}}_{\hat{h}}} \nonumber \\
		&= E\indices{^{\hat{h}_{\mu\nu}}^{\hat{h}}}E\indices{^{\hat{h}}^{\hat{h}_{\alpha\beta}}}\mathcal{I}\indices{_{\hat{h}}_{\hat{h}}}\mathcal{I}\indices{_{\hat{h}_{\mu\nu}}_{\hat{h}_{\alpha\beta}}} \nonumber\\
		&=  \left(\bar{F}_{1\mu\nu}\bar{F}\indices{_1^{\mu\nu}} + \bar{F}_{2\mu\nu}\bar{F}\indices{_2^{\mu\nu}}\right)^2 -8 \bar{F}\indices{_1^{\mu\rho}}\bar{F}\indices{_2^\nu_\rho} \bar{F}\indices{_1_{\mu\sigma}}\bar{F}\indices{_2_\nu^\sigma}\nonumber \\
		&\quad -4\left(\bar{F}\indices{_1^{\mu\rho}}\bar{F}\indices{_1^\nu_\rho} \bar{F}\indices{_1_{\mu\sigma}}\bar{F}\indices{_1_\nu^\sigma} + \bar{F}\indices{_2^{\mu\rho}}\bar{F}\indices{_2^\nu_\rho} \bar{F}\indices{_2_{\mu\sigma}}\bar{F}\indices{_2_\nu^\sigma}\right), \\[3pt]
		E\indices{^{a_{1\alpha}}_{a_{2\beta}}}E\indices{^{a_{2\beta}}_{a_{1\alpha}}}&=  E\indices{^{a_{2\alpha}}_{a_{1\beta}}}E\indices{^{a_{1\beta}}_{a_{2\alpha}}} \nonumber \\
		&= E\indices{^{a_{1\alpha}}^{a_{2\gamma}}}E\indices{^{a_{2\beta}}^{a_{1\delta}}}\mathcal{I}\indices{_{a_{1\alpha}}_{a_{1\delta}}}\mathcal{I}\indices{_{a_{2\beta}}_{a_{2\gamma}}}\nonumber \\
		&= 2\left(\kappa_1\kappa_2 + 4\right)\bar{F}_{1\mu\nu}\bar{F}\indices{_1^{\mu\nu}}\bar{F}_{2\rho\sigma}\bar{F}\indices{_2^{\rho\sigma}}\nonumber \\
		&\quad + \left(\kappa_1^2\kappa_2^2 + 2\kappa_1\kappa_2 + 2\right)\bar{F}\indices{_1^{\mu\rho}}\bar{F}\indices{_2^\nu_\rho} \bar{F}\indices{_1_{\mu\sigma}}\bar{F}\indices{_2_\nu^\sigma}\nonumber \\
		&\quad + \left(2\kappa_1\kappa_2 + 2\right)^2\bar{F}\indices{_1^{\mu\rho}}\bar{F}\indices{_2^\nu^\sigma} \bar{F}\indices{_1_{\mu\nu}}\bar{F}\indices{_2_\rho_\sigma}, \\[3pt]
		E\indices{^{\hat{h}_{\mu\nu}}_{a_{1\alpha}}}E\indices{^{a_{1\alpha}}_{\hat{h}_{\mu\nu}}} &= E\indices{^{a_{1\alpha}}_{\hat{h}_{\mu\nu}}}E\indices{^{\hat{h}_{\mu\nu}}_{a_{1\alpha}}} \nonumber \\
		&=E\indices{^{\hat{h}_{\mu\nu}}^{a_{1\beta}}}E\indices{^{a_{1\alpha}}^{\hat{h}_{\gamma\delta}}}\mathcal{I}\indices{_{a_{1\alpha}}_{a_{1\beta}}}\mathcal{I}\indices{_{\hat{h}_{\mu\nu}}_{\hat{h}_{\gamma\delta}}} \nonumber\\
		&= \left(D_\rho \bar{F}_{1\mu\nu}\right)\left(D^\rho \bar{F}\indices{_1^{\mu\nu}}\right) + \left(D_\mu \bar{F}\indices{_1_\rho^\nu}\right)\left(D_\nu \bar{F}\indices{_1^{\rho\mu}}\right), \\[3pt]
		E\indices{^{\hat{h}_{\mu\nu}}_{a_{2\alpha}}}E\indices{^{a_{2\alpha}}_{\hat{h}_{\mu\nu}}} &= E\indices{^{a_{2\alpha}}_{\hat{h}_{\mu\nu}}}E\indices{^{\hat{h}_{\mu\nu}}_{a_{2\alpha}}} \nonumber \\
		&=E\indices{^{\hat{h}_{\mu\nu}}^{a_{2\beta}}}E\indices{^{a_{2\alpha}}^{\hat{h}_{\gamma\delta}}}\mathcal{I}\indices{_{a_{2\alpha}}_{a_{2\beta}}}\mathcal{I}\indices{_{\hat{h}_{\mu\nu}}_{\hat{h}_{\gamma\delta}}} \nonumber\\
		&= \left(D_\rho \bar{F}_{2\mu\nu}\right)\left(D^\rho \bar{F}\indices{_2^{\mu\nu}}\right) + \left(D_\mu \bar{F}\indices{_2_\rho^\nu}\right)\left(D_\nu \bar{F}\indices{_2^{\rho\mu}}\right).
\end{align}}
We also obtain,
{\allowdisplaybreaks
	\begin{align}\label{calcul9}
		{\left(\Omega_{\rho\sigma}\right)}\indices{^{\hat{h}_{\mu\nu}}_{\hat{h}_{\alpha\beta}}}{\left(\Omega^{\rho\sigma}\right)}\indices{^{\hat{h}_{\alpha\beta}}_{\hat{h}_{\mu\nu}}} &= {\left(\Omega_{\rho\sigma}\right)}\indices{^{\hat{h}_{\mu\nu}}^{\hat{h}_{\gamma\delta}}}{\left(\Omega^{\rho\sigma}\right)}\indices{^{\hat{h}_{\alpha\beta}}^{\hat{h}_{\theta\phi}}}\mathcal{I}\indices{_{\hat{h}_{\alpha\beta}}_{\hat{h}_{\gamma\delta}}}\mathcal{I}\indices{_{\hat{h}_{\mu\nu}}_{\hat{h}_{\theta\phi}}} \nonumber \\
		& =  -6R_{\mu\nu\rho\sigma}R^{\mu\nu\rho\sigma} -36\left(\bar{F}_{1\mu\nu}\bar{F}\indices{_1^{\mu\nu}}\right)^2-36\left(\bar{F}_{2\mu\nu}\bar{F}\indices{_2^{\mu\nu}}\right)^2 \nonumber\\
		&\quad -24\bar{F}_{1\mu\nu}\bar{F}\indices{_1^{\mu\nu}}\bar{F}_{2\rho\sigma}\bar{F}\indices{_2^{\rho\sigma}} -48 \bar{F}_{1\mu\nu}\bar{F}\indices{_2^{\mu\nu}}\bar{F}_{1\rho\sigma}\bar{F}\indices{_2^{\rho\sigma}} \nonumber\\
		&\quad -4 \left(R_{\mu\nu\rho\sigma} - 2R_{\mu\rho\nu\sigma}\right)\left(\bar{F}\indices{_1^{\mu\nu}}\bar{F}\indices{_1^{\rho\sigma}} + \bar{F}\indices{_2^{\mu\nu}}\bar{F}\indices{_2^{\rho\sigma}}\right) \nonumber\\
		&\quad +4R \left(\bar{F}_{1\mu\nu}\bar{F}\indices{_1^{\mu\nu}} + \bar{F}_{2\mu\nu}\bar{F}\indices{_2^{\mu\nu}} \right) + 96 \bar{F}\indices{_1^{\mu\rho}}\bar{F}\indices{_2^\nu_\rho} \bar{F}\indices{_1_{\mu\sigma}}\bar{F}\indices{_2_\nu^\sigma}\nonumber \\
		&\quad + 48 \left(\bar{F}\indices{_1^{\mu\rho}}\bar{F}\indices{_1^\nu_\rho} \bar{F}\indices{_1_{\mu\sigma}}\bar{F}\indices{_1_\nu^\sigma} + \bar{F}\indices{_2^{\mu\rho}}\bar{F}\indices{_2^\nu_\rho} \bar{F}\indices{_2_{\mu\sigma}}\bar{F}\indices{_2_\nu^\sigma}\right), \\[5pt]
		{\left(\Omega_{\rho\sigma}\right)}\indices{^{a_{1\alpha}}_{a_{1\beta}}}{\left(\Omega^{\rho\sigma}\right)}\indices{^{a_{1\beta}}_{a_{1\alpha}}}& = {\left(\Omega_{\rho\sigma}\right)}\indices{^{a_{1\alpha}}^{a_{1\gamma}}}{\left(\Omega^{\rho\sigma}\right)}\indices{^{a_{1\beta}}^{a_{1\delta}}}\mathcal{I}\indices{_{a_{1\beta}}_{a_{1\gamma}}}\mathcal{I}\indices{_{a_{1\alpha}}_{a_{1\delta}}}\nonumber\\ 
		&=  - R_{\mu\nu\rho\sigma}R^{\mu\nu\rho\sigma} -\left(2\kappa_1^4 - 16\kappa_1^2 + 52\right)\left(\bar{F}_{1\mu\nu}\bar{F}\indices{_1^{\mu\nu}}\right)^2  \nonumber \\
		&\quad + 8 R_{\mu\nu\rho\sigma}\bar{F}\indices{_1^{\mu\nu}}\bar{F}\indices{_1^{\rho\sigma}} + \left(4\kappa_1^2 - 16\right)R_{\mu\rho\nu\sigma}\bar{F}\indices{_1^{\mu\nu}}\bar{F}\indices{_1^{\rho\sigma}} \nonumber \\
		&\quad + 8{R}_{\mu\nu}\bar{F}\indices{_1^{\mu\rho}}\bar{F}\indices{_1^\nu_\rho} + \left(2\kappa_1^4 - 40\kappa_1^2 + 88\right)\bar{F}\indices{_1^{\mu\rho}}\bar{F}\indices{_1^\nu_\rho} \bar{F}\indices{_1_{\mu\sigma}}\bar{F}\indices{_1_\nu^\sigma}, \\
		{\left(\Omega_{\rho\sigma}\right)}\indices{^{a_{2\alpha}}_{a_{2\beta}}}{\left(\Omega^{\rho\sigma}\right)}\indices{^{a_{2\beta}}_{a_{2\alpha}}}& = {\left(\Omega_{\rho\sigma}\right)}\indices{^{a_{2\alpha}}^{a_{2\gamma}}}{\left(\Omega^{\rho\sigma}\right)}\indices{^{a_{2\beta}}^{a_{2\delta}}}\mathcal{I}\indices{_{a_{2\beta}}_{a_{2\gamma}}}\mathcal{I}\indices{_{a_{2\alpha}}_{a_{2\delta}}}\nonumber\\ 
		&=  - R_{\mu\nu\rho\sigma}R^{\mu\nu\rho\sigma} -\left(2\kappa_2^4 - 16\kappa_2^2 + 52\right)\left(\bar{F}_{2\mu\nu}\bar{F}\indices{_2^{\mu\nu}}\right)^2  \nonumber \\
		&\quad + 8 R_{\mu\nu\rho\sigma}\bar{F}\indices{_2^{\mu\nu}}\bar{F}\indices{_2^{\rho\sigma}} + \left(4\kappa_2^2 - 16\right)R_{\mu\rho\nu\sigma}\bar{F}\indices{_2^{\mu\nu}}\bar{F}\indices{_2^{\rho\sigma}} \nonumber \\
		&\quad + 8{R}_{\mu\nu}\bar{F}\indices{_2^{\mu\rho}}\bar{F}\indices{_2^\nu_\rho} + \left(2\kappa_2^4 - 40\kappa_2^2 + 88\right)\bar{F}\indices{_2^{\mu\rho}}\bar{F}\indices{_2^\nu_\rho} \bar{F}\indices{_2_{\mu\sigma}}\bar{F}\indices{_2_\nu^\sigma},\\[3pt]
		{\left(\Omega_{\rho\sigma}\right)}\indices{^{\tilde{\Phi}}_{\tilde{\Phi}}}{\left(\Omega^{\rho\sigma}\right)}\indices{^{\tilde{\Phi}}_{\tilde{\Phi}}} &= {\left(\Omega_{\rho\sigma}\right)}\indices{^{\tilde{\Phi}}^{\tilde{\Phi}}}{\left(\Omega^{\rho\sigma}\right)}\indices{^{\tilde{\Phi}}^{\tilde{\Phi}}}\mathcal{I}\indices{_{\tilde{\Phi}}_{\tilde{\Phi}}}\mathcal{I}\indices{_{\tilde{\Phi}}_{\tilde{\Phi}}}= 0, \\[5pt]
		{\left(\Omega_{\rho\sigma}\right)}\indices{^{a_{1\alpha}}_{a_{2\beta}}}{\left(\Omega^{\rho\sigma}\right)}\indices{^{a_{2\beta}}_{a_{1\alpha}}}&=  {\left(\Omega_{\rho\sigma}\right)}\indices{^{a_{2\alpha}}_{a_{1\beta}}}{\left(\Omega^{\rho\sigma}\right)}\indices{^{a_{1\beta}}_{a_{2\alpha}}} \nonumber \\
		&={\left(\Omega_{\rho\sigma}\right)}\indices{^{a_{1\alpha}}^{a_{2\gamma}}}{\left(\Omega^{\rho\sigma}\right)}\indices{^{a_{2\beta}}^{a_{1\delta}}}\mathcal{I}\indices{_{a_{1\alpha}}_{a_{1\delta}}}\mathcal{I}\indices{_{a_{2\beta}}_{a_{2\gamma}}}\nonumber \\
		&= \left(2\kappa_1^2\kappa_2^2 - 44\kappa_1\kappa_2 + 100\right)\bar{F}\indices{_1^{\mu\rho}}\bar{F}\indices{_2^\nu_\rho} \bar{F}\indices{_1_{\mu\sigma}}\bar{F}\indices{_2_\nu^\sigma}\nonumber \\
		&\quad + \left(4\kappa_1\kappa_2 -12 \right)\bar{F}\indices{_1^{\mu\rho}}\bar{F}\indices{_2^\nu^\sigma} \bar{F}\indices{_1_{\mu\nu}}\bar{F}\indices{_2_\rho_\sigma}\nonumber \\
		&\quad -\left(2\kappa_1^2\kappa_2^2 - 12\kappa_1\kappa_2 + 28\right)\bar{F}_{1\mu\nu}\bar{F}\indices{_1^{\mu\nu}}\bar{F}_{2\rho\sigma}\bar{F}\indices{_2^{\rho\sigma}}\nonumber \\
		&\quad +\left(4\kappa_1\kappa_2 - 24\right)\bar{F}_{1\mu\nu}\bar{F}\indices{_1^{\mu\nu}}\bar{F}_{2\rho\sigma}\bar{F}\indices{_2^{\rho\sigma}}, \\[3pt]
		{\left(\Omega_{\rho\sigma}\right)}\indices{^{\hat{h}_{\mu\nu}}_{\tilde{\Phi}}}{\left(\Omega^{\rho\sigma}\right)}\indices{^{\tilde{\Phi}}_{\hat{h}_{\mu\nu}}} &= {\left(\Omega_{\rho\sigma}\right)}\indices{^{\tilde{\Phi}}_{\hat{h}_{\mu\nu}}}{\left(\Omega^{\rho\sigma}\right)}\indices{^{\hat{h}_{\mu\nu}}_{\tilde{\Phi}}} \nonumber \\
		&={\left(\Omega_{\rho\sigma}\right)}\indices{^{\hat{h}_{\mu\nu}}^{\tilde{\Phi}}}{\left(\Omega^{\rho\sigma}\right)}\indices{^{\tilde{\Phi}}^{\hat{h}_{\gamma\delta}}}\mathcal{I}\indices{_{\tilde{\Phi}}_{\tilde{\Phi}}}\mathcal{I}\indices{_{\hat{h}_{\mu\nu}}_{\hat{h}_{\gamma\delta}}} \nonumber\\
		&= 2\kappa_1^2\left(\bar{F}_{1\mu\nu}\bar{F}\indices{_1^{\mu\nu}}\right)^2 + 2\kappa_2^2\left(\bar{F}_{2\mu\nu}\bar{F}\indices{_2^{\mu\nu}}\right)^2 + 4\bar{F}_{1\mu\nu}\bar{F}\indices{_1^{\mu\nu}}\bar{F}_{2\rho\sigma}\bar{F}\indices{_2^{\rho\sigma}}\nonumber\\
		&\quad -8 \kappa_1^2\bar{F}\indices{_1^{\mu\rho}}\bar{F}\indices{_1^\nu_\rho} \bar{F}\indices{_1_{\mu\sigma}}\bar{F}\indices{_1_\nu^\sigma} -8 \kappa_2^2\bar{F}\indices{_2^{\mu\rho}}\bar{F}\indices{_2^\nu_\rho} \bar{F}\indices{_2_{\mu\sigma}}\bar{F}\indices{_2_\nu^\sigma}\nonumber\\
		&\quad -16 \kappa_1\kappa_2\bar{F}\indices{_1^{\mu\rho}}\bar{F}\indices{_2^\nu_\rho} \bar{F}\indices{_1_{\mu\sigma}}\bar{F}\indices{_2_\nu^\sigma}, \\[5pt]
		{\left(\Omega_{\rho\sigma}\right)}\indices{^{\hat{h}_{\mu\nu}}_{a_{1\alpha}}}{\left(\Omega^{\rho\sigma}\right)}\indices{^{a_{1\alpha}}_{\hat{h}_{\mu\nu}}} &= {\left(\Omega_{\rho\sigma}\right)}\indices{^{a_{1\alpha}}_{\hat{h}_{\mu\nu}}}{\left(\Omega^{\rho\sigma}\right)}\indices{^{\hat{h}_{\mu\nu}}_{a_{1\alpha}}} \nonumber \\
		&={\left(\Omega_{\rho\sigma}\right)}\indices{^{\hat{h}_{\mu\nu}}^{a_{1\beta}}}{\left(\Omega^{\rho\sigma}\right)}\indices{^{a_{1\alpha}}^{\hat{h}_{\gamma\delta}}}\mathcal{I}\indices{_{a_{1\alpha}}_{a_{1\beta}}}\mathcal{I}\indices{_{\hat{h}_{\mu\nu}}_{\hat{h}_{\gamma\delta}}} \nonumber\\
		&= 2\left(D_\mu \bar{F}\indices{_1_\rho^\nu}\right)\left(D_\nu \bar{F}\indices{_1^{\rho\mu}}\right) -10\left(D_\rho \bar{F}_{1\mu\nu}\right)\left(D^\rho \bar{F}\indices{_1^{\mu\nu}}\right), \\[5pt]
		{\left(\Omega_{\rho\sigma}\right)}\indices{^{\hat{h}_{\mu\nu}}_{a_{2\alpha}}}{\left(\Omega^{\rho\sigma}\right)}\indices{^{a_{2\alpha}}_{\hat{h}_{\mu\nu}}} &= {\left(\Omega_{\rho\sigma}\right)}\indices{^{a_{2\alpha}}_{\hat{h}_{\mu\nu}}}{\left(\Omega^{\rho\sigma}\right)}\indices{^{\hat{h}_{\mu\nu}}_{a_{2\alpha}}} \nonumber \\
		&={\left(\Omega_{\rho\sigma}\right)}\indices{^{\hat{h}_{\mu\nu}}^{a_{2\beta}}}{\left(\Omega^{\rho\sigma}\right)}\indices{^{a_{2\alpha}}^{\hat{h}_{\gamma\delta}}}\mathcal{I}\indices{_{a_{2\alpha}}_{a_{2\beta}}}\mathcal{I}\indices{_{\hat{h}_{\mu\nu}}_{\hat{h}_{\gamma\delta}}} \nonumber\\
		&= 2\left(D_\mu \bar{F}\indices{_1_\rho^\nu}\right)\left(D_\nu \bar{F}\indices{_1^{\rho\mu}}\right) -10\left(D_\rho \bar{F}_{1\mu\nu}\right)\left(D^\rho \bar{F}\indices{_1^{\mu\nu}}\right), \\[5pt]
		{\left(\Omega_{\rho\sigma}\right)}\indices{^{a_{1\alpha}}_{\tilde{\Phi}}}{\left(\Omega^{\rho\sigma}\right)}\indices{^{\tilde{\Phi}}_{a_{1\alpha}}} &=  {\left(\Omega_{\rho\sigma}\right)}\indices{^{\tilde{\Phi}}_{a_{1\alpha}}}{\left(\Omega^{\rho\sigma}\right)}\indices{^{a_{1\alpha}}_{\tilde{\Phi}}}\nonumber\\
		&= {\left(\Omega_{\rho\sigma}\right)}\indices{^{a_{1\alpha}}^{\tilde{\Phi}}}{\left(\Omega^{\rho\sigma}\right)}\indices{^{\tilde{\Phi}}^{a_{1\beta}}}\mathcal{I}\indices{_{a_{1\alpha}}_{a_{1\beta}}}\mathcal{I}\indices{_{\tilde{\Phi}}_{\tilde{\Phi}}} \nonumber\\
		&= 2\kappa_1^2\left(D_\mu \bar{F}\indices{_1_\rho^\nu}\right)\left(D_\nu \bar{F}\indices{_1^{\rho\mu}}\right)- 2\kappa_1^2\left(D_\rho \bar{F}_{1\mu\nu}\right)\left(D^\rho \bar{F}\indices{_1^{\mu\nu}}\right), \\[5pt]
		{\left(\Omega_{\rho\sigma}\right)}\indices{^{a_{2\alpha}}_{\tilde{\Phi}}}{\left(\Omega^{\rho\sigma}\right)}\indices{^{\tilde{\Phi}}_{a_{2\alpha}}} &=  {\left(\Omega_{\rho\sigma}\right)}\indices{^{\tilde{\Phi}}_{a_{2\alpha}}}{\left(\Omega^{\rho\sigma}\right)}\indices{^{a_{2\alpha}}_{\tilde{\Phi}}}\nonumber\\
		&= {\left(\Omega_{\rho\sigma}\right)}\indices{^{a_{2\alpha}}^{\tilde{\Phi}}}{\left(\Omega^{\rho\sigma}\right)}\indices{^{\tilde{\Phi}}^{a_{2\beta}}}\mathcal{I}\indices{_{a_{2\alpha}}_{a_{2\beta}}}\mathcal{I}\indices{_{\tilde{\Phi}}_{\tilde{\Phi}}} \nonumber\\
		&= 2\kappa_2^2\left(D_\mu \bar{F}\indices{_2_\rho^\nu}\right)\left(D_\nu \bar{F}\indices{_2^{\rho\mu}}\right)- 2\kappa_2^2\left(D_\rho \bar{F}_{2\mu\nu}\right)\left(D^\rho \bar{F}\indices{_2^{\mu\nu}}\right).
\end{align}}
Finally, we utilize all the trace data mentioned above in the definitions \eqref{calcul6a}, \eqref{calcul6b}, and \eqref{calcul6c}. Subsequently, we simplify the results and background invariants with the assistance of the one-shell identities discussed in \cref{iden}. This process essentially yields the irreducible form of trace relations, as presented in \cref{sdc15a,sdc15b,sdc15c}, respectively.

\section{Integrated background invariants in extremal near-horizon limit}\label{enhii}
This section lists the explicit forms of the background curvature invariants $W^2=W_{\mu\nu\rho\sigma}W^{\mu\nu\rho\sigma}$, $E_4$, $R^2$ and $R\bar{F}_{\mu\nu}\bar{F}^{\mu\nu}$ their integrations for the extremal near-horizon (ENH) backgrounds of Kerr-Newman-AdS$_4$, Reissner-Nordstr\"om-AdS$_4$ and Kerr-AdS$_4$ black holes as well as their asymptotically-flat counterparts considered in this paper. These results prove to be essential for the analysis of \cref{elimit,flimit} to derive the $\mathcal{C}_{\text{local}}$ contributions in the extremal limit of the black holes.

\subsection*{Kerr-Newman-AdS$_4$ ENH background}
For the specific structure \eqref{el4} of the ENH geometry of the Kerr-Newman-AdS$_4$ black hole, the necessary four-derivative invariants can be derived in terms of the independent parameters $\left\lbrace r_0, \ell, a\right\rbrace$. Here, the extremality condition is imposed on $\left(r_0, \ell_2\right)$ or $\left(a, q\right)$ via the following constraints
\begin{align}
	\begin{gathered}
		a^2+ q^2  = \frac{r_0^2}{\ell^2}\left(\ell^2 + a^2 + 3r_0^2\right),	\\
		\ell_2^2 =  {\frac{\ell^2\left(a^2 + r_0^2\right)}{\left(a^2 + \ell^2 + 6r_0^2\right)}} =  \frac{\ell^2\left(2r_0^4 + \ell^2 \left(2r_0^2- q^2\right)\right)}{\left(\ell^4 + \ell^2\left(6r_0^2-q^2\right)-3r_0^4\right)}.
	\end{gathered}
\end{align}

Then the ENH invariants take the following forms
{\allowdisplaybreaks
	\begin{subequations}\label{knai}
		\begin{align}
			W^2 &= \frac{48}{\ell ^4 \rho_0^{12}} \bigg[r_0^4 \left(r_0^4-a^2 \ell ^2\right)^2 + a^8\ell^4\cos ^4\theta - a^2r_0^2\Big\lbrace a^4 \left(a^2+2 r_0^2+\ell ^2\right)^2\cos^4\theta \nonumber\\
			&  - a^2\Big(16 r_0^4 \left(a^2+\ell ^2\right)+8 a^2 \ell ^2 \left(a^2+\ell ^2\right)+2 r_0^2 \left(3 a^4+13 a^2 \ell ^2+3 \ell ^4\right)\nonumber \\
			& +9 r_0^6\Big)\cos^2\theta +6 a^4 \ell ^4-4 r_0^6 \left(a^2+\ell ^2\right)+8 a^2 r_0^2 \ell ^2 \left(a^2+\ell ^2\right)\nonumber \\
			&+r_0^4 \left(a^4+6 a^2 \ell ^2+\ell ^4\right)-6 r_0^8\Big\rbrace\cos^2\theta \bigg], \\[5pt]
			E_4 & = \frac{8 }{\ell ^4 \rho_0^{12}}\bigg[a^4 \cos ^4\theta \Big(3a^4\left( a^{4} \cos ^{4}\theta+6 a^{2} r_0^2 \cos ^{2}\theta + 15 r_0^4 \right)\cos ^4\theta \nonumber\\
			& + 5 \big\lbrace a^4 \left(10 r_0^2 \ell ^2+7 r_0^4+\ell ^4\right)+2 a^2 \left(5 r_0^2 \ell ^4+16 r_0^4 \ell ^2+9 r_0^6\right) \nonumber\\
			& + r_0^4\left(7 \ell ^4+18 r_0^2 \ell ^2+18 r_0^4\right) \big\rbrace -6 a^2 r_0^2 \big\lbrace a^4+2 a^2 \left(2 r_0^2+\ell ^2\right) \nonumber\\
			& +4 r_0^2 \ell ^2-6 r_0^4+\ell ^4\big\rbrace\cos ^2\theta \Big) -2 a^2 r_0^2  \Big(a^4 \left(22 r_0^2 \ell ^2+4 r_0^4+19 \ell ^4\right) \nonumber\\
			& +2 a^2 r_0^2 \left(11 \ell ^4+7 r_0^2 \ell ^2-3 r_0^4\right)+ 2r_0^4 \left(2 \ell^4 -3 r_0^2 \ell ^2-9 r_0^4\right)\Big)\cos ^2\theta \nonumber\\
			& -r_0^4 \Big(a^4 \left(r_0^4 -2 r_0^2 \ell ^2 -5 \ell ^4\right)+a^2 r_0^2\left(8 r_0^2 \ell ^2+6 r_0^4-2 \ell ^4\right)+r_0^4 \ell ^2 \left(6 r_0^2+\ell ^2\right)\Big)  \bigg],\\[5pt]			
			R^2 & = \frac{144}{\ell^4},\\[5pt]
			R\bar{F}_{\mu\nu}\bar{F}^{\mu\nu} &= \frac{12}{\ell ^4 \rho_0^{8}}\Big(3r_0^4 + \left(a^2 + \ell^2\right)r_0^2 - a^2\ell^2\Big) \Big(r_0^4 - 6a^2r_0^2\cos^2\theta + a^4\cos^4\theta\Big),
		\end{align}
\end{subequations}  }
where $\rho_0^2 = r_0^2 + a^2 \cos^2 \theta$. Next, the integration results of above background invariants over the desired part of ENH geometry, as detailed in the formula \eqref{el9}, are successively obtained as 
{\allowdisplaybreaks
	\begin{subequations}\label{knaii}
		\begin{align}
			\frac{\left(-2\pi\right)}{\left(4\pi\right)^2}\int_{\text{ENH}}\hspace{-0.10in} \mathrm{d}^2y \,G(y)W^2 &= \frac{\ell_2^2}{8 a^4\Xi \ell^4 r_0^5\left(a^2+r_0^2\right)^2} \bigg[ a^2r_0 \Big(9a^{8}\left(2r_0^2\ell^2 - r_0^4 -\ell^4\right)\nonumber\\
			&  +a^6 r_0^2 \left(70 r_0^2 \ell ^2-135 r_0^4+\ell ^4\right) +2 a^4 r_0^4 \left(106 r_0^2 \ell ^2-77 r_0^4+ 27\ell ^4\right) \nonumber\\
			& + 6a^2 r_0^6 \left(18 r_0^2 \ell ^2  - 13 r_0^4 + 3 \ell ^4\right)- 3 r_0^8 \left(2 r_0^2 \ell ^2 + 15 r_0^4 - \ell ^4\right) \Big)  \nonumber\\
			& -3 r_0^{11} \left(3 r_0^2+\ell ^2\right)^2 -12 a^3 \left(a^2+r_0^2\right)^2 \Big(a^2 \left(r_0^2-\ell ^2\right) \nonumber\\
			& +r_0^2 \left(3 r_0^2+\ell ^2\right)\Big)^2 \arctan\left(\frac{a}{r_0}\right) \bigg],\\[5pt]
			\frac{\left(-2\pi\right)}{\left(4\pi\right)^2}\int_{\text{ENH}} \mathrm{d}^2y \,G(y)E_4 & = 4,\\[5pt]			
			\frac{\left(-2\pi\right)}{\left(4\pi\right)^2}\int_{\text{ENH}} \mathrm{d}^2y \,G(y)R^2 & = -\frac{24\ell_2^2}{\Xi \ell^4 \left(a^2+r_0^2\right)}\left(a^4+4 a^2 r_0^2+3 r_0^4\right),\\[5pt]
			\frac{\left(-2\pi\right)}{\left(4\pi\right)^2}\int_{\text{ENH}}\hspace{-0.10in} \mathrm{d}^2y \,G(y)R\bar{F}_{\mu\nu}\bar{F}^{\mu\nu} &= \frac{12\ell_2^2\left(a^2-r_0^2\right)}{\Xi \ell^4 \left(a^2+r_0^2\right)^2} \Big(r_0^2\left(3 r_0^2 + \ell^2\right) - a^2 \left(\ell^2 - r_0^2\right)\Big),
		\end{align}
\end{subequations}  }
where $\Xi = \frac{\left(\ell^2 - a^2\right)}{\ell^2}$. The above extremal near-horizon integrations over $y$ coordinates must be performed in the range $ 0 \leq \theta \leq \pi$ and $0 \leq \tilde{\phi} \leq 2\pi$ for the following typical form of the ENH metric function, 
\begin{align}
	G(y) = \frac{\ell ^2 \ell _2^2 }{(\ell ^2-a^2)}\left(r_0^2 + a^2\cos^2\theta\right) \sin \theta.
\end{align}  
All the derived results \eqref{knaii} can be employed to calculate the $\mathcal{C}_{\text{local}}$ contribution for the extremal Kerr-Newman-AdS$_4$ black hole, as listed in \cref{el10}.


\subsection*{Kerr-AdS$_4$ ENH background}
For a similar analysis of the Kerr-AdS$_4$ black hole, one needs to consider the special case $q=0$ of the ENH geometry \eqref{el4}. This adjustment further revises all extremal charges and parameters, as indicated by the following bounds 
\begin{align}
	a^2 = \frac{r_0^2\left(3r_0^2 + \ell^2\right)}{\left(\ell^2 - r_0^2\right)}, \quad \ell_2^2 = \frac{2\ell^2r_0^2 \left(r_0^2 + \ell^2\right)}{\left(\ell^4 + 6\ell^2r_0^2 - 3r_0^4\right)},
\end{align}  
where one must consider $r_0 < \ell $. We can now express all the background invariants over the extremal ENH geometry of the Kerr-AdS$_4$ black hole in much simpler forms,
{\allowdisplaybreaks
	\begin{subequations}\label{kai}
		\begin{align}
			W_{\mu\nu\rho\sigma}W^{\mu\nu\rho\sigma} &= \frac{12}{\ell ^4 \rho_0^{8}r_0^2}\left(r_0^2 + a^2\right)^2\left(r_0^2 + \ell^2\right)^2 \Big(r_0^6 - a^2\cos^2\theta\left(15r_0^4 - 15 a^2r_0^2\cos^2\theta + a^4\cos^4\theta\right)\Big), \\[5pt]
			E_4 & = \frac{24}{\ell^2} + W_{\mu\nu\rho\sigma}W^{\mu\nu\rho\sigma},\\[5pt]			
			R^2 & = \frac{144}{\ell^4}, \quad
			R\bar{F}_{\mu\nu}\bar{F}^{\mu\nu} = 0.
		\end{align}
\end{subequations}  } 
Subsequently, the ENH integrations of the above curvature invariants to compute the $\mathcal{C}_{\text{local}}$ contribution \eqref{el11} are derived as
{\allowdisplaybreaks
	\begin{subequations}\label{kaii}
		\begin{align}
			\frac{\left(-2\pi\right)}{\left(4\pi\right)^2}\int_{\text{ENH}} \mathrm{d}^2y \,G(y)W_{\mu\nu\rho\sigma}W^{\mu\nu\rho\sigma} &= 4 - \frac{2 \ell_2^2 \left(\ell ^2 -a^2-3 r_0^2 \right)}{\ell^2\left(\ell ^2-a^2\right)},\\[5pt]
			\frac{\left(-2\pi\right)}{\left(4\pi\right)^2}\int_{\text{ENH}} \mathrm{d}^2y \,G(y)E_4 & = 4,\\[5pt]			
			\frac{\left(-2\pi\right)}{\left(4\pi\right)^2}\int_{\text{ENH}} \mathrm{d}^2y \,G(y)R^2 & = \frac{12 \ell_2^2 \left(\ell ^2 -a^2-3 r_0^2 \right)}{\ell^2\left(\ell ^2-a^2\right)},\\[5pt]
			\frac{\left(-2\pi\right)}{\left(4\pi\right)^2}\int_{\text{ENH}} \mathrm{d}^2y \,G(y)R\bar{F}_{\mu\nu}\bar{F}^{\mu\nu} &= 0.
		\end{align}
\end{subequations}}

\subsection*{Reissner-Nordstr\"om-AdS$_4$ ENH background}
The extremal near-horizon geometry of the electrically-charged Reissner-Nordstr\"om-AdS$_4$ black hole is obtained through appropriate coordinate transformations on the background metric \eqref{rn1},
\begin{align}
	r = r_0 + \lambda \tilde{r}, \quad  t =\frac{\ell_2^2}{\lambda} \tilde{t}.
\end{align}
Upon introducing the Euclideanized time $\tilde{t} \to -i\tilde{\tau}$ and taking the limit $\lambda \to 0$, the extremal near-horizon (ENH) geometry can be expressed in the desired form of AdS$_2\times S^2$,
\begin{align}\label{rnai1}
	\tilde{g}_{\mu\nu}\mathrm{d}\tilde{x}^\mu \mathrm{d}\tilde{x}^\nu = \ell_2^2 \left[\left(\tilde{r}^2-1\right)\mathrm{d}\tilde{\tau}^2 + \frac{\mathrm{d}\tilde{r}^2}{\left(\tilde{r}^2-1\right)}\right] + r_0^2 \left(\mathrm{d}\theta^2 + \sin^2\theta\mathrm{d}\phi^2\right),
\end{align}
Notably, for this non-rotating ($a=0$) charged background, the extremality bounds on the $r_0$ and $\ell_2$ parameters are given by,
\begin{align}
	q^2 = r_0^2\left(1 + \frac{3r_0^2}{\ell^2}\right), \quad \frac{1}{\ell_2^2} - \frac{1}{r_0^2} = \frac{6}{\ell^2},
\end{align}  
where we must operate in the range $r_0 > \ell_2 $. The curvature invariants around the ENH geometry \eqref{rnai1} are computed as follows  
{\allowdisplaybreaks
	\begin{subequations}\label{rnai}
		\begin{align}
			W_{\mu\nu\rho\sigma}W^{\mu\nu\rho\sigma} &= \frac{4\left(r_0^2 - \ell_2^2\right)^2}{3r_0^4\ell_2^4}, \\[5pt]
			E_4 & = - \frac{8}{\ell_2 r_0^2},\\[5pt]			
			R^2 & = \frac{4\left(r_0^2 - \ell_2^2\right)^2}{r_0^4\ell_2^4}, \\[5pt]
			R\bar{F}_{\mu\nu}\bar{F}^{\mu\nu} &= \frac{4q^2\left(r_0^2 - \ell_2\right)^2}{r_0^6\ell_2^2}.
		\end{align}
\end{subequations}  }
Finally, we carry out integrations of the mentioned ENH curvature invariants to derive the $\mathcal{C}_{\text{local}}$ relation, as outlined in \cref{el12}. The specific results are presented as
{\allowdisplaybreaks
	\begin{subequations}\label{rnaii}
		\begin{align}
			\frac{\left(-2\pi\right)}{\left(4\pi\right)^2}\int_{\text{ENH}} \mathrm{d}^2y \,G(y)W_{\mu\nu\rho\sigma}W^{\mu\nu\rho\sigma} &= \frac{4}{3} - \frac{2\left( r_0^4 + \ell_2^4\right)}{3r_0^2\ell_2^2},\\[5pt]
			\frac{\left(-2\pi\right)}{\left(4\pi\right)^2}\int_{\text{ENH}} \mathrm{d}^2y \,G(y)E_4 & = 4,\\[5pt]			
			\frac{\left(-2\pi\right)}{\left(4\pi\right)^2}\int_{\text{ENH}} \mathrm{d}^2y \,G(y)R^2 & = 4 - \frac{2\left(r_0^4 + \ell_2^4\right)}{r_0^2\ell_2^2},\\[5pt]
			\frac{\left(-2\pi\right)}{\left(4\pi\right)^2}\int_{\text{ENH}} \mathrm{d}^2y \,G(y)R\bar{F}_{\mu\nu}\bar{F}^{\mu\nu} &= -\frac{\left(r_0^4 -\ell_2^4 \right)}{r_0^2\ell_2^2}.
		\end{align}
\end{subequations}}

\subsection*{ENH backgrounds in flat limit ($\ell \to \infty$)}
We will now investigate the flat-space limit, $\ell \to \infty$, applied to the previously derived integrated extremal near-horizon (ENH) invariants of asymptotically-AdS$_4$ black hole configurations. This limit results in $R = -\frac{12}{\ell^2} = 0$, leading to similar integrated relations for the asymptotically-flat$_4$ black holes. In this flat-space scenario, the extremality conditions for the generic near-horizon parameters $r_0$ and $\ell_2$, as well as for the revised Kerr parameter $b = \frac{a}{q}$, can be expressed as follows
\begin{align}
	r_0^2 = q^2 + a^2, \quad \ell_2^2 = q^2 + 2a^2, \quad b^2 = \frac{r_0^2}{q^2}\left(\frac{\ell_2^2}{r_0^2} - 1\right).
\end{align} 
For the case of a flat extremal Kerr-Newman black hole solution in STU models, the non-vanishing ENH integrated invariants {are}
{\allowdisplaybreaks
	\begin{subequations}
		\begin{align}
			\frac{\left(-2\pi\right)}{\left(4\pi\right)^2}\int_{\text{ENH}} \mathrm{d}^2y \,G(y)W_{\mu\nu\rho\sigma}W^{\mu\nu\rho\sigma} &= \frac{1}{2(b^2+1)^2(2b^2+1)}\bigg[ \left(3 + 24b^2 + 40b^4 + 16b^6 \right)\nonumber\\
			&\quad -\frac{3(2b^2+1)^2}{b\sqrt{b^2 + 1}}\arctan\left(\frac{b}{\sqrt{b^2 + 1}}\right)\bigg], \\[5pt]
			\frac{\left(-2\pi\right)}{\left(4\pi\right)^2}\int_{\text{ENH}} \mathrm{d}^2y \,G(y)E_4 &= 4.
		\end{align}
\end{subequations}}
In the special case of an extremal Kerr black hole ($q=0$, $\ell_2 = \sqrt{2}r_0$, and $b \to \infty$), the same relations are give by
{\allowdisplaybreaks
	\begin{subequations}
		\begin{align}
			\frac{\left(-2\pi\right)}{\left(4\pi\right)^2}\int_{\text{ENH}} \mathrm{d}^2y \,G(y)W_{\mu\nu\rho\sigma}W^{\mu\nu\rho\sigma} &= 4, \\[5pt]
			\frac{\left(-2\pi\right)}{\left(4\pi\right)^2}\int_{\text{ENH}} \mathrm{d}^2y \,G(y)E_4 &= 4.
		\end{align}
\end{subequations}}
Similarly, for the asymptotically-flat extremal Reissner-Nordstr\"om black hole ($a=0$, $\ell_2 = r_0$, and $b = 0$) in STU models, one obtains
{\allowdisplaybreaks
	\begin{subequations}
		\begin{align}
			\frac{\left(-2\pi\right)}{\left(4\pi\right)^2}\int_{\text{ENH}} \mathrm{d}^2y \,G(y)W_{\mu\nu\rho\sigma}W^{\mu\nu\rho\sigma} &= 0, \\[5pt]
			\frac{\left(-2\pi\right)}{\left(4\pi\right)^2}\int_{\text{ENH}} \mathrm{d}^2y \,G(y)E_4 &= 4.
		\end{align}
\end{subequations}}
All the integrated results mentioned above are appropriately employed in deriving the $\mathcal{C}_{\text{local}}$ expressed in \cref{fl3,fl5,fl6}.



\nocite{*}


																						\end{document}